\documentclass[11pt,onecolumn,twoside]{IEEEtran}
\usepackage{graphicx, fullpage, cite, amsmath, amssymb, multirow, rotating, subfigure}

\title{Structural Properties of the\\ \emph{Caenorhabditis elegans} Neuronal Network}
\author{\IEEEauthorblockN{Lav R. Varshney\IEEEauthorrefmark{2}, 
Beth L. Chen\IEEEauthorrefmark{1}, Eric Paniagua\IEEEauthorrefmark{5},
David H. Hall\IEEEauthorrefmark{3}, and Dmitri B. Chklovskii\IEEEauthorrefmark{4}
\thanks{L.R.V.\ was supported by the NSF Graduate Research Fellowship, and NSF Grants 0325774 and 0836720.
B.L.C.\ was the recipient of an Arnold and Mabel Beckman Graduate Student Fellowship of the Watson School of Biological Sciences. 
D.B.C.\ was supported by National Institute of Mental Health Grant 69838, the Swartz Foundation and a Klingenstein Foundation Award. 
The Center for \emph{C.\ elegans} Anatomy is supported by National Institutes of Health Grant RR 12596 (to D.H.H.).}
}
\IEEEauthorblockA{\IEEEauthorrefmark{2}Massachusetts Institute of Technology}\\
\IEEEauthorblockA{\IEEEauthorrefmark{1}Cold Spring Harbor Laboratory}\\
\IEEEauthorblockA{\IEEEauthorrefmark{5}California Institute of Technology}\\
\IEEEauthorblockA{\IEEEauthorrefmark{3}Albert Einstein College of Medicine}\\
\IEEEauthorblockA{\IEEEauthorrefmark{4}Janelia Farm Research Campus, Howard Hughes Medical Institute}}

\begin{document}

\maketitle

\begin{abstract}

Despite recent interest in reconstructing neuronal networks, complete wiring diagrams 
on the level of individual synapses remain scarce and the insights into function they 
can provide remain unclear. Even for \emph{Caenorhabditis elegans}, whose neuronal 
network is relatively small and stereotypical from animal to animal, published wiring 
diagrams are neither accurate nor complete and self-consistent. 
Using materials from White \emph{et al}.\ and new electron micrographs we assemble whole, 
self-consistent gap junction and chemical synapse networks of hermaphrodite \emph{C.\ elegans}.  
We propose a method to visualize the wiring diagram, which reflects network signal flow. 
We calculate statistical and topological properties of the network, such as 
degree distributions, synaptic multiplicities, and small-world properties, that help in 
understanding network signal propagation. We identify neurons that may play 
central roles in information processing and network motifs that could serve as functional 
modules of the network.  We explore propagation of neuronal activity in response to 
sensory or artificial stimulation using linear systems theory and find several activity 
patterns that could serve as substrates of previously described behaviors.  Finally, we 
analyze the interaction between the gap junction and the chemical synapse networks. 
Since several statistical properties of the \emph{C.\ elegans} network, such as multiplicity 
and motif distributions are similar to those found in mammalian neocortex, they likely 
point to general principles of neuronal networks. The wiring diagram reported here can 
help in understanding the mechanistic basis of behavior by generating predictions about 
future experiments involving genetic perturbations, laser ablations, or monitoring 
propagation of neuronal activity in response to stimulation.
\end{abstract}

\section*{Introduction}
Determining and examining base sequences in genomes \cite{Adams_ea2000,Lander_ea2001} has 
revolutionized molecular biology. Similarly, decoding and analyzing connectivity patterns among neurons in nervous systems, 
the aim of the emerging field of connectomics \cite{BriggmanD2006,Smith2007,LichtmanLS2008,SpornsTK2005}, may make a major impact on neurobiology. 
Knowledge of connectivity wiring diagrams alone may not be sufficient to understand the function
of nervous systems, but it is likely necessary. Yet because of the scarcity of reconstructed
connectomes, their significance remains uncertain. 

The neuronal network of the nematode \emph{Caenorhabditis elegans} is 
a logical model system for advancing the connectomics program.  It 
is sufficiently small that it can be reconstructed and analyzed as a whole.
The $302$ neurons in the hermaphrodite worm 
are identifiable and consistent across individuals \cite{WhiteSTB1986}.
Moreover the connections between neurons, consisting of 
chemical synapses and gap junctions, are stereotypical 
from animal to animal with more than $75\%$ reproducibility 
\cite{WhiteSTB1986, Durbin1987, Bargmann1993, HallR1991}.

Despite a century of investigation \cite{Goldschmidt1908, Goldschmidt1909}, 
knowledge of nematode neuronal networks is incomplete. 
The basic structure of the \emph{C.\ elegans} nervous system had been reconstructed using electron micrographs 
\cite{WhiteSTB1986}, but a major gap in the connectivity of ventral cord 
neurons remained.  Previous attempts to assemble the whole wiring 
diagram made unjustified assumptions that several reconstructed 
neurons were representative of others \cite{AchacosoY1992}. 
Much previous work analyzed the properties of the 
neuronal network (see e.g.~\cite{WattsS1998, AmaralSBS2000, 
LatoraM2003, MiloSIKCA2002, SpornsK2004, MoritaOOFOK2001, KaiserH2006} 
and references therein and thereto)  based on these incomplete 
or inconsistent wiring diagrams \cite{AchacosoY1992, WhiteSTB1986}.  

In this paper, we advance the experimental phase of the connectomics program 
\cite{SpornsTK2005, HagmannKGTWMT2007} by reporting a near-complete wiring 
diagram of \emph{C.\ elegans} based on original data from 
White \emph{et al}.\ \cite{WhiteSTB1986} but also including new serial section electron microscopy 
reconstructions and updates.  Although this new wiring diagram has not been published before now,
it has been freely shared with the community through the WormAtlas \cite{Wormatlas} and has also been
used in studies such as \cite{ChenHC2006}.\footnote{See {\sc Methods} section for details on
freely obtaining the wiring diagram in electronic form.}

We advance the theoretical phase of connectomics \cite{StamR2007,SpornsHK2007}, 
by characterizing signal propagation through the reported neuronal network and its relation
to behavior. We compute for the first time, local properties that may play a computational 
purpose, such as the distribution of multiplicity and the number of terminals, as well as
global network properties associated with the speed of signal propagation. 
Unlike the conventional ``hypothesis-driven'' mode 
of biological research, our work is primarily ``hypothesis-generating'' in the tradition of 
systems biology.

Our results should help investigate the function of the \emph{C.\ elegans} neuronal
network in several ways.  A full wiring diagram, especially when conveniently visualized 
using a method proposed here, helps in designing maximally informative 
optical ablation \cite{Zhang_ea2007} or genetic inactivation \cite{LuoCS2008} experiments. 
Our eigenspectrum analysis characterizes the 
dynamics of neuronal activity in the network, which should help predict 
and interpret the results of experiments using sensory and artificial stimulation
and imaging of neuronal activity. 

Organization of the {\sc Results} section reflects the duality of contribution 
and follows the tradition laid down by genome sequencing 
\cite{Adams_ea2000,Lander_ea2001}. We start by describing and visualizing the 
wiring diagram. Next, we analyze the non-directional gap junction network
and the directional chemical synapse network separately.  There are two primary reasons for separate analysis. 
First, understanding the parts before the whole provides didactic benefits. 
Second, separate consideration is valuable since we do not 
know the relative weight of gap junctions and chemical synapses and so any combination of the two 
involves additional assumptions.  Finally, we analyze the combined 
network of gap junctions and chemical synapses.  

\section*{Results}
\subsection{Reconstruction}
\subsubsection{An Updated Wiring Diagram}

The \emph{C.\ elegans} nervous system contains $302$ neurons and is  
divided into the pharyngeal nervous system containing $20$ neurons 
and the somatic nervous system containing $282$ neurons.  We updated the wiring diagram 
(see {\sc Methods}) of the larger somatic nervous system.  Since neurons CANL/R and VC06
do not make synapses with other neurons, we restrict our
attention to the remaining $279$ somatic neurons. The wiring diagram consists of 
$6393$ chemical synapses, $890$ gap junctions, and $1410$ 
neuromuscular junctions.  

The new version of the wiring diagram incorporates original data from White \emph{et al}.\
\cite{WhiteSTB1986}, Hall and Russell \cite{HallR1991}, updates based upon later 
work \cite{HobertHu, Durbin1987}, as well as new reconstructions.  Although neuronal 
circuitry in the head and tail was previously documented \cite{WhiteSTB1986, HallR1991},
the connection details for 58 motor neurons in the ventral cord of the worm were lacking.  
We compiled most of the missing data using original electron micrographs and 
handwritten notes from White and coworkers. The dorsal side of the worm around the midbody,
however, was not previously documented. Using original thin worm sections of animal \emph{N2U} 
prepared by White \emph{et al}.~\cite{WhiteSTB1986}, we generated new micrographs 
and reconstructed neurons with processes in this region. In total, over $3000$ synaptic contacts, 
including chemical synapses, gap junctions, and neuromuscular junctions were either 
added or updated from the previous version of the \emph{C.\ elegans} wiring diagram.

From our compilation of wiring data, including new reconstructions of ventral 
cord motor neurons, we applied self-consistency criteria to isolate records 
with mismatched reciprocal records.  The discrepancies were reconciled by checking 
against electron micrographs and the laboratory notebooks of White \emph{et al}.  
Connections in the posterior region of the animal were also cross-referenced with 
reconstructions published by Hall and Russell \cite{HallR1991}. Reconciliation 
involved $561$ synapses for $108$ neurons ($49$\% chemical `sends,' $31$\% chemical 
`receives,' and $20$\% electrical junctions).  The current wiring diagram is 
considered self-consistent under the following criteria:
\begin{enumerate}
\item A record of Neuron $A$ sending a chemical synapse to Neuron $B$ must be paired 
with a record of Neuron $B$ receiving a chemical synapse from Neuron $A$.
\item A record of gap junction between Neuron $C$ and Neuron $D$ must be paired 
with a separate record of gap junction between Neuron $D$ and Neuron $C$.  
\end{enumerate}

Although the updated wiring diagram represents a significant advance, it is only about $90$\% 
complete because of missing 
data and technical difficulties. Due to sparse sampling along lengths of the sublateral, canal-associated 
lateral, and midbody dorsal cords, about $5$\% of the total chemical synapses are missing, as
concluded from antibody staining for synapses\cite{DuerrHRu}.  Many gap junctions are 
likely missing due to the difficulty in identifying them in electron micrographs using
conventional fixation and imaging methods. Hopefully, application of high-pressure freezing 
techniques and electron tomography will help identify missing gap junctions \cite{HallLA2005}.
Finally, it should be noted that this reconstruction combined partial imaging of three worms, with images
for the posterior midbody being from the male \emph{N2Y}.

The basic qualitative properties of the updated \emph{C.\ elegans} nervous system 
remain as reported previously \cite{WhiteSTB1986, Durbin1987, Bargmann1993}. 
Neurons are divided into $118$ classes, based on morphology, dendritic specialization, 
and connectivity.  Based on neuronal structural and functional properties,
the classes can be divided into three categories: sensory neurons, interneurons, and motor neurons.  
Neurons known to respond to specific environmental conditions, either anatomically, by 
sensory ending location, or functionally, are classified as sensory neurons. They 
constitute about a third of neuron classes. Motor neurons are recognized by the presence 
of neuromuscular junctions. Interneurons are the remainder of the neuron classes and 
constitute about half of all classes.  A few of the neurons could have dual classification, 
such as sensory/motor neurons. Some interneurons are much more important for developmental function 
than for function in the final neuronal network \cite{HallLA2005}.

The majority of sensory 
neuron and interneuron categories contain pairs of bilaterally symmetric neurons.  Motor 
neurons along the body are organized in repeating groups whereas motor neurons in the head 
have four- or six-fold symmetry.  A large fraction of neurons send long processes to
the nerve ring in the circumpharyngeal region to make synapses with other neurons \cite{WhiteSTB1986}. 

The neurons in \emph{C.\ elegans} are structurally simple: most neurons 
have one or two unbranched processes and form \emph{en passant} synapses along them. 
Dendrites are recognized by being strictly ``postsynaptic'' or by containing a specialized
sensory apparatus, such as amphid and phasmid sensory neurons. Interneurons lack
clear dendritic specialization. It is interesting to note
that a given worm neuron has connections with only about $15$\% of neurons with 
which it has physical contact \cite{WhiteSTB1986,Durbin1987}, 
a similar number to the connectivity fraction in other nervous systems 
\cite{StepanyantsC2005, VarshneySC2006}.  

\subsubsection{Wiring Diagram as Adjacency Matrices}
\begin{figure}
  \centering
  \includegraphics[width=5in]{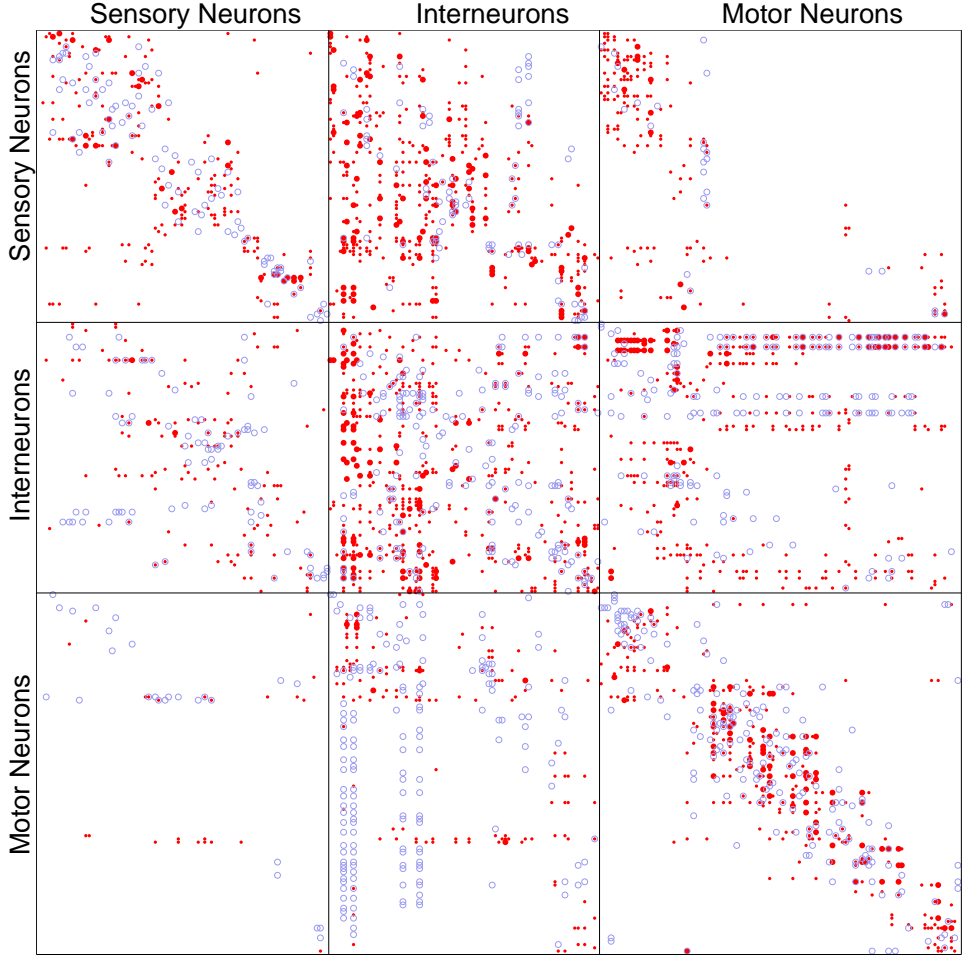}
  \caption{Adjacency matrices for the gap junction network (blue circles) and the chemical 
	synapse network (red points) with neurons grouped by category
	(sensory neurons, interneurons, motor neurons).  Within each category, neurons are 
	in anteroposterior order.  Among chemical synapse connections, small points indicate less 
	than $5$ synaptic contacts, whereas large points indicate $5$ or more synaptic contacts.
	All gap junction connections are depicted in the same way, irrespective
	of number of gap junction contacts. 
  }
  \label{fig:1}
\end{figure}

In the remainder of the paper, we describe and analyze the connectivity of gap 
junction and chemical synapse networks of \emph{C.\ elegans} neurons. Gap junctions 
are channels that provide electrical coupling between neurons, whereas chemical 
synapses use neurotransmitters to link neurons. The network of gap junctions and 
the network of chemical synapses are initially treated separately, with each represented 
by its own adjacency matrix, Figure~\ref{fig:1}.  In an adjacency matrix $A$, the element in the $i$th row and 
$j$th column, $a_{ij}$, represents the total number of synaptic contacts from the $i$th 
neuron to the $j$th. If neurons are unconnected, the corresponding element of the 
adjacency matrix is zero.   An adjacency matrix may be used due to self-consistency 
in the gathered data. 

Although gap junctions may have directionality, i.e.\ conduct 
current in only one direction, this has not been demonstrated in \emph{C.\ elegans}.  
Even if directionality existed, such information cannot 
be extracted from electron micrographs.  Thus we treat the gap junction network as an 
undirected network with a symmetric adjacency matrix.  Weights in both $a_{ij}$ and $a_{ji}$ 
represent the total number of gap junctions between neurons $i$ and $j$.  

Since chemical synapses possess clear directionality that can be extracted from 
electron micrographs, we represent the chemical network as a directed network with
an asymmetric adjacency matrix. The elements of the adjacency matrix take 
nonnegative values, which reflect the number of synaptic contacts between 
corresponding neurons.  Contacts are given equal weight, regardless of 
the apparent size of the synaptic apposition. We 
use nonnegative values for most of the paper because we cannot determine whether a synapse is 
excitatory, inhibitory, or modulatory from electron micrographs of \emph{C.\ 
elegans}. For the linear systems analysis, we do however make a rough guess of the signs of synapses
based on neurotransmitter gene expression data.

Electron micrographs for \emph{C.\ elegans} have a further limitation
that causes some synaptic ambiguity.  When a presynaptic terminal makes contact 
with two adjacent processes of different neurons (send\_joint in Durbin's notation \cite{Durbin1987}), 
it is not known which of these processes acts as a postsynaptic terminal; both might be involved. 
We count all polyadic synaptic connections.  Polyadic connections are briefly
revisited in the {\sc Discussion}.

\subsubsection{Visualization}
\begin{figure*}
  \includegraphics[width=7in]{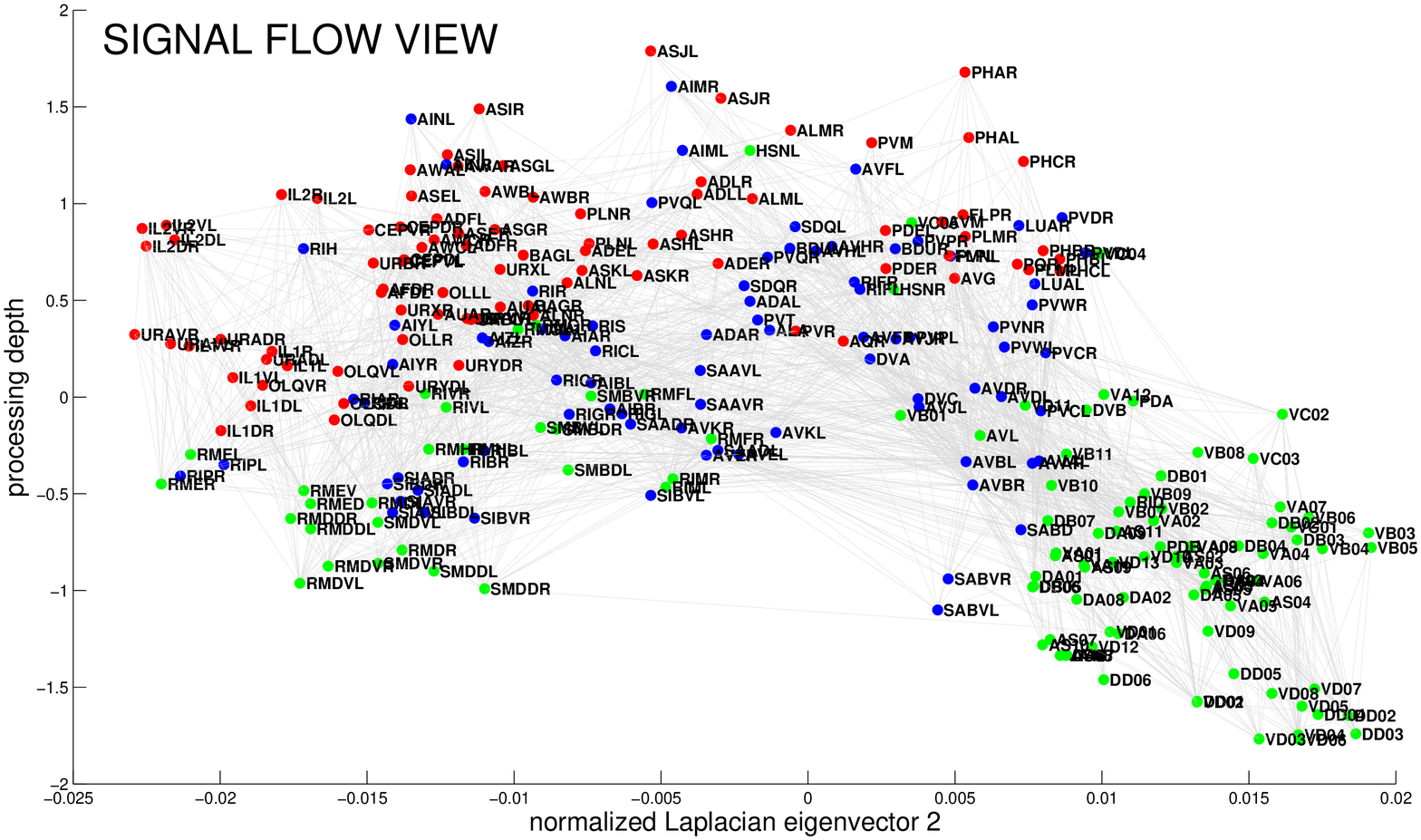} 
  \includegraphics[width=7in]{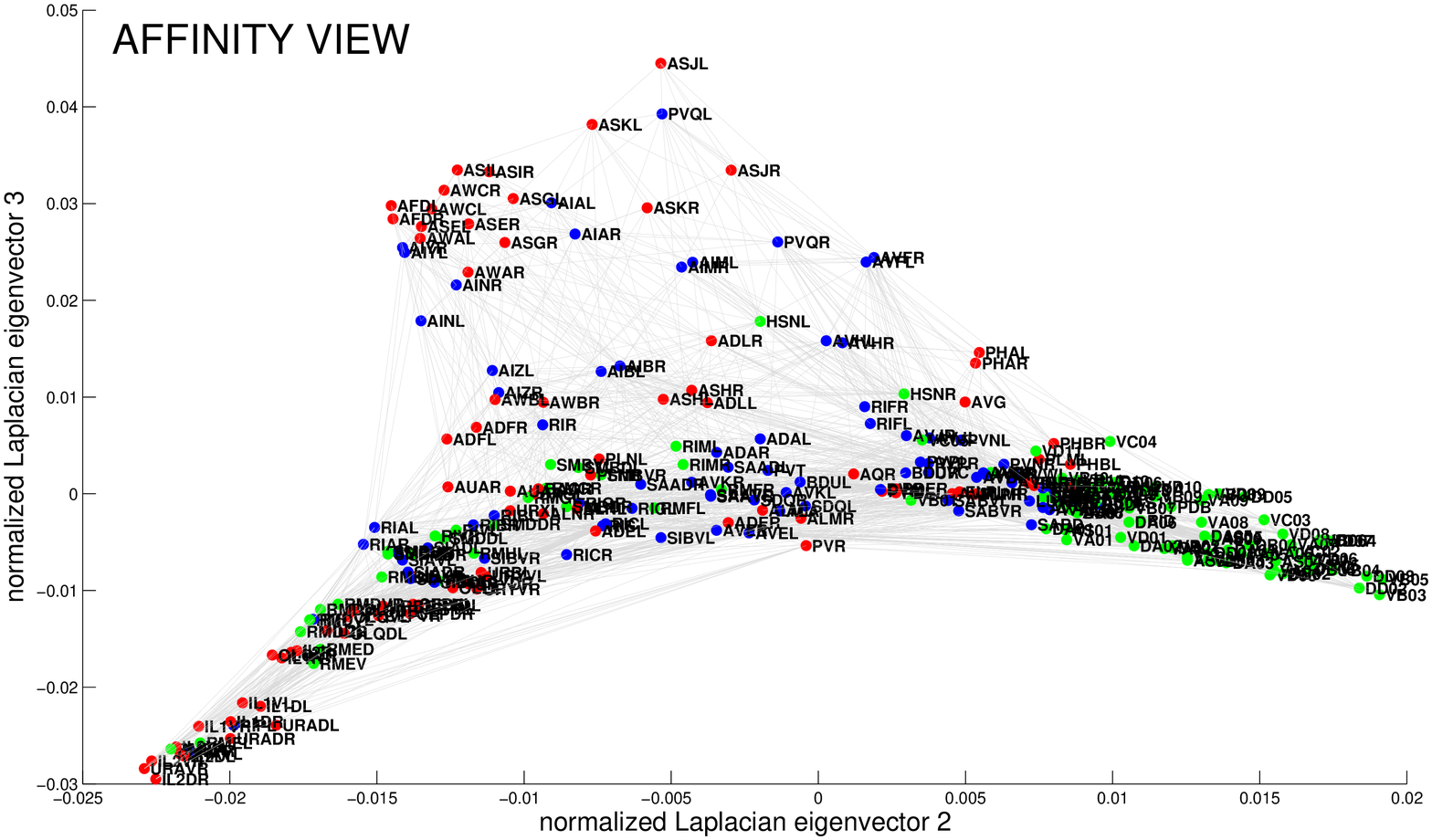}\\
  \vspace{-0.4in}
  \caption{The \emph{C.\ elegans} wiring diagram is a network of identifiable, labeled 
	neurons connected by chemical and electrical synapses.  Red, sensory neurons; 
	blue, interneurons; green, motorneurons.  (a). Signal flow view shows neurons arranged so 
	that the direction of signal flow is mostly downward. (b). Affinity view shows structure 
	in the horizontal plane reflecting weighted non-directional adjacency of neurons in the network.
  }
  \label{fig:D}
\end{figure*}

Although statistical measures that we investigate later in this paper provide significant 
insights, they are no substitute to exploring detailed connectivity in the neuronal
network. As the number of connections between neurons is large even for relatively
simple networks, such analysis requires a convenient way to visualize the wiring diagram. 
Previously, various fragments of the wiring diagram were drawn to illustrate specific pathways 
\cite{Durbin1987,GrayHB2005,ChalfieSWSTB1985}. Here, we propose a method to visualize the whole wiring diagram in a way
that reflects signal flow through the network as well as the closeness of neurons in the network, Figure~\ref{fig:D}. To this end,
we use spectral network drawing techniques because they have certain optimality properties \cite{Hall1970} 
and aesthetic appeal. Next, we give an intuitive description of our visualization method; mathematical details 
can be found in Appendix~\ref{app:drawing}.

The vertical axis in Figure~\ref{fig:D}(a), represents the position of neurons 
in the signal flow hierarchy \cite{Koren2005,Seung2009} of the chemical
synapse network with sensory neurons at the top and motor neurons at the bottom,
with interneurons in between.  We want the vertical coordinate of pre- and post-synaptic neurons 
to differ by one, however due to ``frustration'' this is not always possible.  Frustration 
happens when distances measured along network connections cannot be made to correspond 
to the hierarchy distances: there are two different hierarchical paths that require 
a particular neuron to appear in two different places.  We look for the layout 
that has smallest deviation from this condition and find a closed form solution 
\cite{CarmelHK2004,Koren2005}.  The number of synapses from
sensory to motor neurons---the signal flow depth of the network---can be read off
the vertical coordinate.  Depending on the specific neurons considered, the 
depth is typically 2--3 \cite{Durbin1987}.  

Neuronal position on the horizontal plane, Figure~\ref{fig:D}(b), represents 
the connectivity closeness of neurons in the combined chemical and electrical 
synapse network. Neuronal coordinates are given by the second and third 
eigenmodes of the symmetrized network's graph Laplacian (see below).
In this representation, pairs of synaptically coupled neurons with larger number
of connections in parallel tend to be positioned closer in space. 

Thus, Figure~\ref{fig:D} represents not the physical placement of neurons in the worm but signal flow 
and closeness in the network.  Such visualization reveals that motorneurons 
and some interneurons segregate into two lobes along the first horizontal axis: the right lobe contains 
motorneurons in the ventral cord and the left lobe consists of neck/tail neurons. The bi-lobe structure 
suggests partial autonomy of motorneurons in the ventral cord and neck/tail. Interneurons that could coordinate 
the function of the two lobes can be easily identified by their central location.

\subsection{Gap Junction Network}

For quantitative characterization, we first consider the gap junction network. 

\subsubsection{Basic Structure and Connectivity}

The gap junction network that we analyze consists of $279$ neurons and $514$ gap junction connections,
consisting of one or more junctions. The network is not fully connected, but is 
divided into a giant component containing $248$ neurons, two smaller components of $2$ 
and $3$ neurons, and $26$ isolated neurons with no gap junctions (Table~\ref{tab:3}). 
The giant component has $511$ connections.  An Erd\"{o}s-R\'{e}nyi random 
network\footnote{
Construction of an (unweighted) Erd\"{o}s-R\'{e}nyi random network  requires
a single parameter, the probability of a connection between two neurons.}
with $279$ neurons and connection probability $0.0133$ (thus with $514$ expected 
connections) would be expected to have $271$ neurons in the giant component.  The true gap junction
giant component is much smaller; the probability of finding such a small giant component in a random
network is on the order of $10^{-14}$ (see {\sc Methods}).  A better comparison,
however, can be made to random networks with degree distributions that match
the degree distribution of the gap junction network \cite{MaslovS2002}. Here, the degree of a neuron 
is the number of neurons with which it makes a gap junction.  
The giant component in a degree-matched random network would be expected to be $251$ 
neurons (see {\sc Methods}), about the same size as the measured giant component.
Using connectivity data from \cite{AchacosoY1992}, Majewska and Yuste
had previously pointed out that most neurons in 
\emph{C.\ elegans} belong to the giant component \cite{MajewskaY2001}.  
Our results agree roughly with \cite{MajewskaY2001}, although our dataset excludes 
non-neuronal cells and places certain neurons in different connected 
components.  

The adjacency matrix of the network, $A$, is depicted in Figure~\ref{fig:1} (the number of gap junctions 
in a connection is not depicted).  The matrix is symmetric since the network is undirected. We may explore the 
utility of representing the wiring diagram as a three-layer network by grouping 
neurons by category (sensory neurons, interneurons, motor neurons). As shown 
in Tables~\ref{tab:1}A and \ref{tab:1}B, each category has many recurrent connections;
with the exception of connections between sensory and motor neurons, there are also
many connections between categories.  In particular, Table~\ref{tab:1}B indicates that 
motor neurons send to interneurons roughly the same number of connections as  
recurrently sent back to motor neurons.  These observations suggest that on the level
of gap junctions, the value of a three-layer network abstraction is questionable.

\subsubsection{Distributions of Degree, Multiplicity and the Number of Terminals}

In this section, we analyze statistical properties of individual neurons and 
synaptic connections. To characterize the ability of individual neurons to propagate or collect signals,
we compute the degree $d_i$ of neuron $i$, 
which is the number of neurons that are coupled to $i$ by at least one gap 
junction. The mean degree is $3.68$, however this value is not representative as
the degree varies in a wide range, from $0$ to $40$. Thus, 
it is important to look at the degree distribution, which has been used to characterize and classify other
networks previously 
\cite{Strogatz2001, BarabasiA1999, ItzkovitzMKZA2003,ClausetSN2009}.

\begin{figure}[h]
  \centering
\subfigure[]{
	\includegraphics[width=2in]{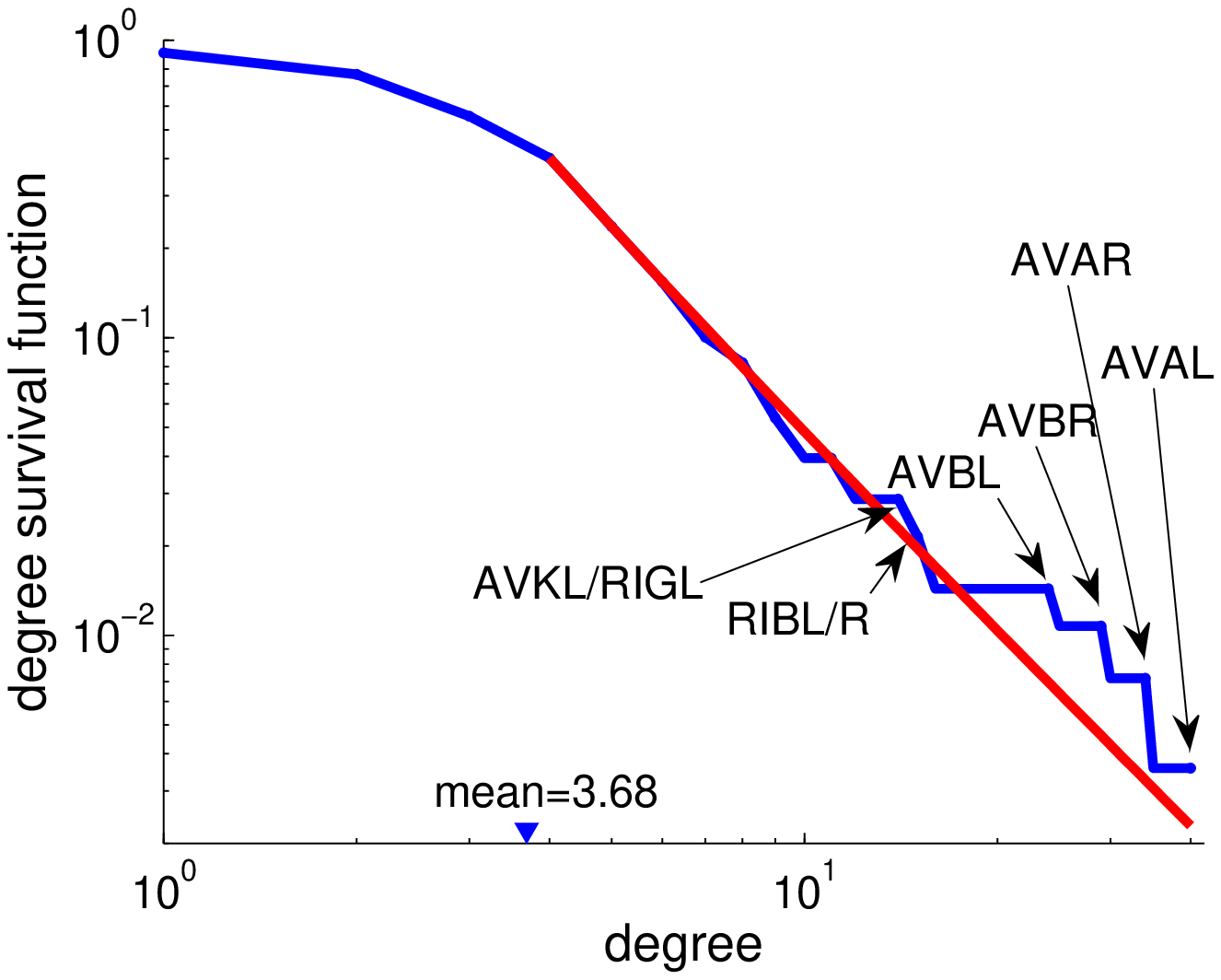} 
	\label{fig:3}
	}
\subfigure[]{
	\includegraphics[width=2in]{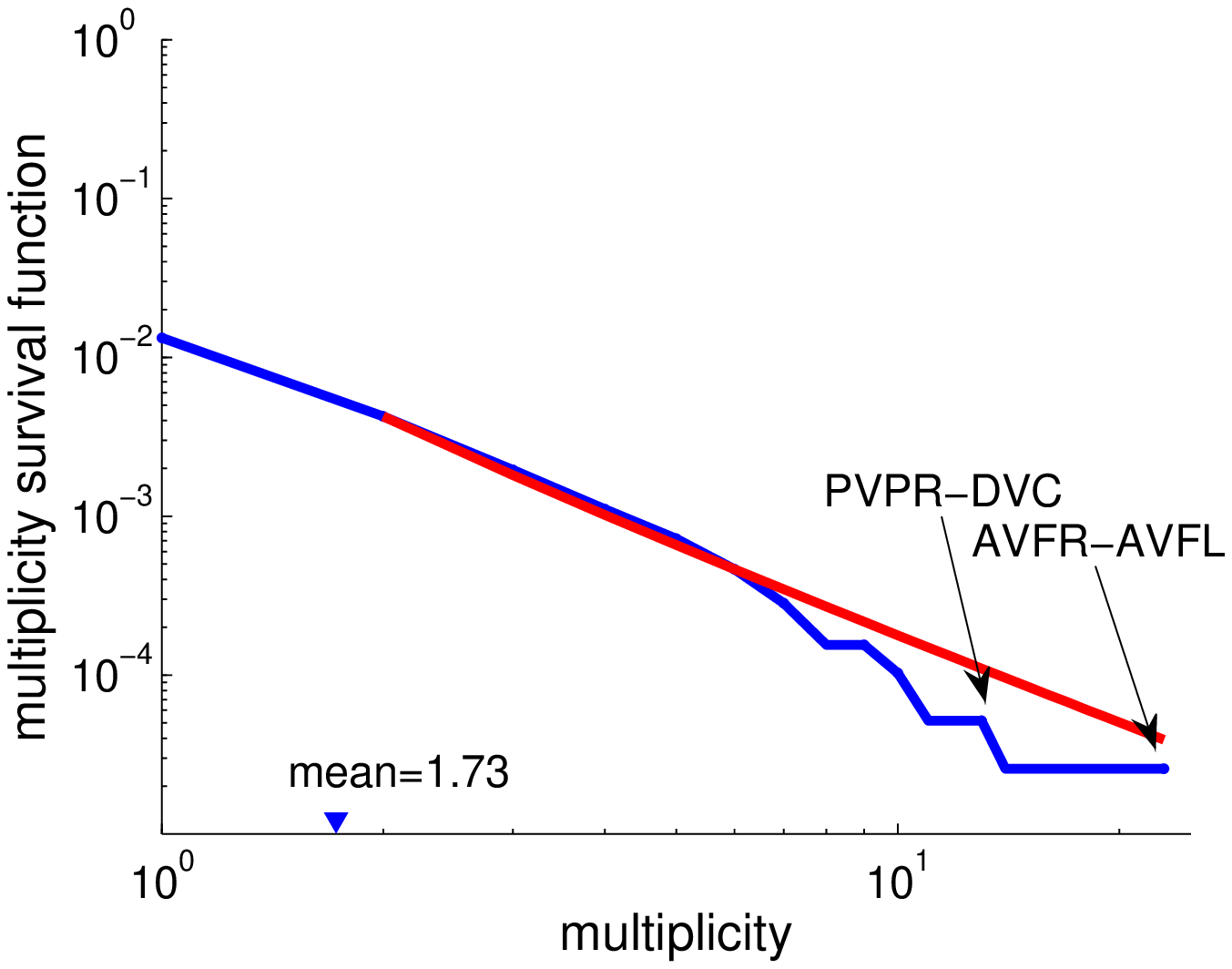} 
	\label{fig:4}
	}
\subfigure[]{
	\includegraphics[width=2in]{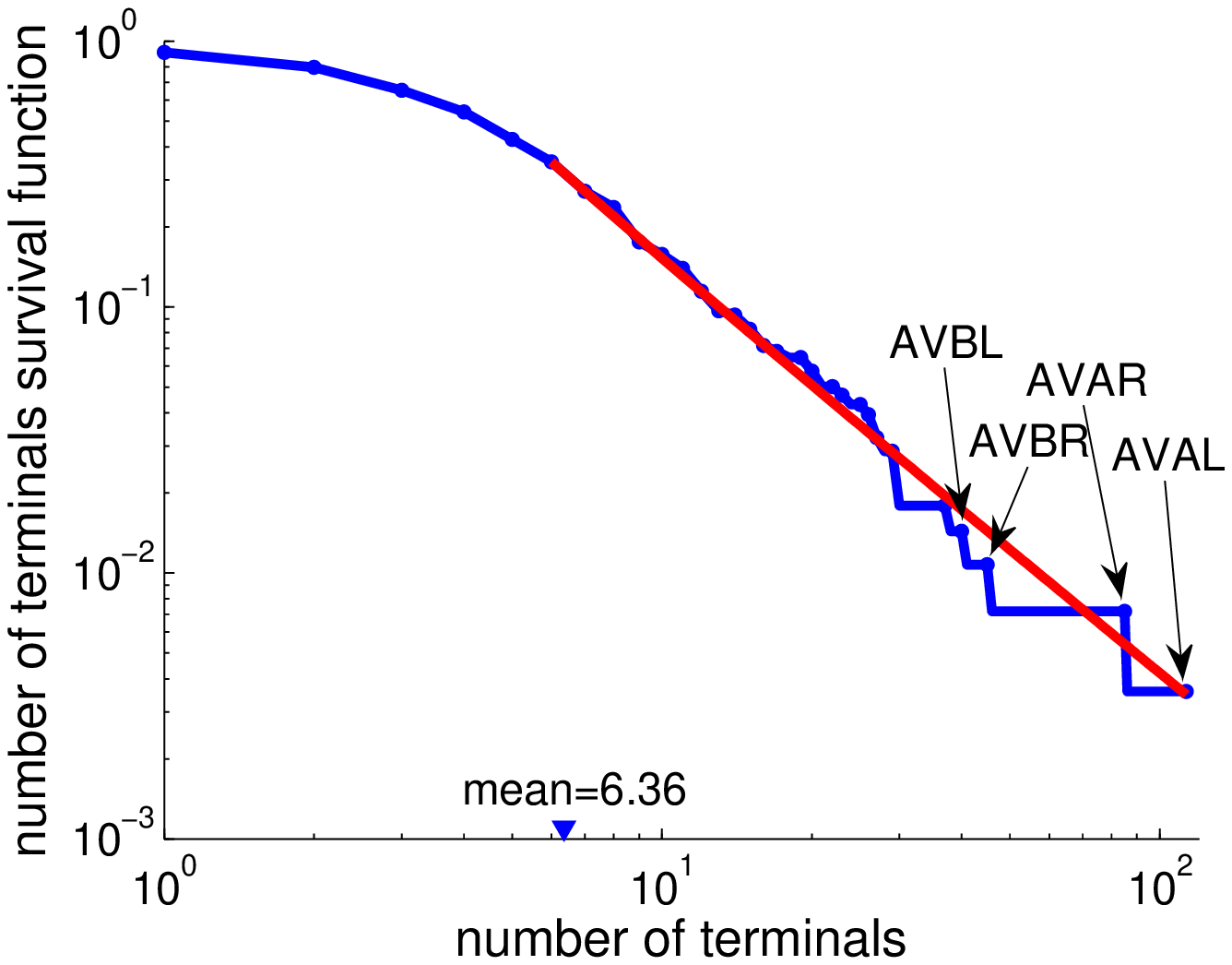}
	\label{fig:5}
	}
  \caption{ Survival functions for the distributions of degree, multiplicity, 
    and number of synaptic terminals in the gap junction network. Neurons or 
    connections with exceptionally high statistics are labeled. The tails of the 
	distributions can be fit by a power law with the exponent $3.14$ for
    the degrees \subref{fig:3}; $2.76$ for the multiplicity distribution 
    \subref{fig:4}; $2.53$ for the number of synaptic terminals \subref{fig:5}. 
    The exponents for the power law fits of the corresponding survival functions 
    are obtained by subtracting one.
}
\end{figure}

To visualize the discrete degree distribution, $p(d)$, we use the survival function:
\begin{equation}
P(d) = \sum_{k=d}^{\infty} p(k) \mbox{,}
\end{equation}
which is the complement of the cumulative distribution function, Figure~\ref{fig:3}.
The advantages of looking at the survival function rather than the degree 
distribution directly are that histogram binning is not required and 
that noise in the tail is reduced \cite{Newman2003}. The survival function is also later applied 
to visualize other statistics.  Various commonly
encountered distributions and their corresponding survival functions are
given in Appendix~\ref{app:survival}.  

We perform a fitting procedure for the tail of the gap junction degree distribution \cite{ClausetSN2009} (see {\sc Methods}).
We find that the tail ($d \geq 4$) can be fit by the power law with exponent $\gamma = 3.14$, Figure~\ref{fig:3}, 
but not by the exponential decay ($p$-value $<0.1$). This result is
consistent with the view that the gap junction network is scale-free \cite{BarabasiA1999}. 

To characterize the direct impact that one neuron can have on another, we quantify 
the strength of connections by the multiplicity, $m_{ij}$, between neurons $i$ and $j$, 
which is the number of synaptic contacts (here gap junctions) connecting $i$ to 
$j$.  The degree treats synaptic connections as binary, whereas the 
multiplicity quantifies the number of contacts. The multiplicity distribution 
for the gap junction network is shown in Figure~\ref{fig:4}. We find that the multiplicity distribution 
for $m \geq 2$ obeys a power law with exponent $\gamma = 2.76$. 
Although the exponential decay fit to the tail passes the $p$-value test, the
log-likelihood is significantly lower than for the power law.

Finally, the number of terminals that lie on a given neuron is the sum of the 
multiplicities of all gap junction connections. The tail of 
the distribution of the number of synaptic terminals, Figure~\ref{fig:5}, is 
adequately fit by a power law with exponent $\gamma = 2.53$.

Identifying neurons that play a central or special role in the transmission
or processing of information may also prove useful \cite{Scott2000, BrandesU2005, GuimeraA2005, ChatterjeeS2008, PanCS2009}.
To rank neurons according to their roles, we introduce several centrality indices. 
Perhaps the simplest centrality index is \emph{degree centrality} $c_d(i)$.  
Degree centrality is simply the degree of a neuron, $c_d(i) = d_i$, 
and is motivated by the idea that a neuron with connections to many other neurons
has a more important or more central role in the network than a neuron connected 
to only a few other neurons.  Neurons that have
unusually high degree centrality include AVAL/R and 
AVBL/R.  The same neurons lie in the tail of the distribution of the number of synaptic
terminals, Figure~\ref{fig:5}, suggesting strong coupling to the network.  
These neuron pairs are command interneurons responsible for coordinating backward and forward 
locomotion, respectively \cite{ChalfieSWSTB1985,RiddleBMP1997,Wormatlas}.  The high degree 
centralities of RIBL/R suggest a similarly central function for those neurons, 
though they each only have $19$ gap junction terminals, in the middle of the distribution
of number of terminals, suggesting weaker coupling to the network.

\subsubsection{Small World Properties}
Having described statistical properties of individual neurons and connections,
such as the degree and multiplicity distributions, we now investigate properties that 
may describe the efficiency of signal transmission across the gap 
junction network.  Traditionally \cite{WattsS1998}, this analysis does not consider 
multiplicity of gap junctions but treats them as binary.  
We analyze signal propagation when including multiplicities in the next subsection. 

The geodesic distance, $d_{ij}$, between two 
neurons in the network is the length of the shortest network path between them.  
The network path is measured by the number of connections that are crossed 
rather than by physical distance.  The average geodesic distance over all 
pairs of neurons is the characteristic path length \cite{WattsS1998}:
\begin{equation}
L = \frac{1}{N(N-1)}\sum_{i,j: i\neq j} d_{ij} \mbox{,}
\end{equation}
where $N$ is the number of neurons.  This global measure describes how readily 
or rapidly a signal can travel from one neuron to another since it is simply the average 
distance between all neurons.  Clearly, the measure $L$
requires the network to be connected (otherwise $L$ diverges), so we restrict 
attention to the giant component. 

A signal originating in one neuron in the giant component 
must cross $L = 4.52$ gap junction connections on average to reach another neuron 
of the giant component.  For an Erd\"{o}s-R\'{e}nyi random network with $248$ neurons and 
$511$ connections, the characteristic path length is approximately $\log(248)/\log(511/248) = 7.63$
\cite{WattsS1998}.
When the actual degree distribution of the gap junction network is taken into 
account, a random network from that ensemble would be expected to have characteristic
path length $3.05$ (see {\sc Methods}).
The distribution of geodesic distances $d_{ij}$ 
in the giant component is shown in Figure~S\ref{fig:6}.  

A second measure for signal propagation is the clustering coefficient 
$C$, which measures the density of connections among an average neuron's 
neighbors.  It is defined as \cite{WattsS1998}:
\begin{equation}
C = \frac{1}{N}\sum_i C_i \hspace{2cm} C_i = \frac{2E(\mathcal{N}_i)}{k_i(k_i-1)}\mbox{,}
\end{equation}
where $E(\mathcal{N}_i)$ is the number of connections between neighbors of $i$, $k_i$ is 
the number of neighbors of $i$, and $C_i$ measures the density of connections 
in the neighborhood of neuron $i$ (we set $C_i = 1$ when $k_i = 1$).  
We find the clustering coefficient $C = 0.21$. The clustering coefficient for 
an Erd\"{o}s-R\'{e}nyi random network with $248$ neurons and $511$ connections is approximately
$511/(248^2) = 0.0083$ \cite{WattsS1998}. For a degree-matched random network, we computed 
the clustering coefficient $C = 0.05$. Thus, the giant component of the gap 
junction network is strongly clustered relative to random networks, both Erd\"{o}s-R\'{e}nyi 
and degree-matched. 

Small world networks are more clustered than Erd\"{o}s-R\'{e}nyi random networks
and yet have smaller average distances \cite{CanchoJS2001}.  Thus, the giant component of the 
gap junction network may be 
classified as a small world network.  Table~\ref{tab:SMN} shows a comparison of 
the gap junction network of \emph{C.\ elegans} with other networks that have been 
characterized as small world networks.

Next we consider how quickly individual neurons reach all other neurons in the network. 
The normalized closeness of a neuron $i$ is the average geodesic distance $d_{ij}$
across all neurons $j$ that are reachable from $i$ \cite{BrandesU2005}:
\begin{equation}
d_{\rm{avg}}(i) = \frac{1}{N-1}\sum_{j: j\neq i} d_{ij} \mbox{.}
\end{equation}
The normalized closeness centrality, which takes higher
values for more central neurons, is defined as the inverse,
$c_c(i) = 1/d_{\rm{avg}}(i)$. 

Restricting to the giant component of the gap junction network,
the six most central neurons are AVAL, AVBR, RIGL, AVBL, RIBL, and AVKL.
In addition to command interneuron classes AVA and AVB, these include
RIBL and RIGL, both ring interneurons, and AVKL, an interneuron in 
the ventral ganglion of the head. The set of neurons that are closeness 
central mostly overlaps with the set of neurons that are degree central. 

The Spearman rank correlation coefficient \cite{Spearman1904} 
between degree centrality $c_d(i)$ and closeness centrality $c_c(i)$ 
for the entire giant component, however, is only $0.036$.  Since correlation between the
two centrality measures does not extend to peripheral neurons, ordering of importance is different. 

\subsubsection{Spectral Properties}

Global network properties discussed in the previous section characterize 
signal transmission while ignoring connection weights. As weights affect
the effectiveness of signal transmission and vary among connections, we
now analyze the weighted network by using linear systems theory. Although neuronal 
dynamics can be nonlinear, spectral properties nevertheless provide important 
insights into function. For example, the initial success of the Google search engine
is largely attributed to linear spectral analysis of the World Wide Web \cite{BryanL2006}. 

We characterize the dynamics of the gap junction network by the following system of 
linear differential equations, which follow from charge conservation \cite{Koch1999,FerreeL1999}:
\begin{equation}
C_i\tfrac{dV_i}{dt} = \sum_j (V_j - V_i)g_{ij}-g^m_i V_i\mbox{,}
\end{equation}
where $V_i$ is the membrane potential of neuron $i$, $C_i$ is the membrane capacitance of neuron $i$,
$g_{ij}$ is the conductance of gap junctions between neurons $i$ and $j$, and $g^m_i$ is the membrane
conductance of neuron $i$. Assuming that each neuron has the same capacitance $C$ and each
gap junction has the same conductance $g$, i.e. $g_{ij} = g A_{ij}$, we can rewrite this equation 
in terms of the time constant $\tau = C / g$  as:
\begin{equation}
\label{eq:diffeq_gap}
\tau\tfrac{dV_i}{dt} = \sum_j (V_j - V_i) A_{ij} - \tfrac{g^m_i}{g} V_i\mbox{.}
\end{equation}
 
Assuming that gap junction conductance is greater than the membrane conductance, we
temporarily neglect the last term and rewrite this equation in matrix form: 
\begin{equation}
\tau  \tfrac{dV}{dt} = - L V\mbox{,}
\end{equation}
where $L$ is the Laplacian matrix of the weighted network, $L = D-A$, $D$ contains 
the number of neuron gap junctions on the diagonal and zeros elsewhere,
and $V$ is a column vector of the membrane potentials.  A different 
plausible differential equation model is discussed in Appendix~\ref{app:eigenratio}.

\begin{figure}[h]
  \centering
\subfigure[]{
	\includegraphics[width=3in]{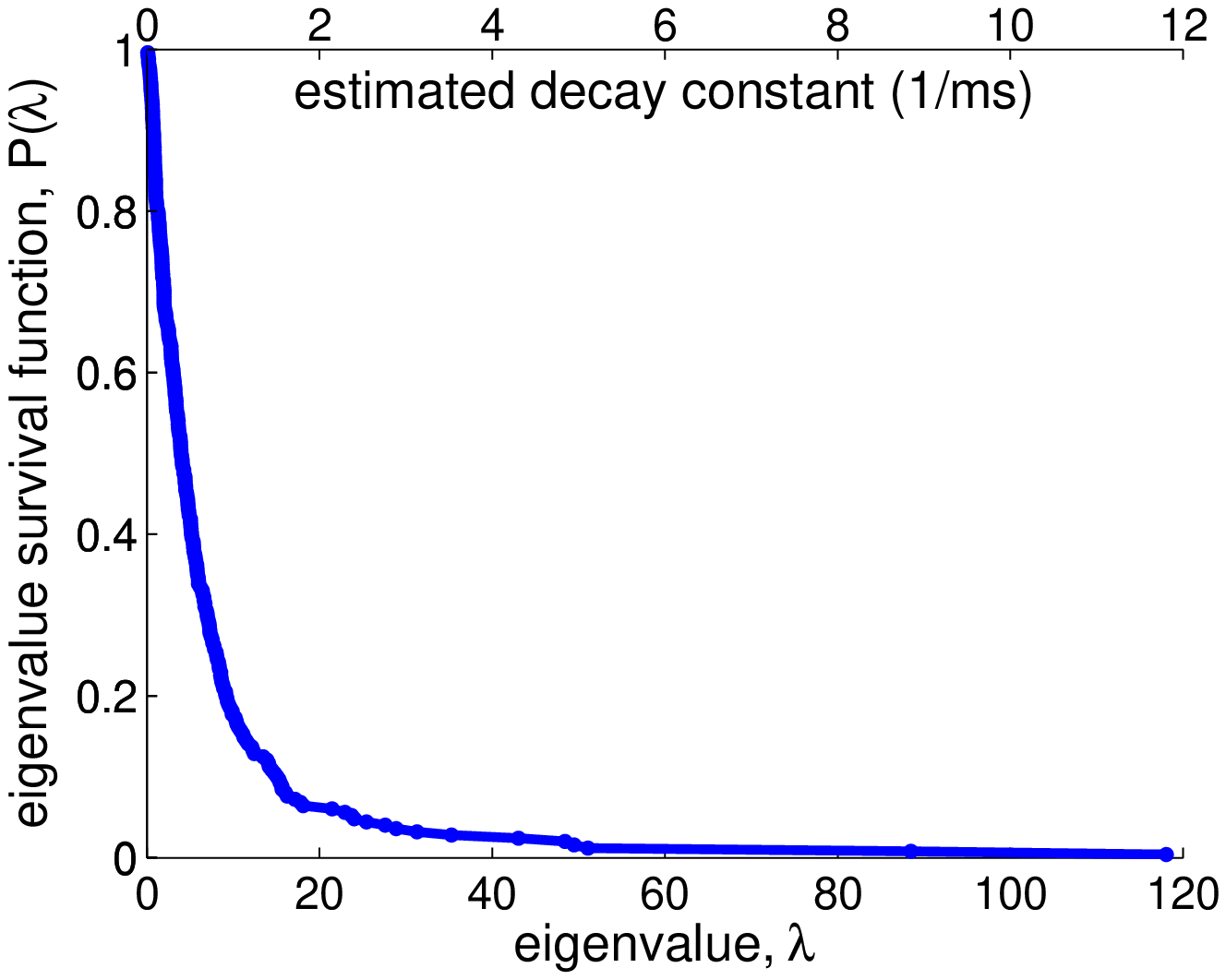}
	\label{fig:LSdist}
	}
\subfigure[]{
	\includegraphics[width=3in]{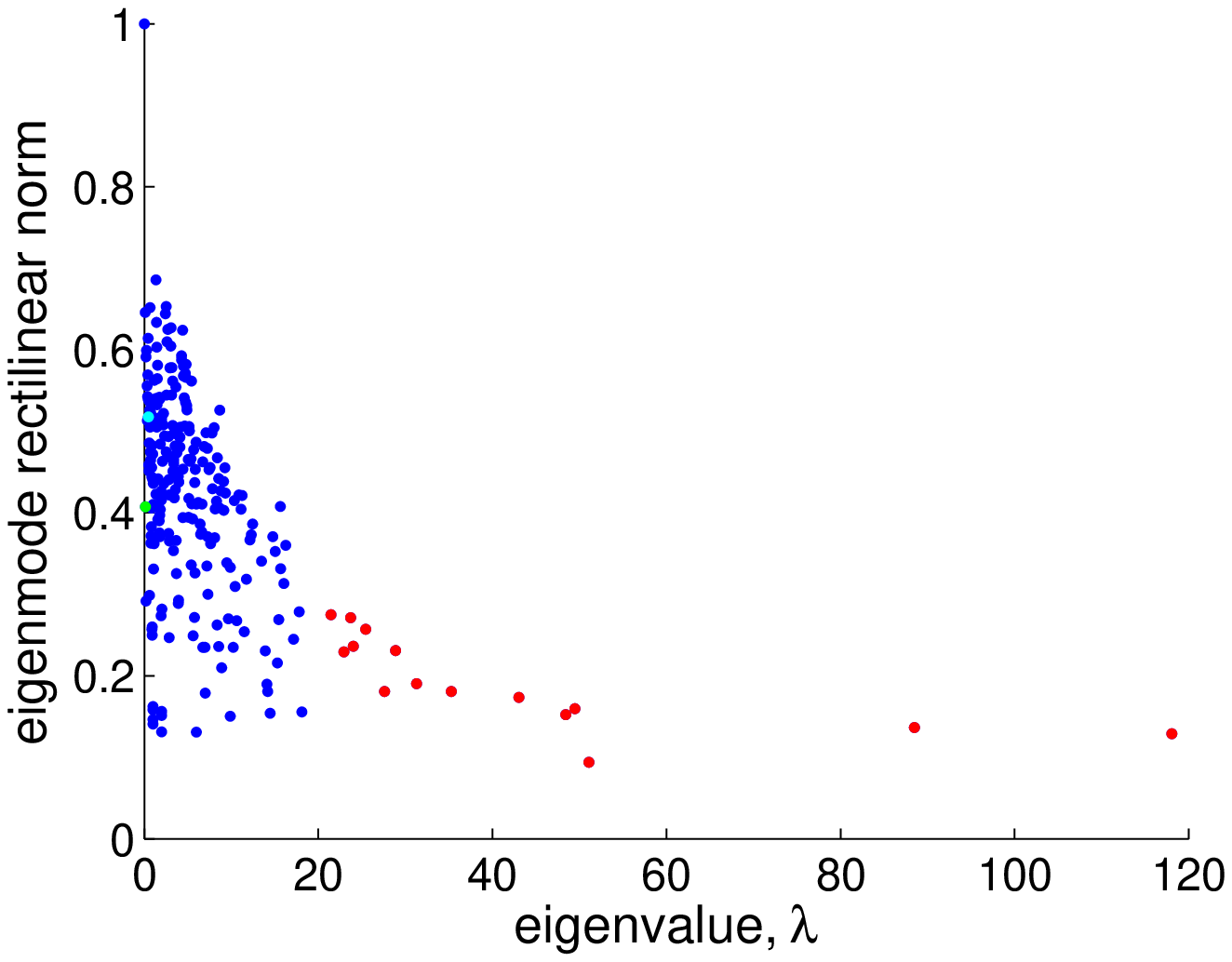}
	\label{fig:SparseSpec}
	}
\subfigure[]{
	\includegraphics[width=3in]{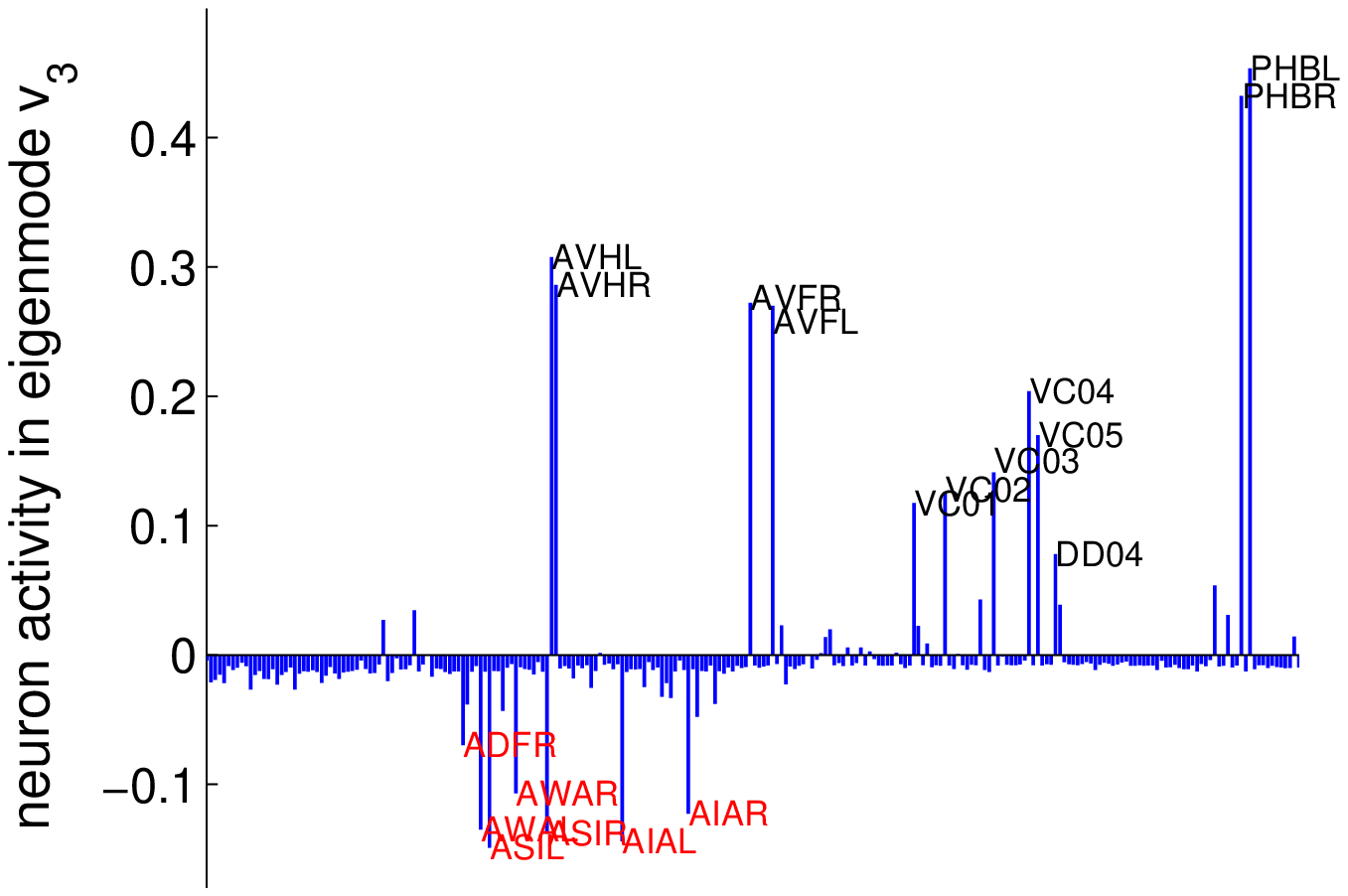}
	\label{fig:eigsl3_gap}
	}
\subfigure[]{
	\includegraphics[width=3in]{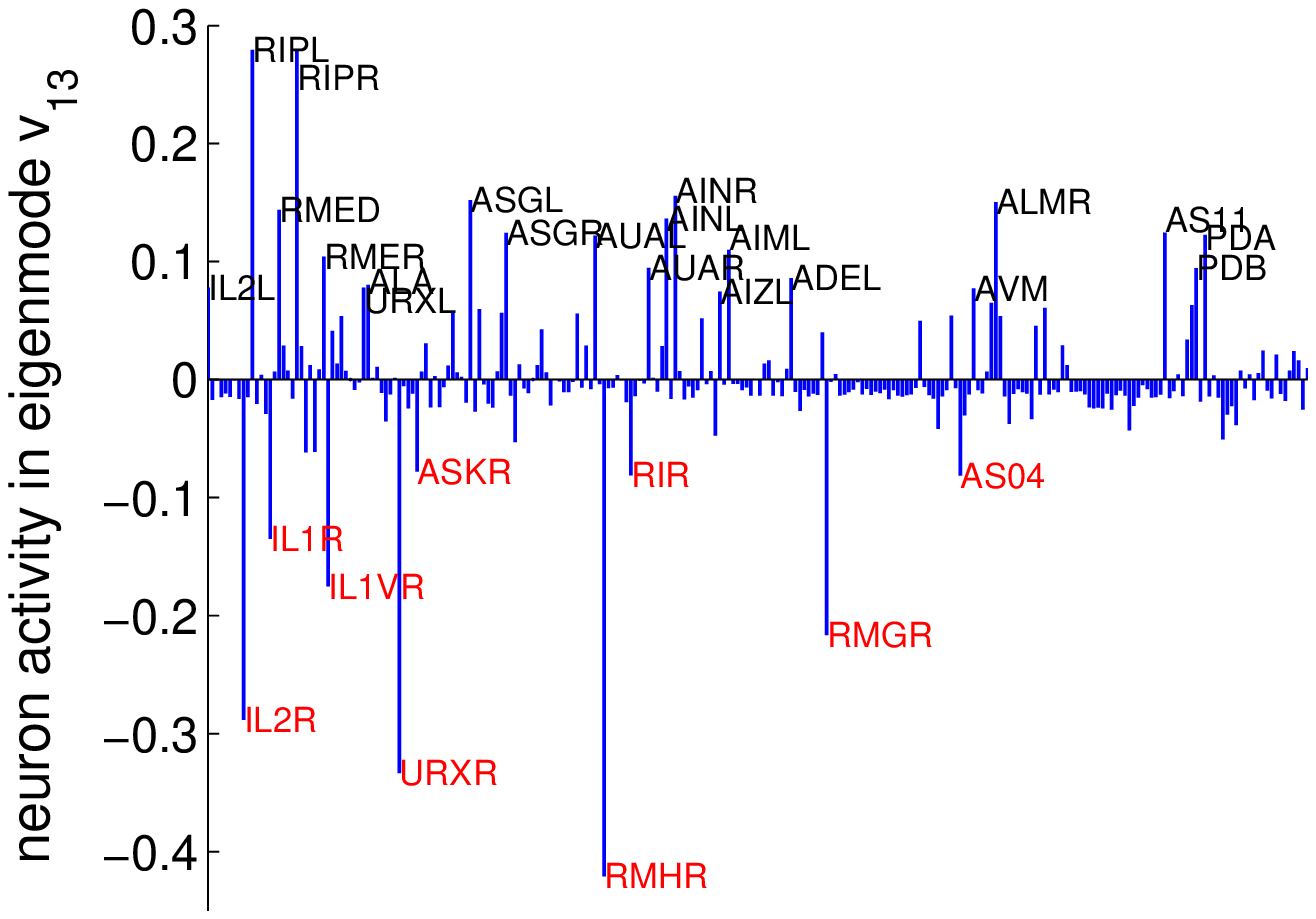}
	\label{fig:eigsl13_gap}
	}
  \caption{Linear systems analysis of the giant component of the gap junction network.  \subref{fig:LSdist}. 
	Survival function of the eigenvalue spectrum (blue).  The algebraic connectivity, $\lambda_2$, is $0.12$ and the 
	spectral radius, $\lambda_{248}$, is $118$.  A time scale associated with the decay 
	constant is also given.  
	\subref{fig:SparseSpec}. Scatterplot showing the rectilinear norm and decay constant of the eigenmodes of
	the Laplacian. The fastest modes from Figure~\ref{fig:fasteigs_gap} are marked in red.  
	The sparsest and slowest modes, most amenable to biological analysis,
	are located in the lower-left corner of the diagram.
	\subref{fig:eigsl3_gap}. Eigenmode of Laplacian corresponding to $\lambda_{3}$ (marked green in panel~\subref{fig:SparseSpec}).
	\subref{fig:eigsl13_gap}. Eigenmode of Laplacian corresponding to $\lambda_{13}$ (marked cyan in panel~\subref{fig:SparseSpec}).
}
\end{figure}

This system of coupled linear differential equations can be solved by performing a coordinate
transformation to the Laplacian  eigenmodes.  Since the Laplacian eigenmodes are decoupled
and evolve independently in time, performing an eigendecomposition of initial conditions 
leads to a full description of the system dynamics. We show the survival function of the 
eigenspectrum of the Laplacian in Figure~\ref{fig:LSdist}. 

What insight can be gained from inspection of the Laplacian eigenmodes? 
The gap junction network is equivalent to a network of resistors, where each gap 
junction acts as a resistor. The eigenmodes give intuition about experiments 
where a charge is distributed among neurons of the network and the spreading 
charge among the neurons is monitored in time. If the charge 
is distributed among neurons according to an eigenmode, the relative shape of the distribution 
does not change in time. The charge magnitude decays with a time constant specified by the eigenvalue.
The smallest eigenvalue of the Laplacian is always zero, corresponding to the infinite 
relaxation time. In the corresponding eigenmode each neuron is charged equally.

If the charge is distributed according to eigenmodes corresponding to small eigenvalues,
the decay is rather slow. Thus, these eigenmodes correspond to long-lived excitation.
The existence of slowly decaying modes often indicates that the network contains weakly coupled 
subnetworks, in which neurons are strongly coupled among themselves.
The corresponding charge distribution usually has negative values on one subnetwork and 
positive values on the other subnetwork. Because of the relatively slow equilibration of
charge between the subnetworks, such eigenmode decays slowly.  

For example, one might speculate that the eigenmode associated with $\lambda_3$ 
(Figure~\ref{fig:eigsl3_gap}) on the `black' side reflects a coupling of chemosensory
neurons in the tail (PHBL/R) along with interneurons (AVHL/R, AVFL/R) and 
motor neurons (VC01-05) involved in egg-laying behavior. At the level of gap junctions, 
these neurons are weakly coupled with chemosensory neurons in the head (ADFR, ASIL/R,
AWAL/R) and related interneurons (AIAL/R) on the `red' side. 

Another interesting example is the eigenmode associated with $\lambda_{13}$ (Figure~\ref{fig:eigsl13_gap}). 
Neurons on the `red' side overlap significantly with those identified previously
in a hub-and-spoke circuit mediating pheromone attraction, oxygen sensing, and social behavior 
\cite{MacoskoPFCBCB2009}. Such overlap is consistent with the view \cite{MacoskoPFCBCB2009}
that this network of neurons solves a consensus problem \cite{Olfati-SaberFM2007}. 

The above two examples demonstrate that spectral analysis can uncover 
circuits that have been described using experimental studies. 
The probability of a known functional circuit to appear in an eigenmode by chance is small (see {\sc Methods}).
It would be interesting to see whether other eigenmodes have a biological interpretation
and therefore generate predictions for future experiments.  

To prioritize further analysis of eigenmodes for biological significance, it may be advantageous to 
focus on the slow and sparse modes, where few neurons exhibit significant activity.  We can quantify 
sparseness of normalized eigenmodes by the sum of absolute values (rectilinear norm) of the eigenmode components; sparser eigenmodes have smaller
rectilinear norms \cite{ZouHT2006}.  Figure~\ref{fig:SparseSpec} is a scatterplot of eigenmodes showing 
both their decay constant and their rectilinear norms. 

The full set of eigenmodes of the connected component is shown in Figure~\ref{fig:eigs_gap}.
The eigenmodes corresponding to large eigenvalues decay fast, suggesting that corresponding
neurons have the same membrane potential on relevant time scales and act effectively as a 
single unit. Many such eigenmodes peak (with opposite signs) for left-right 
neuronal pairs (Figure~\ref{fig:fasteigs_gap}), often known to be functionally identical, 
which therefore act as a single unit.  

What is the absolute value of decay constants for various eigenmodes? 
Current knowledge of electrical parameters for \emph{C.\ elegans} neurons allows us to estimate
the decay times only approximately.  Assuming neuron capacitance of $2$pF \cite{GoodmanHAL1998} 
and gap junction conductance of $200$pS, we find a time constant $\tau=10$ms. This implies that 
the slowest non-trivial mode corresponding to the second lowest eigenvalue, $\lambda_2 = 0.12$ 
has decay time of about $83$ms, Figure~\ref{fig:LSdist}.  This eigenvalue, $\lambda_2$, is known as the algebraic 
connectivity of a network \cite{Mohar1991b} and is discussed further in Appendix~\ref{app:eigenratio}.

What is the effect of the dropped term corresponding to the membrane current in \eqref{eq:diffeq_gap}?  As this term
would correspond to adding a scaled identity matrix to the Laplacian, the spectrum should 
uniformly shift to higher values by the corresponding amount. Thus, even the eigenmode corresponding 
to the zero eigenvalue would now have a finite decay time. Assuming the membrane conductance of 
about $100$pS \cite{GoodmanHAL1998}, we find $20$ms decay time. This leads to a $0.5$ increase
in the values of $\lambda$. Now, the slowest non-trivial mode corresponds to 
a decay time of about $16$ms. 

In addition to highlighting groups of neurons that could be functionally related, spectral analysis 
allows us to predict, under linear approximation, the outcome of experiments that study the spread of 
an arbitrarily generated excitation in the neuronal network. Such excitation can be generated in sensory 
neurons by presenting a sensory stimulus \cite{ChalasaniCTGRGB2007} or in any neuron by expressing 
channelrhodopsin in that cell \cite{NagelSHKABOHB2003,NagelBLABG2005,Zhang_ea2007}.  The spread of 
activity can be monitored electrophysiologically or using calcium-sensitive indicators.  

To predict the spread of activity, we may decompose the excitation pattern into the eigenmodes 
and, by taking advantage of eigenmode independence, express temporal evolution
as a superposition of the independently decaying eigenmodes.  The initial redistribution of 
charge would correspond to the fast eigenmodes, whereas the long-term evolution of charge distribution 
would be described by the slow eigenmodes.  Appendix~\ref{app:eigendecomposition}
further discusses eigendecomposition and the interpretation of eigenmodes. 

\subsubsection{Motifs}

\begin{figure}[h]
  \centering
\subfigure[]{
  \includegraphics[width=3in]{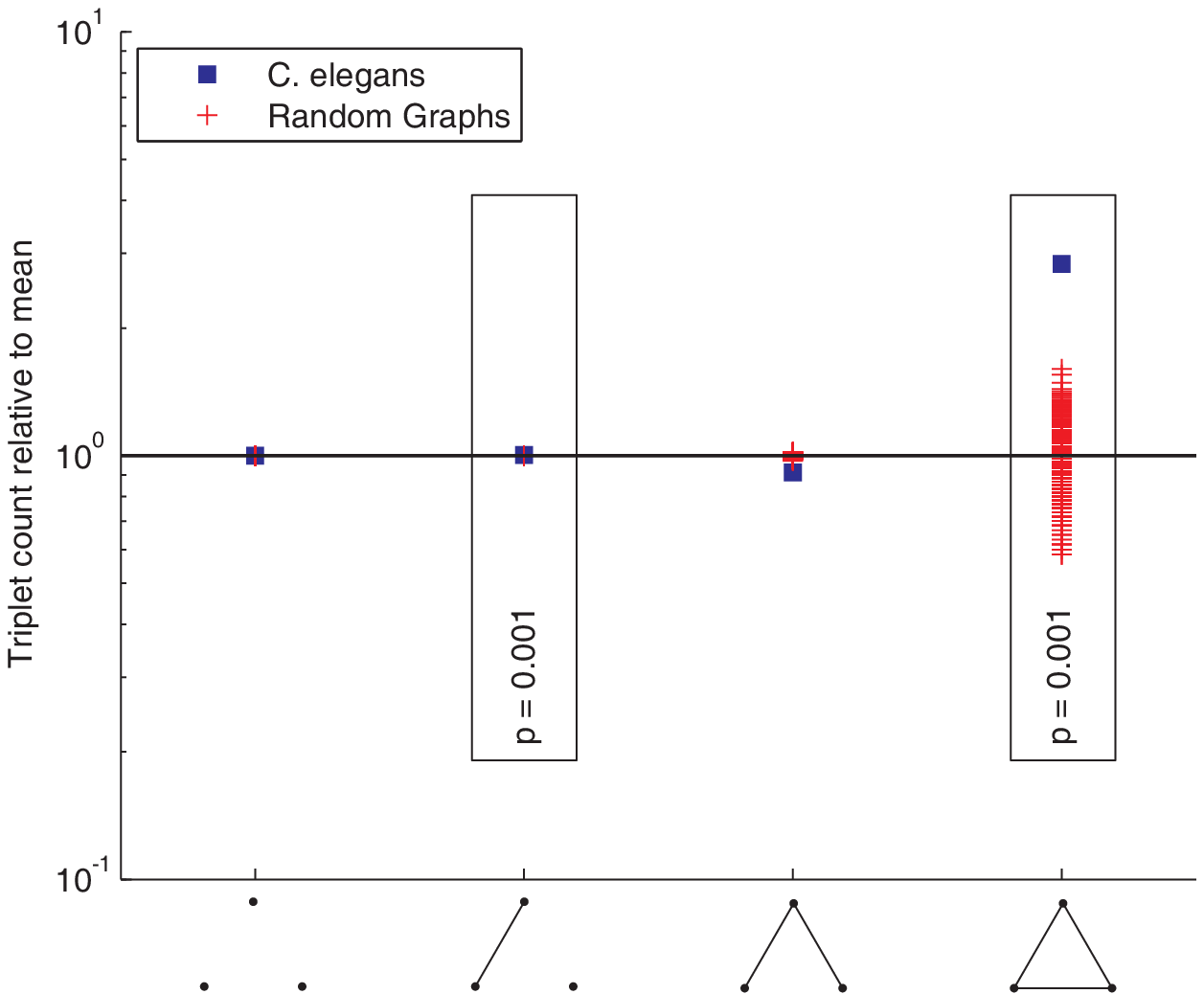}
  \label{fig:SGg3}
}
\subfigure[]{
  \includegraphics[width=3in]{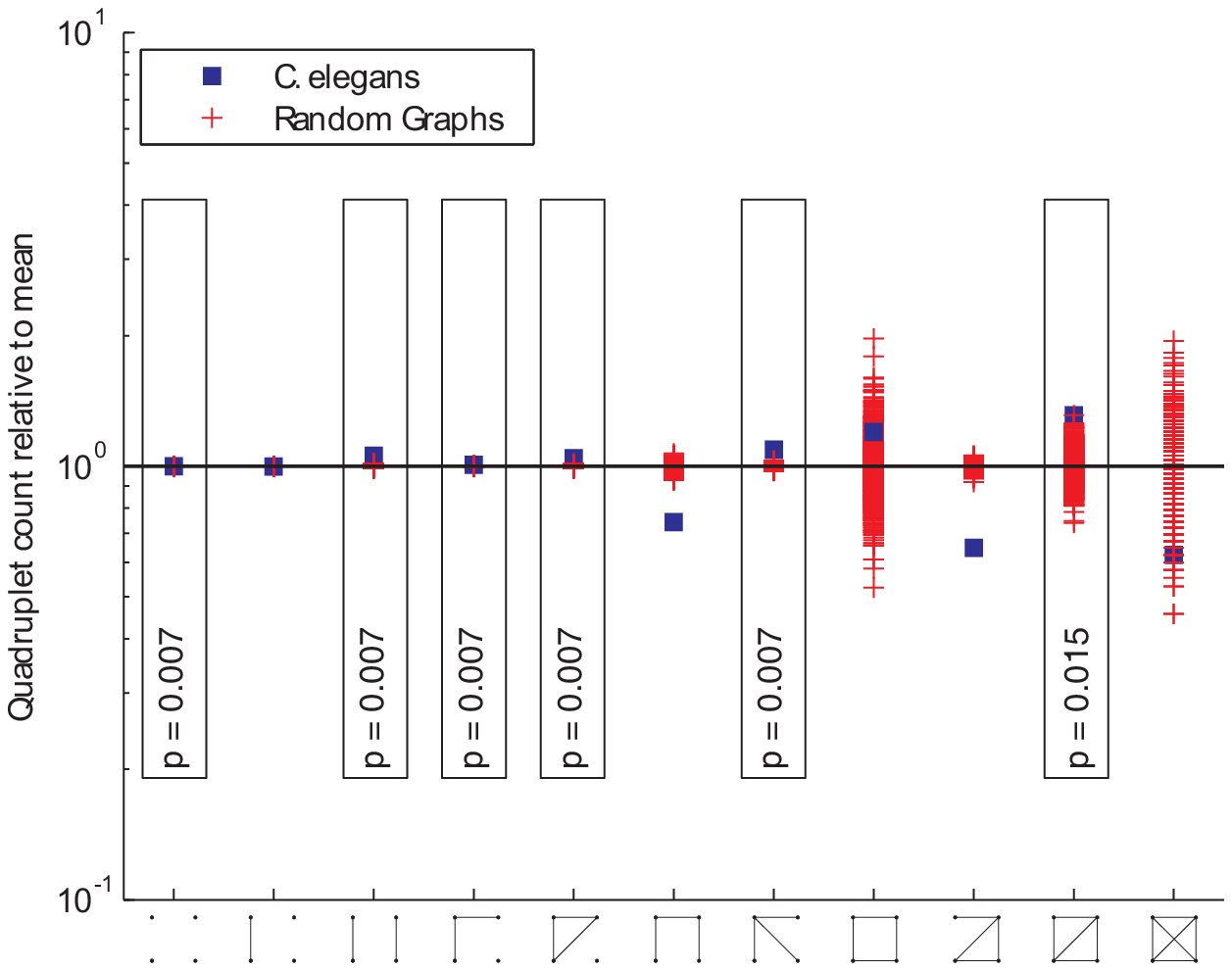}
  \label{fig:SGg4}
}
  \caption{Subnetwork distributions for the gap junction network.  Overrepresented subnetworks are boxed, with the 
	$p$-value from the step-down min-P-based algorithm for multiple-hypothesis correction \cite{ReiglAC2004,SongSRNC2005}
	($n = 1000$) shown inside.  \subref{fig:SGg3}. The ratio of the $3$-subnetwork 
	distribution and for the mean of a degree-preserving ensemble of random networks ($n = 1000$).  The counts
	for the particular random networks that appeared in the ensemble are also shown. 
	\subref{fig:SGg4}. The ratio of the $4$-subnetwork distribution and for 
	the mean of a degree and triangle-preserving ensemble of random networks ($n = 1000$).  The counts
	for the particular random networks that appeared in the ensemble are also shown.
  }
\end{figure}

Several of the quantitative properties computed thus far measure 
global network structure and may determine aspects of system 
operation.  Now we examine the network locally and analyze the frequency
of various connectivity subnetworks among small groups of neurons. Overrepresentation 
in the subnetwork distribution often displays building blocks of the network such as 
computational units \cite{MiloSIKCA2002, Przulj2007}. Since the gap junction network is 
undirected, there are four kinds of subnetworks that can appear over three 
neurons; this distribution is shown in Figure~\ref{fig:SGg3}. As a null-hypothesis we 
use random network ensembles that preserve the degree distribution. We find that fully
connected triplets are overrepresented.

Four neurons can be wired into 11 kinds of subnetworks; this distribution is 
shown in Figure~\ref{fig:SGg4}. In the case of quadruplets, the null-hypothesis preserves 
the degree for each neuron and the number of triangles.  A numerical rewiring procedure is used
to generate samples from these random network ensembles \cite{MaslovS2002,ReiglAC2004}, 
since no analytical expression for expected subnetwork counts is extant
\cite{FosterFGP2007}. We find that a ``fan'' (motif \#7) and a ``diamond'' (motif \#10)
are overrepresented.

Note that neurons participating in motifs also make connections with neurons
outside of the motif, which are traditionally not drawn in putative functional 
circuits \cite{Durbin1987, ChalasaniCTGRGB2007}.  Such putative functional
circuit diagrams may even omit connections within the motif 
\cite{Durbin1987, ChalasaniCTGRGB2007}, which we do not allow.

\subsection{Chemical Synapse Network}

Now we consider the chemical synapse network.  Recall 
that due to structural differences between presynaptic and 
postsynaptic ends of a chemical synapse, electron micrographs can be used to 
determine the directionality of connections.  Hence the adjacency matrix is 
not symmetric as it was for the gap junction network.  

\subsubsection{Basic Structure and Connectivity}

The network that we analyze consists of $279$ neurons and $2194$ directed connections 
implemented by one or more chemical synapses.  
The adjacency matrix of the network shown in Figure~\ref{fig:1} is suggestive of a 
three-layer architecture.  
Table~\ref{tab:2} shows the distribution of connections 
between categories in the three-layer architecture.  Each chemical subnetwork is 
characterized by a high number of recurrent connections, just as for the gap junction.
However, the majority of connections with other subnetworks is consistent with
feedforward information processing (sensory to interneuron and interneuron to motorneurons).
Therefore, a three-layer network abstraction may be more valuable for chemical synapses
than for gap junctions. 

There are two different definitions of connectivity for directed networks.  A 
weakly connected component is a maximal group of neurons which are mutually 
reachable by possibly violating the connection directions, whereas a strongly connected 
component is a maximal group of neurons that are mutually reachable without 
violating the connection directions.  The whole chemical synapse network is 
weakly connected and can be divided into a giant strongly connected component 
with $237$ neurons, a smaller strongly connected component of $2$ neurons, and 
$40$ neurons that are not strongly connected (Table~\ref{tab:NSCCc}).
  
The random directed network corresponding to the chemical network is fully 
weakly connected, even when the degree distribution is taken into account
(see {\sc Methods}). A strongly connected giant component as small as
in the chemical network is not likely in a random network (see \cite{Karp1990}).  
Thus, the chemical network is more segregated than would be expected for a random network.

\subsubsection{Distributions of Degree, Multiplicity and the Number of Terminals}

\begin{figure}
  \begin{minipage}{3in}
\subfigure[]{
  \includegraphics[width=2.6in]{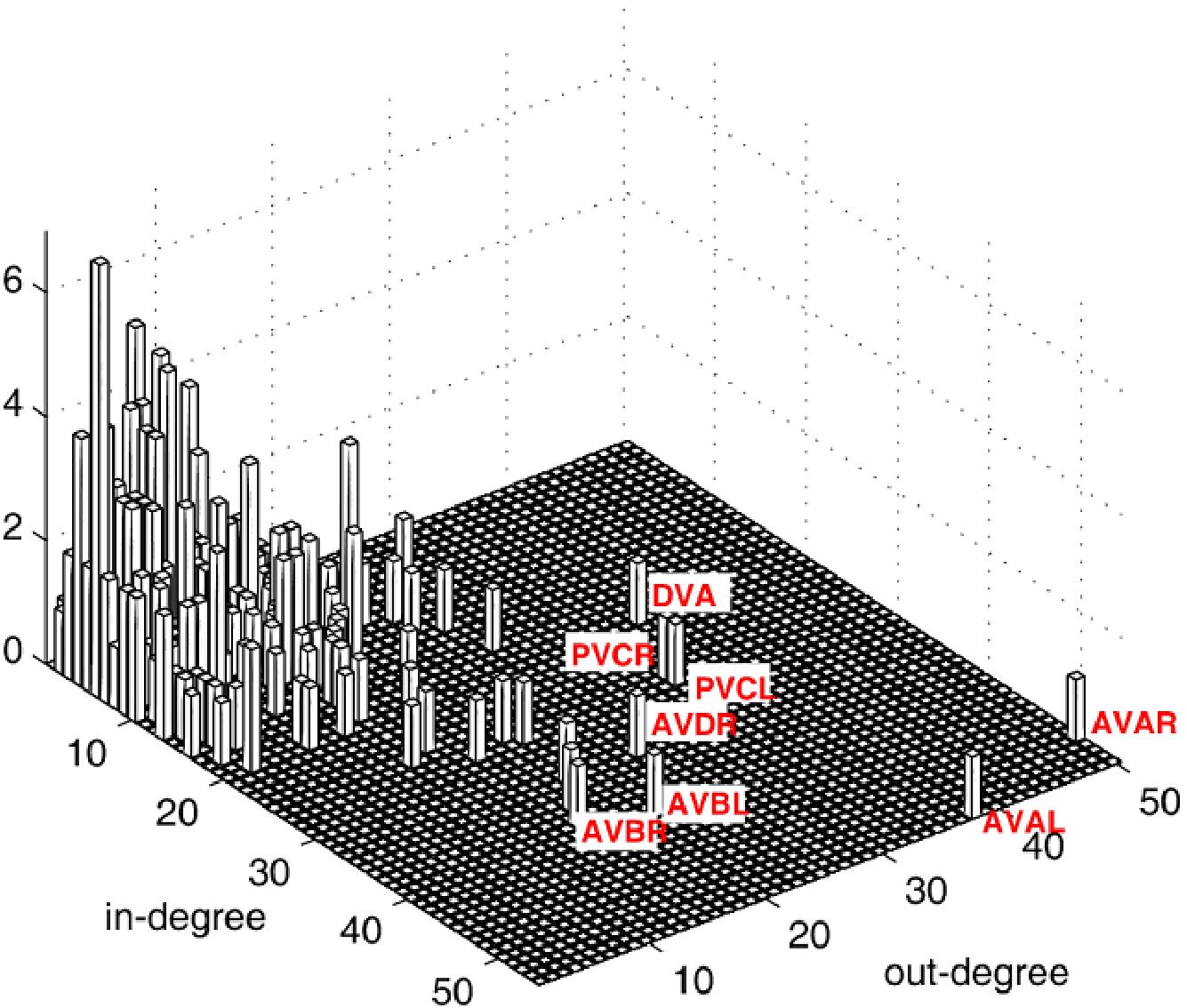}
  \label{fig:DDjo}
}
\subfigure[]{
  \includegraphics[width=2.6in]{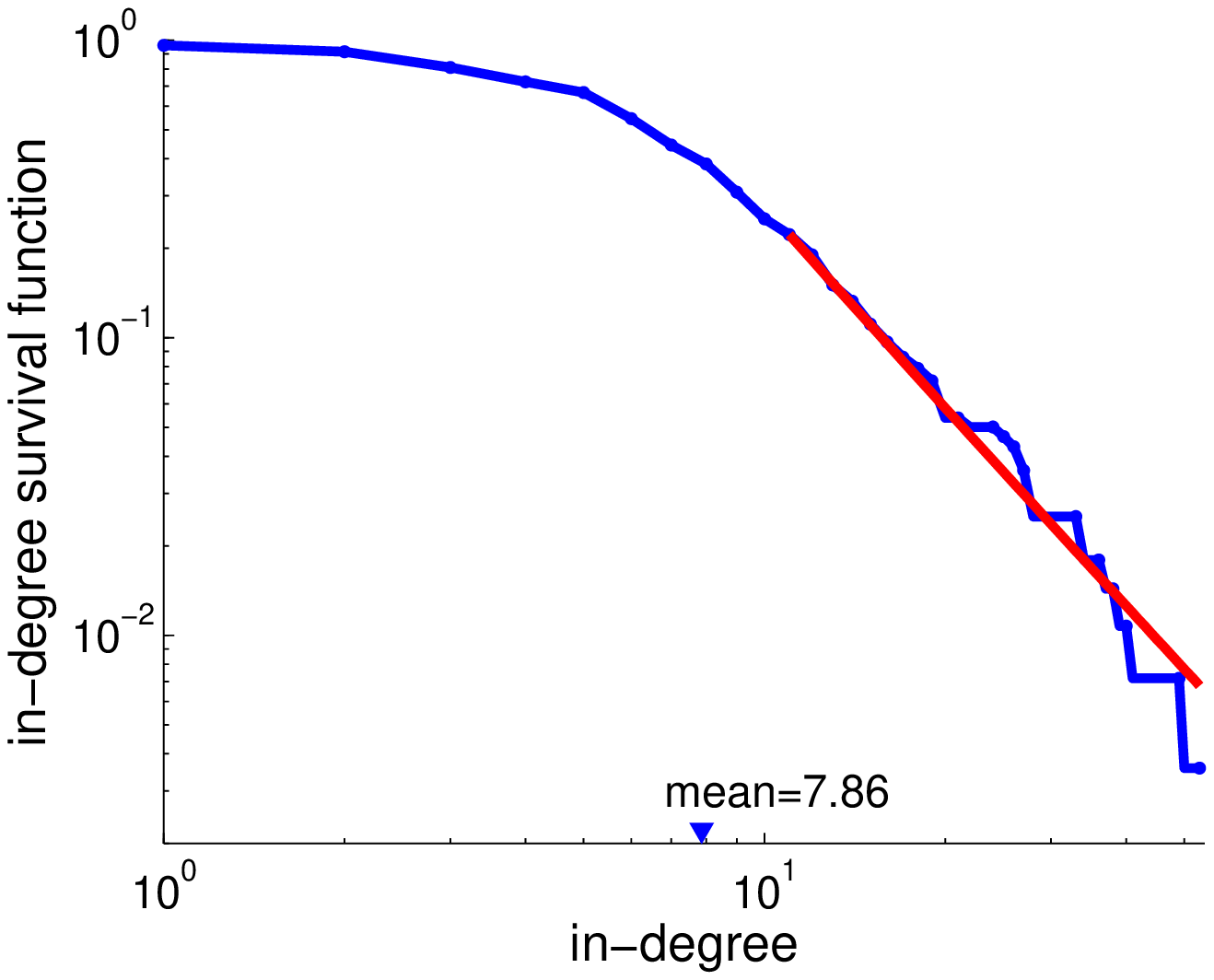}
  \label{fig:DDin}
}
\subfigure[]{
  \includegraphics[width=2.6in]{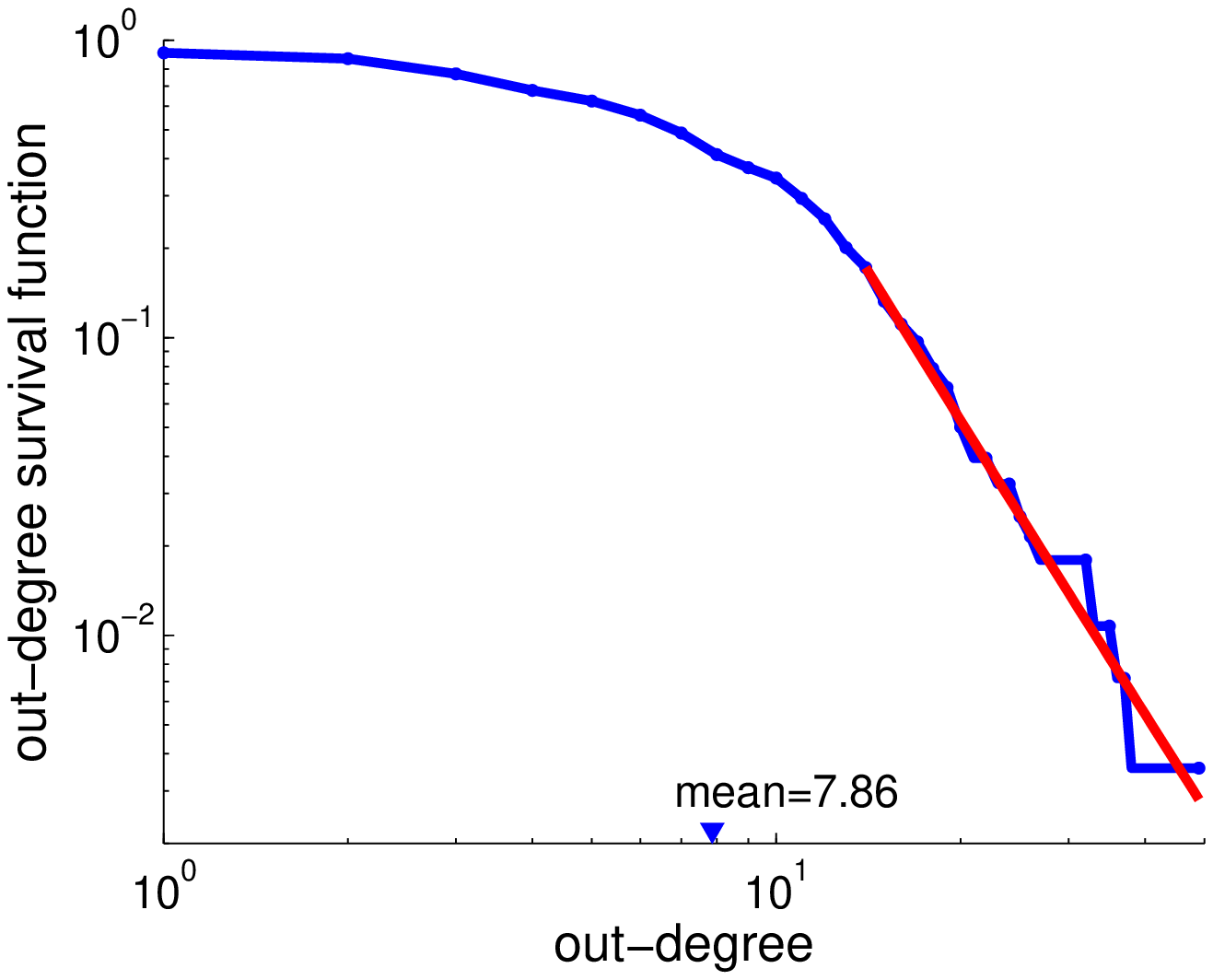}
  \label{fig:DDout}
}
\end{minipage}
\begin{minipage}{3in}
\subfigure[]{
  \includegraphics[width=2.6in]{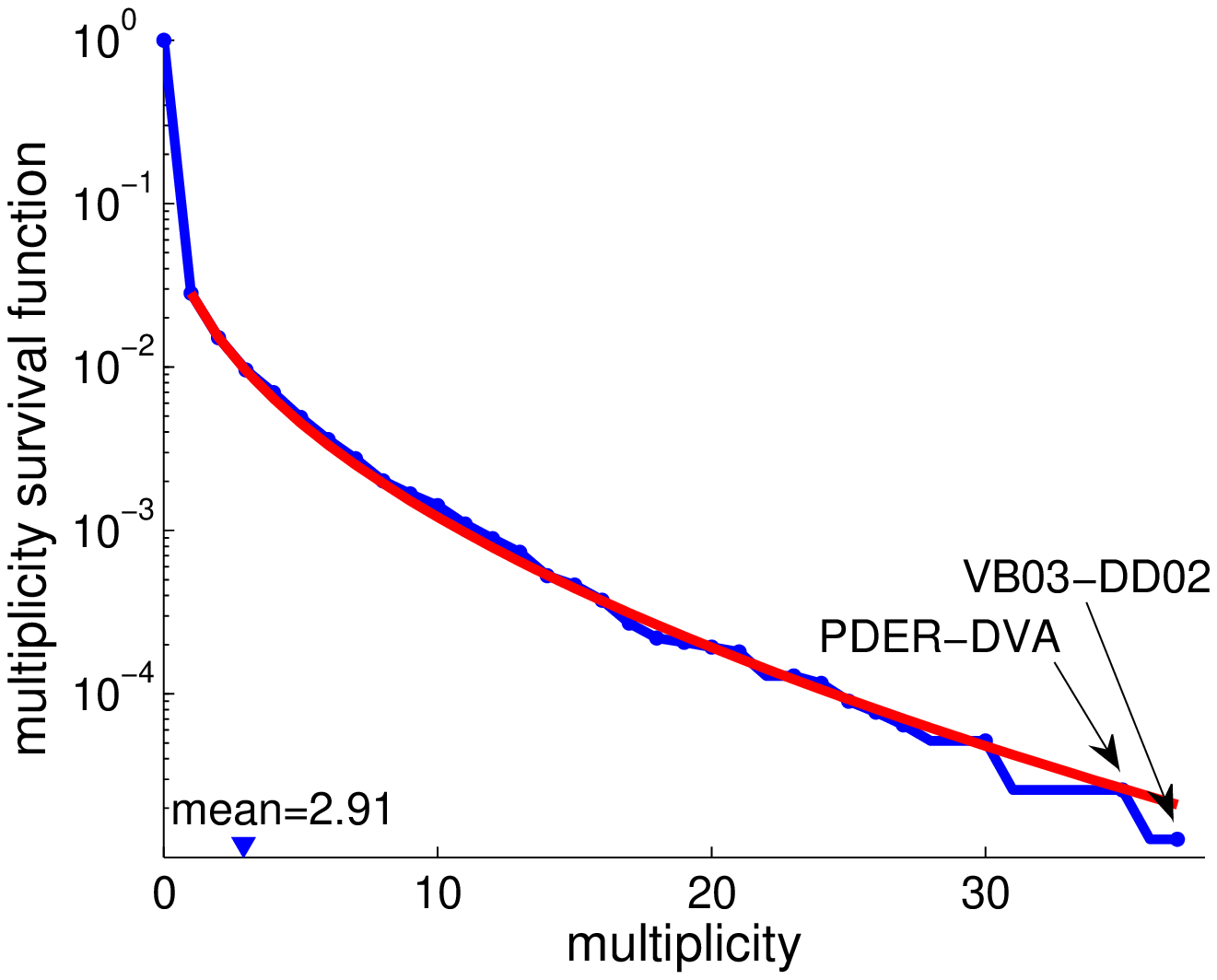}
  \label{fig:MDc}
}
\subfigure[]{
  \includegraphics[width=2.6in]{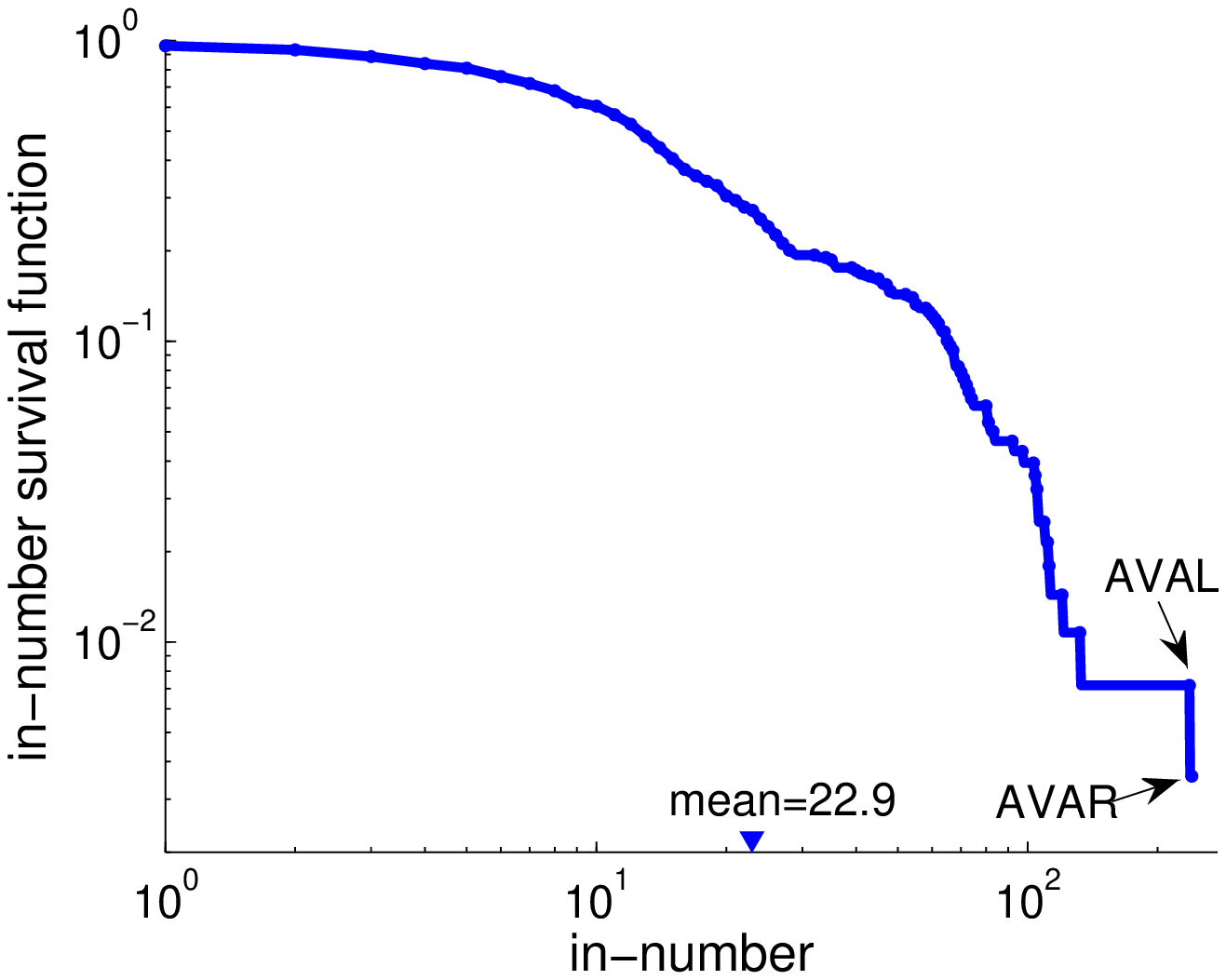}
  \label{fig:NDin}
}
\subfigure[]{
  \includegraphics[width=2.6in]{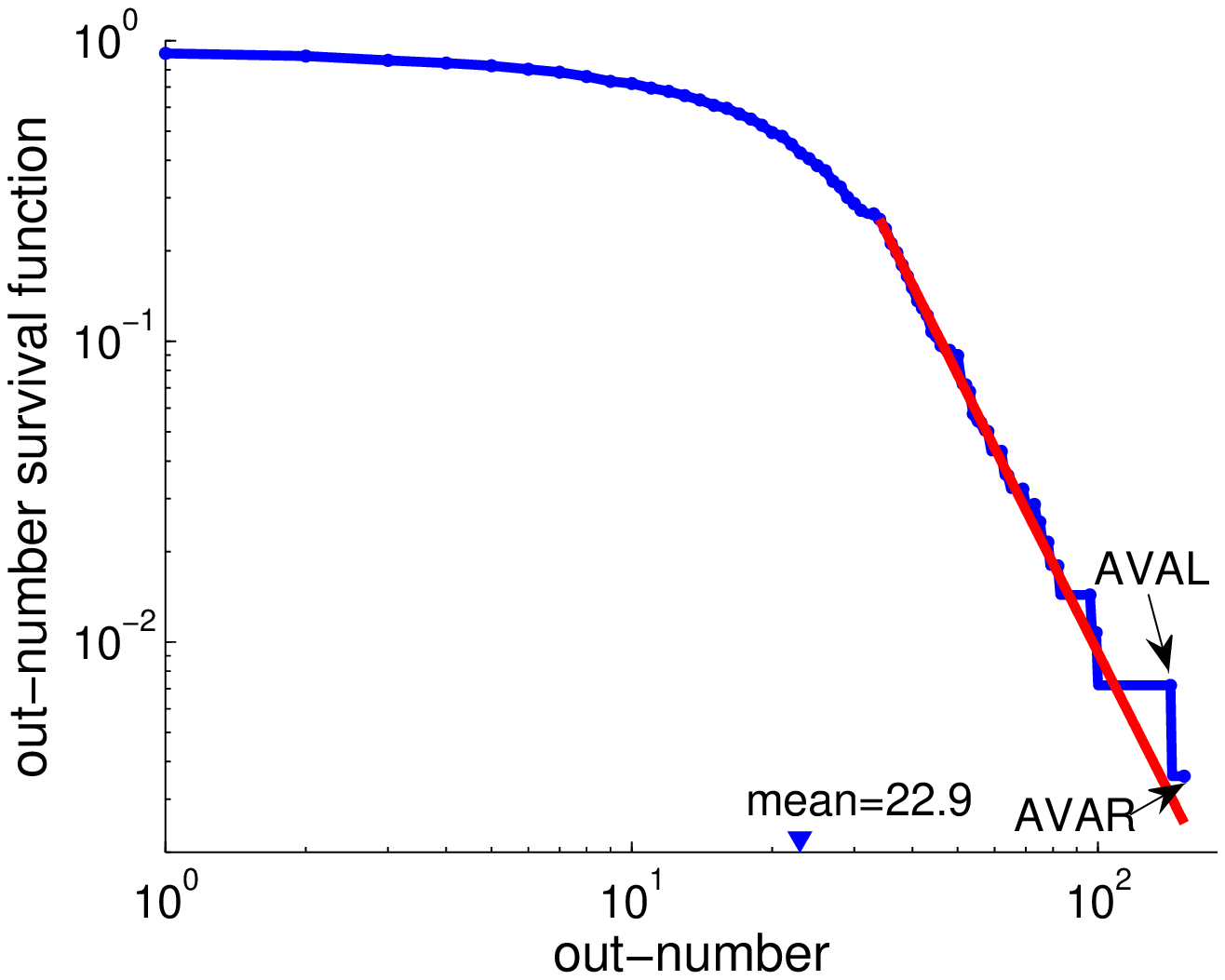}
  \label{fig:NDout}
}
\end{minipage}

\caption{Degree distribution \subref{fig:DDjo} and survival functions for the 
    distributions of in-/out-degree, multiplicity, and in-/out-number of synaptic 
    terminals in the chemical synapse network \subref{fig:DDin}--\subref{fig:NDout}.
    Neurons or connections with unusually high statistics are labeled.   
    The tails of the distributions can be fit by a power law with exponents
    $3.17$ for in-degree \subref{fig:DDin}; $4.22$ for out-degree \subref{fig:DDout};
    and $4.05$ for out-number \subref{fig:NDout}. The exponents for the survival
    function fits can be obtained by subtracting one. The survival function of
    the multiplicity distribution for $m \geq 1$ can be fit by a stretched exponential 
    of the form $e^{-(m/\beta)^{\gamma}}$ where $\beta = 0.36$ and
    $\gamma = 0.47$ \subref{fig:MDc}. No satisfactory fit was found for 
    the distribution of in-numbers \subref{fig:NDin}.  
  }
\end{figure}

Since chemical synapses form a directed network, neuron connectivity 
is characterized by in-degrees (the number of incoming connections) 
and out-degrees (the number of outgoing connections) rather than simply 
degrees. The joint distribution of in-degrees and out-degrees is shown 
in Figure~\ref{fig:DDjo}.  As can be seen by the distribution clustering 
around the diagonal line, the in-degrees and out-degrees are correlated; 
the correlation coefficient is $0.52$, very similar to the 
correlation coefficient of email networks, $0.53$ \cite{NewmanFB2002}.
  
The survival functions associated with the marginal distributions of 
in-degrees and out-degrees are shown in Figures~\ref{fig:DDin} and \ref{fig:DDout} 
respectively. The mean number of incoming and outgoing connections is $7.86$ each. 
We attempt to fit these distributions. The tails of the two distributions
can be satisfactorily fit by power laws with exponents $3.17$ and $4.22$
respectively. Exponential fit is ruled out ($p$-value $<0.1$) for the in-degree
but not for the out-degree distribution. In the latter case, the log-likelihood 
is insignificantly lower for the exponential decay than for the power law.

Multiplicity of connection, $m_{ij}$, is the number of synapses in parallel 
from neuron $i$ to neuron $j$. The corresponding survival function (including 
unconnected pairs) is shown in Figure~\ref{fig:MDc}. The mean number of 
synapses per connection (excluding unconnected pairs) is $2.91$. The tail of 
the distribution can be fitted by an exponential, but not by a power law 
($p$-value $<0.1$). In addition, the whole distribution ($m \geq 1$) can be fit 
by a stretched exponential (or Weibull) distribution,
$p(m) \sim (m/\beta)^{\gamma-1} e^{-(m/\beta)^\gamma}$
with the scale parameter $\beta=0.36$ and the shape parameter $\gamma=0.47$.
A stretched exponential applied to the whole distribution has the same 
number of fitting parameters as an exponential decay fit to the tail 
starting with an adjustable $m$. 
Log-likelihood comparison of the tail exponential and the stretched exponential 
favors the latter.

As for the gap junction network, we can also study the distribution of 
number of synaptic terminals on a neuron.  This involves adding the 
multiplicities of the connections, rather than just counting the number of pre- or
post-synaptic partners. The joint histogram (not shown) exhibits similar correlation as for 
the degree distribution, with correlation coefficient $0.42$. 

Figures~\ref{fig:NDin} and \ref{fig:NDout} show the marginal survival 
functions for the number of post-synaptic terminals (in-number) and 
the number of pre-synaptic terminals (out-number). The mean number of
pre- and post-synaptic terminals is $22.9$ each. We were unable to 
find a satisfactory simple fit to the in-number distribution, Figure~\ref{fig:NDin}. 
The tail of the out-number distribution could be fit by a power law with
exponent $4.05$, but not by an exponential, Figure~\ref{fig:NDout}.

As for the gap junction network, we can identify central neurons (cf.\ \cite{MasudaKK2009,ChatterjeeS2008})
for the chemical network. The degree centrality in a directed network may 
be defined with respect to the in-degree or the out-degree.  
Interestingly, neuron AVAL has the best in-degree, whereas AVAR has
the best out-degree and AVAR has the best out-degree and AVAL has the second
best out-degree, Figure~\ref{fig:DDjo}.

\subsubsection{Small World Properties}

In the strongly connected component, we can define the directed geodesic distance 
as the shortest path between two neurons that respects the direction of the connections.  
This distribution is shown in Figure~S\ref{fig:gdDc}.  The directed characteristic
path length, $L$, is the average directed geodesic distance over 
all pairs of neurons in the strongly connected component and is 
computed to be $3.48$ steps. For a random network degree-matched 
to the chemical network, one would expect $L = 2.91$. The similarity of
the geodesic distances suggests that signals diffuse as quickly as in a random network.   

Although there are several definitions of clustering for 
directed graphs in the literature \cite{Fagiolo2007}, we use the clustering of the out-connected
neighbors since it captures signal flow emanating from a given neuron.  This is:
\begin{equation}
C = \frac{1}{N} \sum_i C_i \hspace{2cm} C_i = \frac{E(\mathcal{N}_i)}{k_i(k_i - 1)} \mbox{,}
\end{equation}
where $E(\mathcal{N}_i)$ is the number of connections between out-neighbors of neuron $i$, 
$k_i$ is the number of out-neighbors of $i$, and $C_i$ measures the density 
of connections in the neighborhood of neuron $i$.  For the chemical network, the 
clustering coefficient is $0.22$.  Using the Watts-Strogatz approximations 
to $L$ and $C$, the clustering coefficient for a random network is 
$C_r \approx \tfrac{1}{N} \exp(\tfrac{\ln(N)}{L})$; so for $N = 279$ and $L = 3.48$, 
a random network would have $C \sim 0.018$. For a degree-matched random network we computed
the clustering coefficient $C = 0.079$. Since the clustering coefficient 
for the chemical network is much more than a similar random directed network, 
it may be considered a small-world network, cf.\ Table~\ref{tab:SMN}.

For directed networks, measures of in-closeness and out-closeness may be defined 
using the average directed geodesic distance.  In particular, the normalized in-closeness 
is the average geodesic distance from all other neurons to a given neuron:
\begin{equation}
d_{i\rm{avg}}(i) = \frac{1}{N-1}\sum_{j: j\neq i} d_{ji} \mbox{,}
\end{equation}
and the out-closeness is the average geodesic distance from a given neuron to all other neurons:
\begin{equation}
d_{o\rm{avg}}(i) = \frac{1}{N-1}\sum_{j: j\neq i} d_{ij} \mbox{,}
\end{equation}
where $N$ is the number of neurons.  Normalized centralities are the inverses:
$c_{ic}(i) = 1/d_{i\rm{avg}}(i)$ and $c_{oc}(i) = 1/d_{o\rm{avg}}(i)$.
The motivation behind these indices is similar to that in the gap junction case. In-closeness 
central neurons can be easily reached from all other neurons in the network. Out-closeness
central neurons can easily reach all other neurons in the network. Normalized in-closeness 
centrality $c_{ic}(i)$ and normalized out-closeness centrality $c_{oc}(i)$ are weakly 
anti-correlated, with correlation coefficient $-0.12$.

For the giant component of the chemical network, the most in-closeness central
neurons include AVAL, AVAR, AVBR, AVEL, AVER, and AVBL.  All are
command interneurons involved in the locomotory circuit; these neurons
are also central in the gap junction network. The in-closeness centrality 
of command interneurons may indicate that in the \emph{C.\ elegans} nervous system,
signals can propagate efficiently from various sources towards these 
neurons and that they are in a good position to integrate it. 

The most out-closeness central neurons include DVA, ADEL, ADER, PVPR, AVJL, HSNR, PVCL,
and BDUR.  Only PVCL is a command interneuron involved in locomotion. The neuron DVA 
is an interneuron that performs mechanosensory integration; 
ADEL/R are sensory dopaminergic neurons in the head; and the other central neurons
are interneurons in several parts of the worm. The out-closeness centrality
of these neurons may indicate that signals can propagate efficiently from
these neurons to the rest of the network and that they are in a good position for broadcast.

\subsubsection{Spectral Properties}

Although chemical synapses are likely to introduce more nonlinearities than
gap junctions, linear systems analysis can provide interesting insights, especially 
in the absence of other tools. Such an approach has additional merit in 
\emph{C.\ elegans}, where neurons do not fire classical action potentials \cite{GoodmanHAL1998}
and have chemical synapses that likely release neurotransmitters tonically \cite{FerreeL1999}.
To justify such analysis, a system of linear equations may be derived by approximating
sigmoidal synaptic transmission functions with linear dependencies. This can be done by 
expanding synaptic transmission functions into a Taylor series around an equilibrium point \cite{FerreeL1999}.

A major source of uncertainty in linear systems analysis of the chemical network is the unknown
sign of connections, i.e.\ excitatory or inhibitory, due to the difficulty 
in performing electrophysiology experiments. We use a rough approximation that GABAergic 
synapses are inhibitory, whereas glutamergic and cholinergic synapses are excitatory 
\cite{BrownleeF1999}, but see \cite{ChalasaniCTGRGB2007}. Thus inhibitory
neurons are identified by looking at GABA expression \cite{McIntireJKH1993}.\footnote{The 
$26$ GABAergic neurons are DVB, AVL, RIS, DD01--DD06, VD01--VD13, and the four RME neurons.}

Similarly to the gap junction network, we write the system of linear differential equations
for the chemical synapse network \cite{Koch1999,FerreeL1999}:
\begin{equation}
C_i\tfrac{dV_i}{dt} = \sum_j V_j g_{ji}-g^m_i V_i\mbox{,}
\end{equation}
where $V_i$ is the membrane potential of neuron $i$ measured relative to the equilibrium, 
$C_i$ is the membrane capacitance of neuron $i$, $g_{ji}$ is the conductance in neuron $i$
contributed by a chemical synapse in response to voltage $V_j$ measured
relative to the equilibrium and $g^m_i$ is the membrane
conductance of neuron $i$. Assuming that each neuron has the same capacitance $C$ and each
chemical synapse contact has the same conductance $g$, i.e. $g_{ij} = g A_{ij}$, we can rewrite this equation 
in terms of the time constant $\tau = C / g$  as:
\begin{equation}
\label{eq:diffeq_chem}
\tau\tfrac{dV_i}{dt} = \sum_j V_j A_{ji} - \tfrac{g^m_i}{g} V_i\mbox{.}
\end{equation}
 
To avoid redundancy we defer analyzing this system of differential equations to 
the next section, where we consider the combined system including both gap junctions and chemical synapses.

\subsubsection{Motifs}

We also find subnetwork distributions for the chemical synapse network.  
Since the network is directed, there are many more possible subnetworks.  
In particular there are $3$ possible subnetworks on two neurons and $16$ 
possible subnetworks on three neurons.  We identify overrepresented 
subnetworks by comparing to random networks, generated with a rewiring 
procedure \cite{MaslovS2002,ReiglAC2004}.
Such random network ensembles preserve in-degree and
out-degree in the case of doublets and, additionally, the numbers of 
bidirectional and unidirectional connections for each neuron
in the case of triplets.

\begin{figure}[h]
  \centering
  \subfigure[]{
    \includegraphics[width=3in]{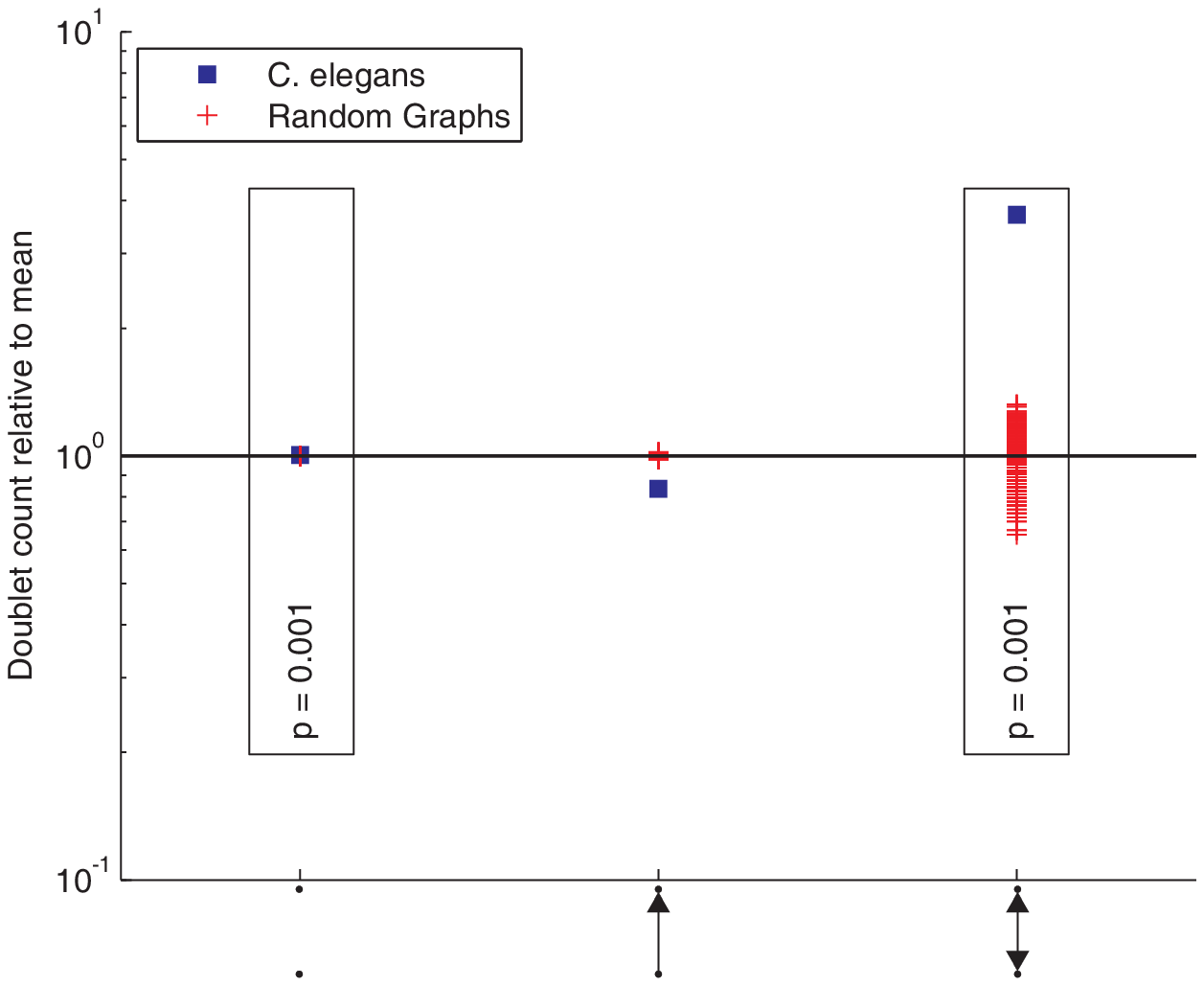}
    \label{fig:SGc2}
  }
  \subfigure[]{
    \includegraphics[width=3in]{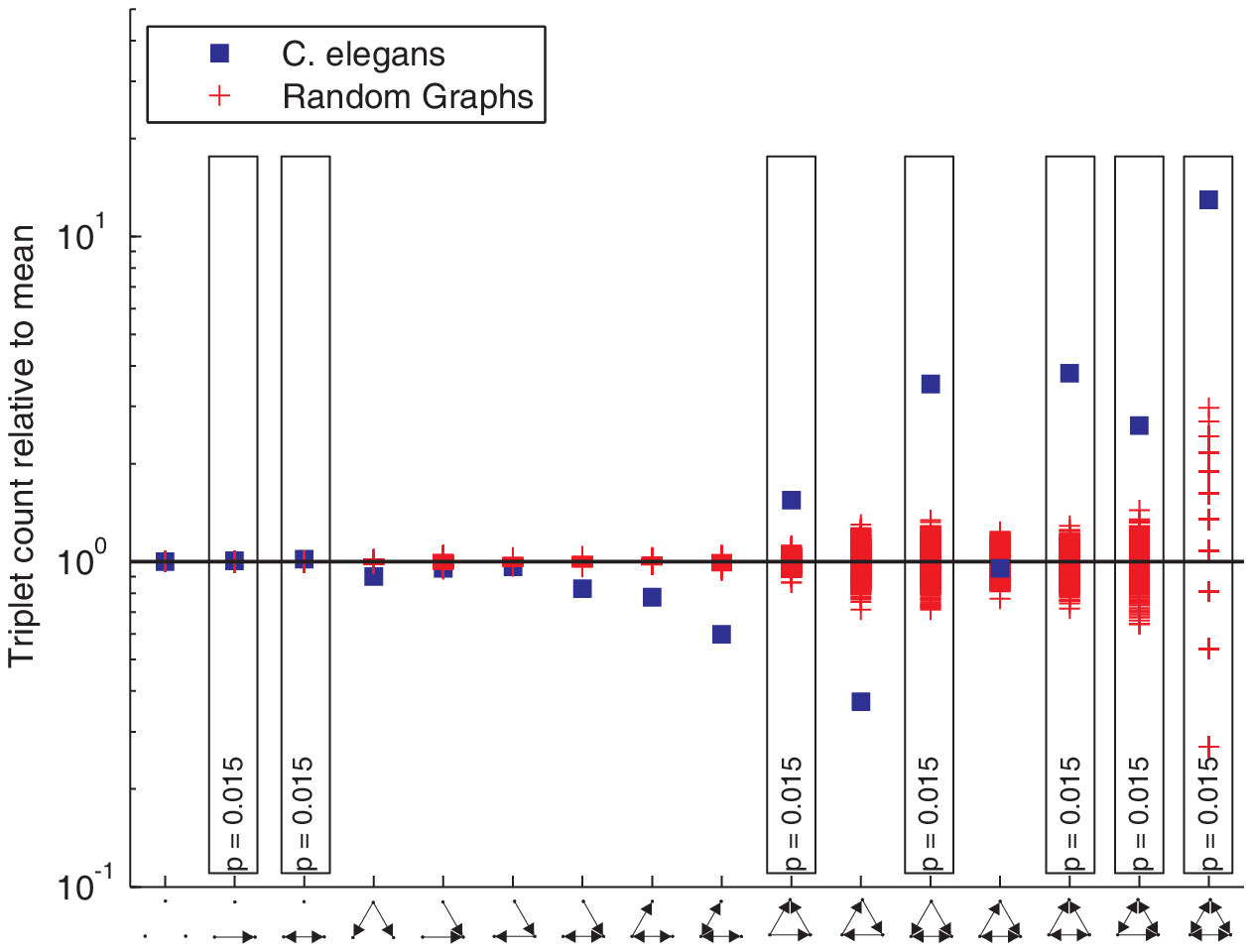}
    \label{fig:SGc3}
  }
  \caption{Subnetwork distributions for the chemical synapse network. 
    Overrepresented subnetworks are boxed, with the $p$-value from the 
    step-down min-P-based algorithm for multiple-hypothesis correction 
    \cite{ReiglAC2004,SongSRNC2005}	($n = 1000$) shown inside.  \subref{fig:SGc2}. 
    The ratio of the $2$-subnetwork distribution and the mean of a random 
    network ensemble ($n = 1000$).  Realizations of the random network 
    ensemble are also shown.  \subref{fig:SGc3}. The ratio of the $3$-subnetwork 
    distribution and the mean of a random network ensemble ($n = 1000$).  
    Realizations of the	random network ensemble are also shown.
    }
\end{figure}

Figures~\ref{fig:SGc2} and \ref{fig:SGc3} 
show the subnetwork distributions on two and three neurons, respectively. 
We find that the \emph{C.\ elegans} 
network contains similar overrepresented subnetworks as
found by analyzing incomplete data \cite{MiloSIKCA2002,ReiglAC2004}.  
For example, there is greater reciprocity 
in the chemical network than would be expected in a random network.  Similarly, 
triplets with connections (of any direction) between each pair of neurons
(seven rightmost triplets in Figure~\ref{fig:SGc3})
collectively occur with much greater frequency 
than would be expected for a random network.

Overrepresentation of reciprocal \cite[Ch.~7]{Durbin1987} 
and triangle \cite{WhiteSTB1986} motifs were previously noted.  
Such overrepresentation
would arise naturally if proximity was a limiting factor for connectivity,
however there is no evidence for this limitation.  
Rather we believe motifs have a functional role.

\subsection{Full Network}

Having considered the gap junction network and the chemical synapse network 
separately, we also examine the two networks collectively.  To study the two 
networks, one may either look at a single network that takes the union of the connections 
of the two networks or one may look at the interaction between the two networks.

\subsubsection{Single Combined Network}

First we look at a combined network, which is produced by simply adding 
the adjacency matrices of the gap junction and chemical networks together, while
ignoring connection weights.  Thus we implicitly treat gap junction connections 
as double-sided directed connections.  This new network consists of 
$279$ neurons and $2990$ directed connections.  It has one large strongly connected 
component of $274$ neurons and $5$ strongly isolated neurons.    The five 
isolated neurons are IL2DL/R, PLNR, DD06, and PVDR; this set is simply the 
intersection of the isolated neurons in the gap junction and chemical networks 
and does not seem to have any commonalities among members.  Of course, 
it follows that since the chemical network is a single weakly connected component, 
this combined network is also a single weakly connected component.

Naturally, the combined network is more compact than the individual networks.
The mean path length $L = 2.87$, the geodesic distance distribution (Figure~S\ref{fig:geoDb}) 
becomes narrower. For a random network degree-matched to the combined network, 
one would expect $L = 2.62$, not significantly different. The clustering coefficient 
for the combined network is $C = 0.26$.  The clustering coefficient for a 
similar random network would have been $C = 0.026$ \cite{WattsS1998}, and
for a degree-matched random network $C = 0.10$. Therefore,
the combined network, just like the individual networks, may be classified as small world.  
Turning to closeness centrality, the most in-close central neurons are AVAL/R, AVBR/L, and AVEL/R,
as would be expected from the individual networks.  The most out-close central neurons are
DVA, ADEL, AVAR, AVBL, and AVAL, which include the top out-close neurons for both individual
networks.

We can also calculate the degree distribution of this combined network.  
The correlation coefficient between the in-degree and out-degree is $0.71$; 
it is not surprising that the coefficient is so large considering that 
the gap junctions introduce an in- and out-connection simultaneously.  
Similar to the chemical synapse network, the tails of both 
the in-degree and the out-degree survival functions 
(Figures~S\ref{fig:inDb} and S\ref{fig:outDb}) can be fit with power laws.
The tail of the out-degree could also be fit by an exponential decay, albeit
with lower likelihood.
 
The neurons with the greatest degree centrality are AVAL and AVAR.  
As for the chemical synapse network, neuron AVAL has the best in-degree 
and AVAR has the second best in-degree, 
whereas AVAR has the best out-degree and AVAL has the 
second best out-degree (Figures~S\ref{fig:inDb} and S\ref{fig:outDb}).   
The next two neurons are AVBL/R in both in-degree and out-degree senses.  

As for the chemical synapse network, the tail of the out-number distribution
was fit by a power law and the tail of the in-number distribution could not 
be fit satisfactorily. The tail of the out-number distribution could also
be fit by an exponential, albeit with lower likelihood. The multiplicity can
be fit satisfactorily by a stretched exponential.

\subsubsection{Spectral properties}

In this section we apply linear systems analysis to the combined network
of chemical synapses and gap junctions taking into account multiplicities
of individual connections. Due to our ignorance about the relative
conductance of a single gap junction and of a single chemical synapse, we
assume that they are equal. By combining equations \eqref{eq:diffeq_gap} and 
\eqref{eq:diffeq_chem} we arrive at:
\begin{equation}
\tau\tfrac{dV_i}{dt} = \sum_j [-V_j L^{\rm{(gap)}}_{ij}+ V_j A^{\rm{(chem)}}_{ji}]- \tfrac{g^m_i}{g} V_i\mbox{,}
\end{equation}
where $A^{\rm{(chem)}}_{ji}$ is negative if neuron $j$ is GABAergic and positive otherwise.  

\begin{figure}[h]
  \centering
  \subfigure[]{
	\includegraphics[width=3in]{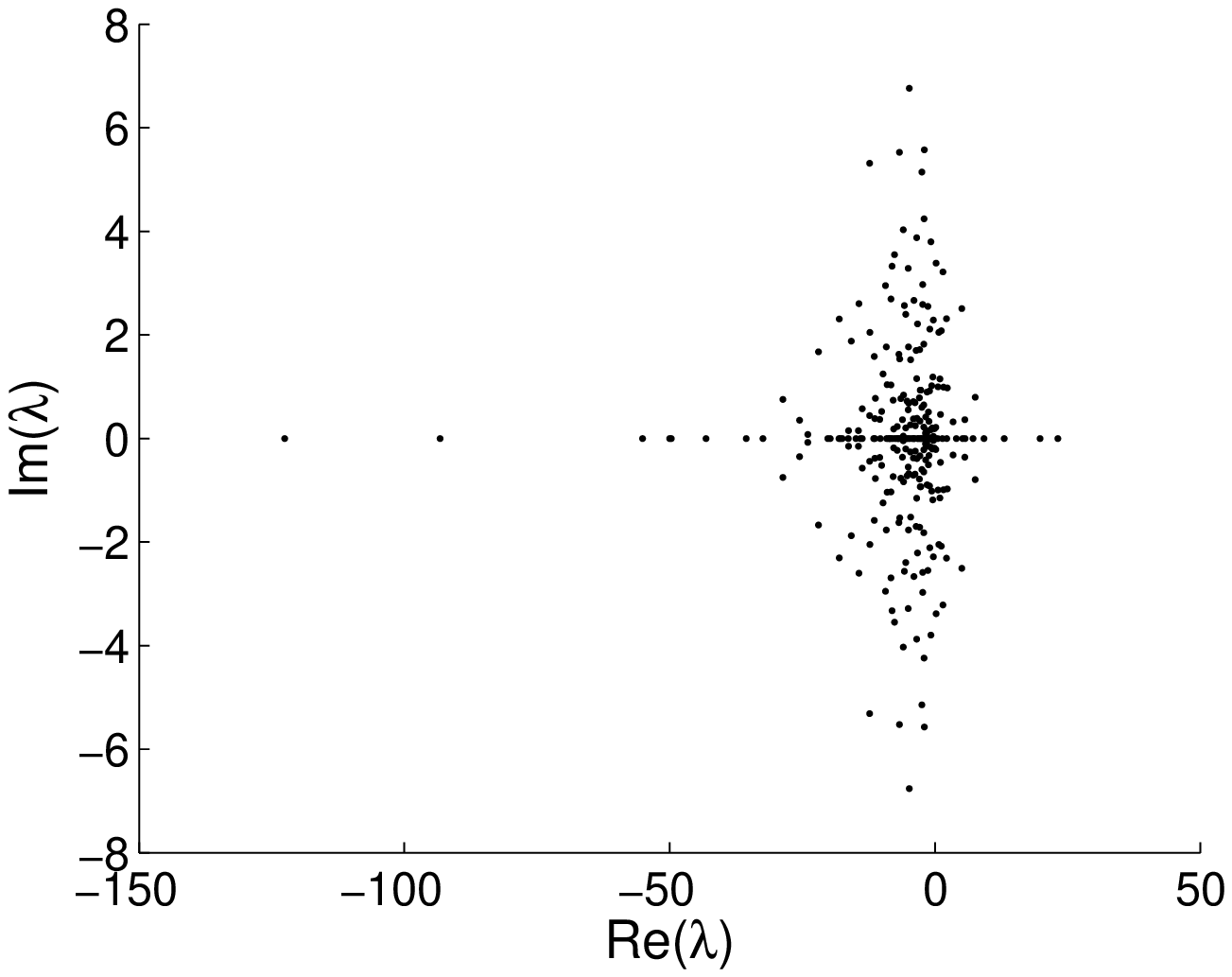}
	\label{fig:comb_eigs_plane}
  }
  \subfigure[]{
	\includegraphics[width=3in]{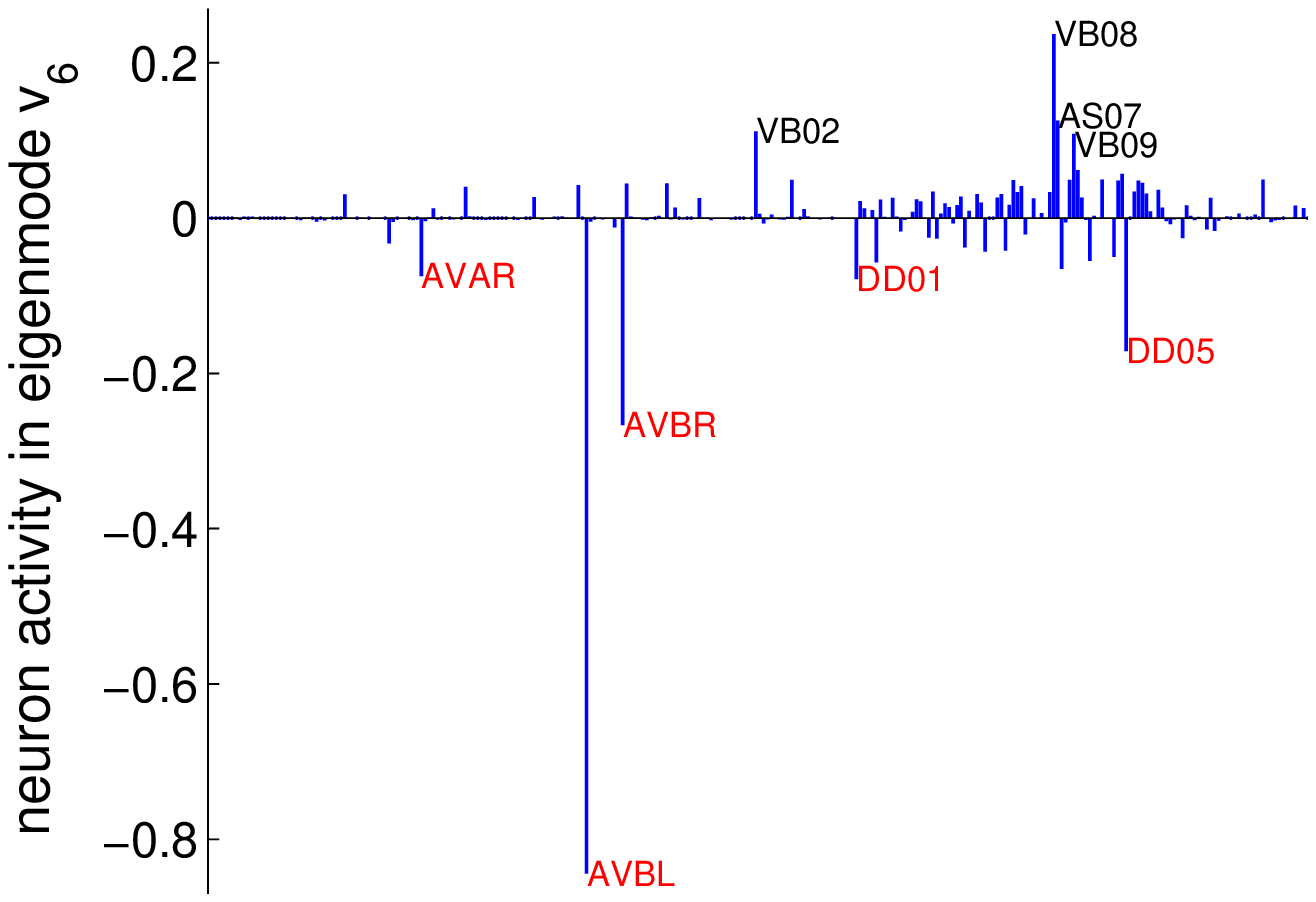}
	\label{fig:speceig_comb_v6}
  }
  \subfigure[]{
	\includegraphics[width=2in]{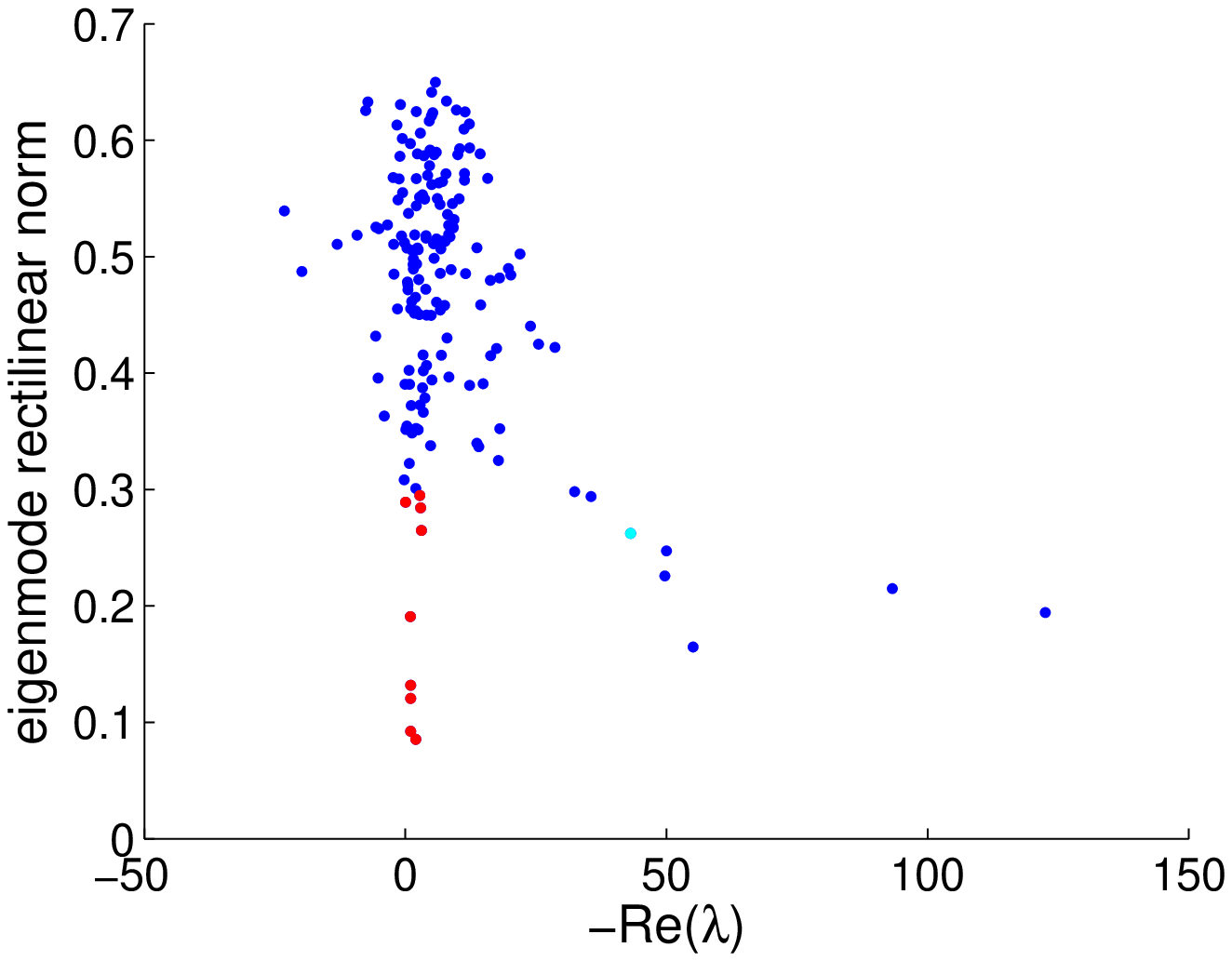}
	\label{fig:chemeigs}
  }
  \subfigure[]{
	\includegraphics[width=4.2in]{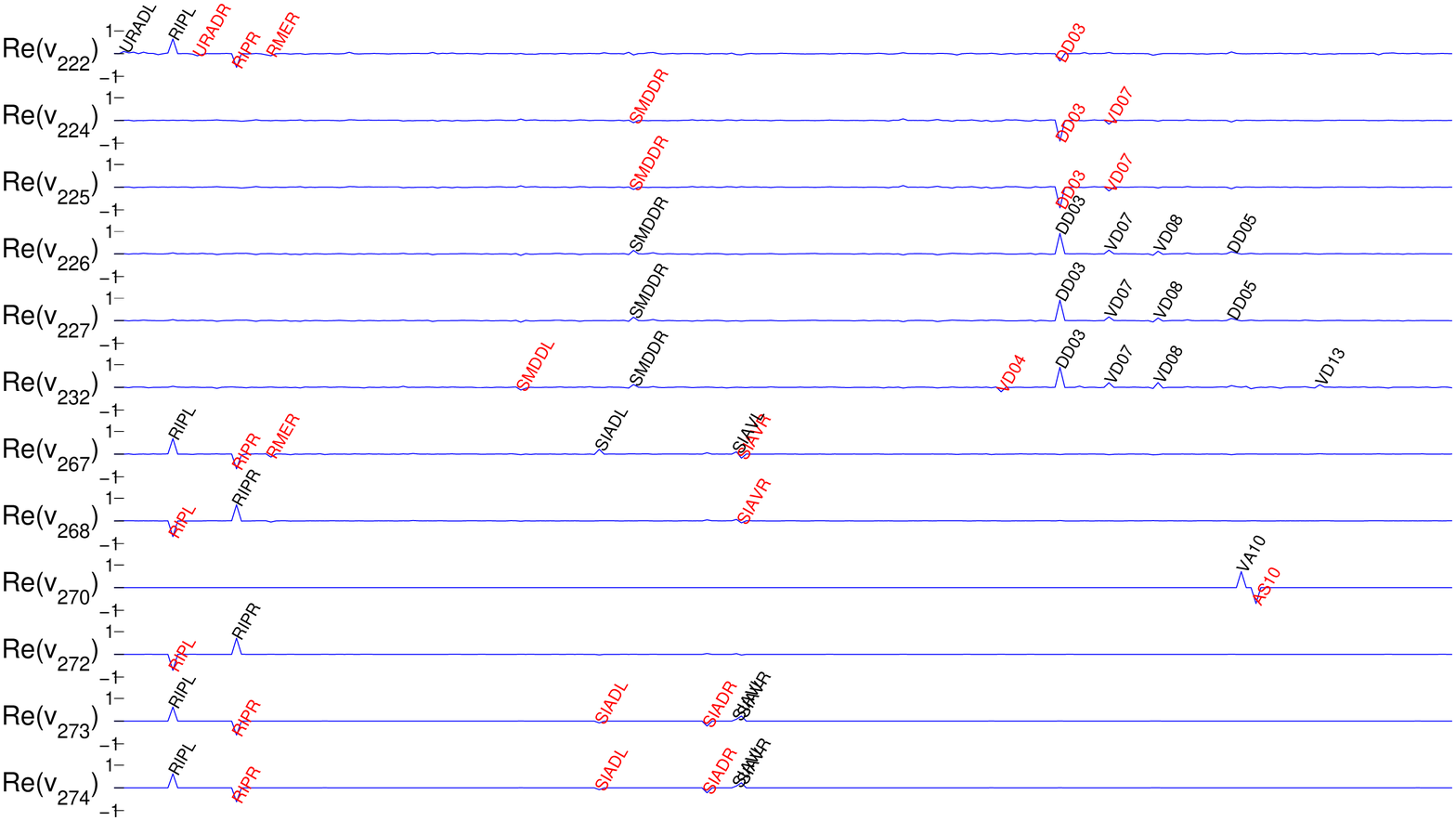}
	\label{fig:results}
  }
  \caption{Linear systems analysis for the strong giant component of the combined network.  
	\subref{fig:comb_eigs_plane}. Eigenvalues plotted in the complex plane.  
	\subref{fig:speceig_comb_v6}. The eigenmode associated with eigenvalue $\lambda_6$ (marked cyan in panel~\subref{fig:chemeigs}).
	\subref{fig:chemeigs}. Scatterplot showing the sparseness and decay constant of the eigenmodes.  
	\subref{fig:results}. Sparse and slow eigenmodes of the combined network (marked red in panel~\subref{fig:chemeigs}).  
	The real parts of the eigenmodes corresponding to $\lambda_{222}, \lambda_{224}, \lambda_{225}, \lambda_{226}, 
	\lambda_{227}, \lambda_{232}, \lambda_{267}, \lambda_{268}, \lambda_{270}, \lambda_{272}, 
	\lambda_{273}$, and $\lambda_{274}$ are shown. 
	The eigenmodes are labeled with neurons that take value above a fixed absolute value threshold.
	Neurons with negative values are in red, whereas neurons with positive values are in black.
  }
\end{figure}

We proceed to find a spectral decomposition for the combined network. To avoid trivial
eigenmodes, we restrict our attention to the strongly connected component of the combined 
network containing $274$ neurons. As before, we ignore the $\tfrac{g^m_i}{g} V_i$ term and
only study the matrix $\Phi = -L^{\rm{(gap)}} + A^{T\rm{(chem)}}$.  Since $\Phi$ is not symmetric,
eigenvalues and eigenmodes may be complex-valued, occurring in complex conjugate pairs.  
Eigenvalues are plotted in the complex plane in Figure~\ref{fig:comb_eigs_plane}.

What is the meaning of complex eigenvalues? The imaginary part of an 
eigenvalue is the frequency at which the associated eigenmode oscillates. 
The real part of an eigenvalue determines
the amplitude of the oscillation as it varies with time. Eigenmodes 
that have an eigenvalue with a negative real part decay with time, whereas eigenmodes that have an eigenvalue 
with a positive real part grow with time. When examining the temporal evolution 
of the eigenmodes whose eigenvalues are shown in Figure~\ref{fig:comb_eigs_plane}, 
one should keep in mind that the ignored $\tfrac{g^m_i}{g} V_i$
term would shift the real part of the eigenvalues towards more negative values.

As for the gap junction network alone, we can look for eigenmodes
that may have functional significance.  For example, the sixth eigenmode
of the combined network, Figure~\ref{fig:speceig_comb_v6}, includes neurons
that are involved in sinusoidal body movement.  As before, one may focus on 
sparse and slow eigenmodes for ease of investigation. The distribution of 
rectilinear norm and real part of eigenvalues is shown in Figure~\ref{fig:chemeigs},  
and twelve of the sparsest and slowest eigenmodes of the combined network
are plotted in Figure~\ref{fig:results}.

Having the eigenspectrum of the combined network allows one to calculate the response
of the network to various perturbations. By decomposing sensory stimulation
among the eigenmodes and following the evolution of each eigenmode, one could
predict the worm's response to the stimulation. A similar calculation could be done
for artificial stimulation of the neuronal network, induced for example, using 
channelrhodopsin \cite{NagelSHKABOHB2003,NagelBLABG2005,Zhang_ea2007}.

\subsubsection{Interaction Between Networks}

We have measured the structural properties of the combined network formed 
by adding together the adjacency matrices of the gap junction and chemical 
synapse network, however it is unclear how they interact.  The two networks
could be independent, or their connections could overlap more or less often than by chance.

\begin{figure}[h]
  \centering
    \includegraphics[width=4in]{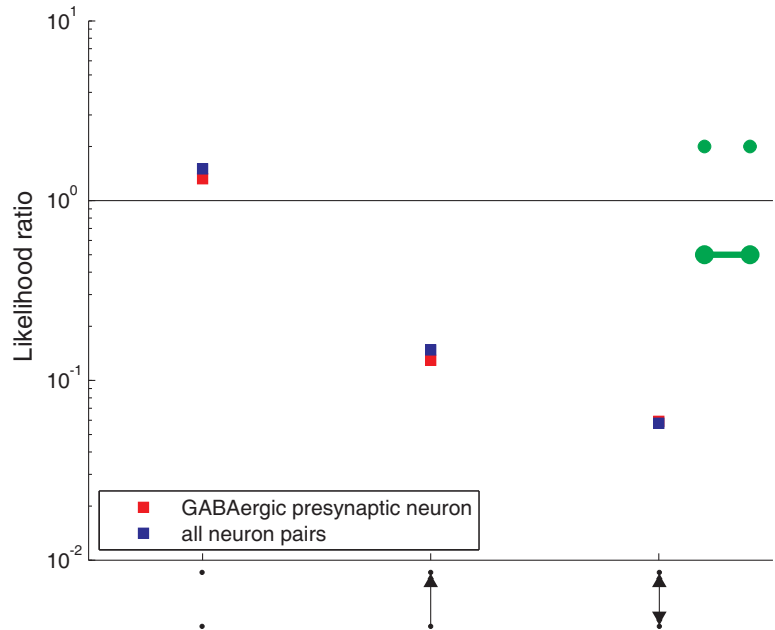}
  \caption{Likelihood ratio for chemical network doublets given the absence/presence of a gap junction between the two neurons.}
  \label{fig:likeli2}
\end{figure}

To investigate how the two networks overlap, we 
look at local structure.  Figure~\ref{fig:likeli2} shows the likelihood ratios 
of chemical synapse connections being absent, being unidirectional, and being
bidirectional given the presence or absence of a gap junction between the
same pair of neurons (see {\sc Methods}).  As can be seen, chemical synapses are 
more likely to be absent when there is no gap junction than when there is one.
Unidirectional, and especially bidirectional, chemical synapses are more 
likely when there is a gap junction between given neurons.  In this sense, 
the two networks are correlated, however it should be noted that when 
there is a gap junction, about $60$\% of the time there is no chemical 
synapse in either direction either.

Durbin had found that chemical and gap junction networks are essentially
independent when imposing physical adjacency restrictions \cite[Ch.~7]{Durbin1987},
but as noted above, there is no evidence that proximity is a limiting factor for 
connectivity.  We believe there may be a functional role for 
correlation/anticorrelation of the joint presence of gap junction and chemical
connections.

Why might the presence of connections in two networks either be correlated or 
anticorrelated?  One possibility is that correlated
connections simultaneously perform different functions \cite{GargSLZLWK2005}
whereas anticorrelation yields connections between distinct kinds of 
neuronal pairs \cite{TavoularisW,ReznickKV2004,SharmaM2008}.

What are the different functions performed by chemical synapses and gap junctions
that could lead to correlation?  One possibility is that the two different functions 
are sign-inverting and non-inverting coupling. Gap junctions are non-inverting:
higher potential in a neuron raises the potential in other gap-junction-coupled neurons.
Chemical synapses, on the other hand, may be either excitatory (non-inverting) or
inhibitory (inverting).  When the likelihood computations are repeated considering 
only neuron pairs where the presynaptic 
neuron is known to be GABAergic  \cite{McIntireJKH1993}, there is not much change, 
see Figure~\ref{fig:likeli2}.  This suggests that the primary purpose of
overlapping inhibitory chemical synapses is not to counter excitatory
gap junctions.  This result, however, is  only suggestive since the neurotransmitters
and their action on postsynaptic receptors in many neurons have not been determined.
Some other reason, such as differing temporal properties or robustness from redundancy, 
is needed to explain correlation.

Another measure of the interaction between the two networks is the correlation between 
the degree sequences.  The correlation coefficient between the gap junction degree and the chemical network 
in-degree is greater than and the correlation coefficient between the gap 
junction degree and the chemical network out-degree is less than the correlation coefficient
between the chemical network in-degree and out-degree, as shown in Table~\ref{tab:corrcoef} where
comparisons to correlation coefficients between randomly permuted degree sequences (see {\sc Methods})
are shown.  Large correlation coefficients imply that neurons are ordered in similar ways according 
to degree centrality.

The two networks seem to primarily reinforce each other with
correlated structure rather than augment each other with anticorrelated connections.

\begin{table}
\centering
\caption{Degree Sequence Correlation Coefficients}
\begin{tabular}{|r|r|r|r|r|}
\hline
& gap/in & gap/out & in/out & email \cite{NewmanFB2002} \\ \hline
correlation coefficient $\rho$ & $0.64$ & $0.44$  & $0.52$ & $0.53$ \\ \hline 
avg.~rand.~perm.\ $\rho$ & $-0.00 \pm 0.06$ & $0.00 \pm 0.06$ & $0.00 \pm 0.06$ & \\ \hline
\end{tabular}
\label{tab:corrcoef}
\end{table}

\subsection{Robustness Analysis}
Although the reported wiring diagram corrects errors in previous work and 
is considered self-consistent, one might wonder how remaining ambiguities and errors
in the wiring diagram might affect the quantitative results presented.
For network properties that are defined locally, such as degree, 
multiplicity, and subnetwork distributions, clearly small errors in 
the measured wiring diagram lead to small errors in the calculated 
properties.  For global properties such as characteristic path length and
eigenmodes, things are less clear.

To study the robustness of global network properties to errors in the 
wiring diagram, we recalculate these properties in the wiring
diagrams with simulated errors. We simulate errors by removing randomly 
chosen synaptic contacts with a certain probability and assigning them to 
a randomly chosen pair of neurons. Then, we calculate the global network
properties on the ensemble of edited wiring diagrams. The variation 
of the properties in the ensemble gives us an idea of robustness.

First, we explore the robustness of the small world properties and the giant
component calculations. We edit wiring diagrams by moving each gap junction 
contact with $10$\% probability and chemical synapse contact with $5$\% probability. 
Tables~\ref{tab:robustness_gap} and \ref{tab:robustness_chem} show the global 
properties for $1000$ random networks obtained by editing the experimentally measured 
network.  These tables suggest that our quantitative results are reasonably 
robust to ambiguities and errors in the wiring diagram.

Properties for the neuronal network from prior work in
\cite{AchacosoY1992} are also shown for comparison.  The number of synaptic
contacts that must be moved to achieve this network (editing distance) roughly
corresponds to that with $25.6$\% probability. 

\begin{figure}
  \centering
  \includegraphics[width=5in]{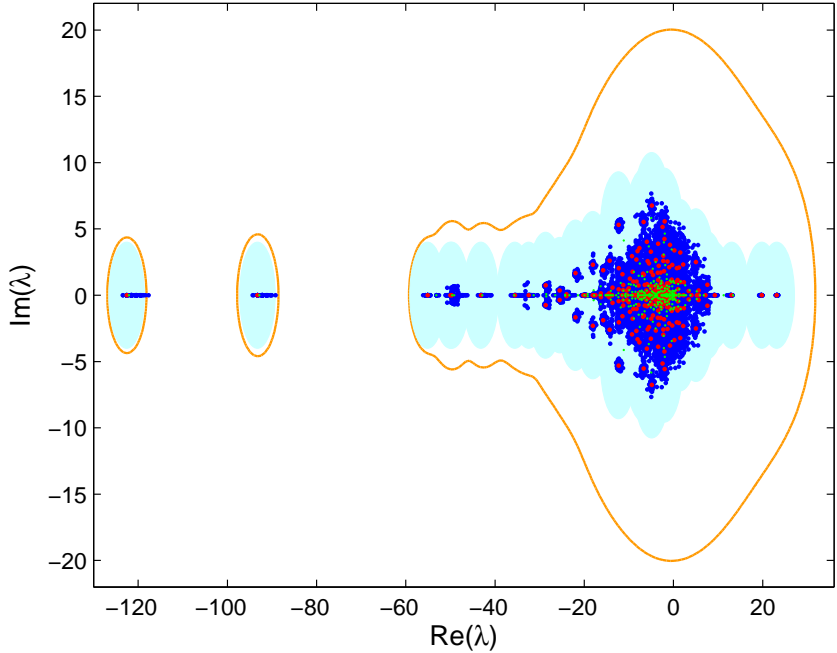}
  \caption{The spectrum of the giant component of the combined network matrix 
    $\Phi$ (red), $\epsilon$-disks around the spectrum (light blue), spectra 
    of $100$ randomly edited networks (blue), and the $\epsilon$-pseudospectrum 
    (orange). The value $\epsilon = 4$ is used (the average spectral norm of 
    the $100$ editing matrices was $3.4 \pm 0.9$).  The spectrum of the giant 
	component of the combined network matrix $\Phi$ under an alternate quantitation
	of send\_joint synapses is also shown (green).
  }
  \label{fig:pseudospectrum}
\end{figure}

Second, we characterize robustness for the linear systems analysis. Because of
greater sensitivity of the eigenvalues to errors, we edit wiring diagrams 
by moving each gap junction contact with $1$\% probability and a chemical synapse 
contact with $0.5$\% probability.  The spectra for $100$ randomly edited networks 
along with the spectrum for the measured network (Figure~\ref{fig:comb_eigs_plane}) 
are shown in Figure~\ref{fig:pseudospectrum}. Although the locations of 
eigenvalues shift in the complex plane, many of them move less than the nearest
neighbor distance and remain isolated.

In addition to considering the effect of typical random edits, we can characterize 
the effect of worst-case errors on the eigenvalues using the 
$\epsilon$-pseudospectrum \cite{Trefethen1992}, which gives the eigenvalue loci 
$\Lambda_{\epsilon}$ for all perturbations by matrices of norm $\epsilon$ 
(Figure~\ref{fig:pseudospectrum}). For the gap junction, $\Lambda_{\epsilon}(L)$ 
is simply the set of disks of radius $\epsilon$ around the eigenvalues, but 
for the chemical and combined networks, $\Lambda_{\epsilon}(A^T)$ and 
$\Lambda_{\epsilon}(\Phi)$ are larger. In the worst case scenario, most 
eigenmodes become mixed up. 

Electron micrographs of chemical synapses have a further ambiguity when more
than one postsynaptic partner receives input at a release site.  We treated such 
polyadic (send\_joint) synapses no differently than other synapses, but one might alternatively determine
multiplicity by counting such synapses at $50$\% strength.  This alternate quantitation 
clearly does not change statistics that ignore multiplicity; the change in the spectrum is depicted
in Figure~\ref{fig:pseudospectrum}.

\section*{Discussion}

We have presented a corrected and more comprehensive version of the neuronal 
wiring diagram of hermaphrodite \emph{C.\ elegans} using materials from 
White \emph{et al}.\ \cite{WhiteSTB1986} and new electron micrographs. We 
also proposed a convenient way to visualize the neuronal wiring diagram.  
The corrected wiring diagram and its visualization should help in planning
experiments, such as neuron ablation.

Next, we performed several statistical analyses of the corrected wiring, which
should help with inferring function from structure.

By using several different centrality indices, we found central neurons, which may
play a special role in information processing. In particular, command 
interneurons responsible for worm locomotion have high degree centrality
in both chemical and gap junction networks. Interestingly, command 
interneurons are also central according to in-closeness, implying that they
are in a good position to integrate signals. However, most command 
interneurons do not have highest out-closeness, meaning that other out-closeness
central neurons, such as DVA, ADEL/R, PVPR, etc., are in a good position to deliver signals 
to the rest of the network.

Linear systems analysis yielded a principled methodology to hypothesize
functional circuits and to predict the outcome of both sensory and 
artificial stimulation experiments. We have identified several modes that
map onto previously identified behaviors.

Networks with similar statistical structural properties may share functional 
properties thus providing insight into the function of the \emph{C.\ elegans} 
nervous system. To enable comparison of the \emph{C.\ elegans} network with 
other natural and technological networks \cite{BoccalettiLMCH2006}, we computed several 
structural properties of the neuronal network. In particular, the 
gap junction network, the chemical synapse network, and the combined 
neuronal network may all be classified as small world networks because 
they simultaneously have small average path lengths and large clustering 
coefficients \cite{WattsS1998}.  

The tails of the degree and terminal number distributions for the gap, chemical 
and combined networks (with the exception of the in-numbers) follow a power law 
consistent with the network being scale-free in the sense of Barab\'{a}si 
and Albert \cite{BarabasiA1999}. The tails of some distributions
can also be fit by an exponential decay, consistent with a previous
report \cite{AmaralSBS2000}. However, we found that exponential fits for 
the tails have (sometimes insignificantly) lower log-likelihoods than power 
laws, making the exponential decay a less likely alternative. For whole 
distributions, neither distribution passes the $p$-value test; if 
one is forced to choose, the exponential decay may be a less poor alternative.

Several statistical properties of the \emph{C.\ elegans} network are similar
to those of the mammalian cortex. In particular, the whole distribution of 
\emph{C.\ elegans} chemical synapse multiplicity is well-fit by a stretched 
exponential (or Weibull) distribution (Figure~\ref{fig:MDc}).   
Taking multiplicity as a proxy of synaptic connection strength, this is 
reminiscent of the synaptic strength distribution in mammalian cortex, which
was measured electrophysiologically, \cite{SongSRNC2005, VarshneySC2006}. 
The definition of stretched exponential distribution is slightly
different \cite{VarshneySC2006}, but has the same tail behavior.
The stretch factor is $\sim 0.5$, close to that in the cortical 
network. 

In addition, we found that motif frequencies in the chemical synapse network
are similar to those in the mammalian cortex \cite{SongSRNC2005}. Both reciprocally 
connected neuron pairs and triplets with a connection between every pair of neurons 
(regardless of direction) are over-represented. The similarity of the connection
strength and the motif distributions may reflect similar constraints in the two networks. 
Since proximity is unlikely to be the limiting factor, we suggest that these constraints
may reflect functionality.
We found that the chemical synapse and the gap junction networks are correlated, 
which may provide insight into their relative roles.

To conclude the paper, let us note that our scientific development was not
hypothesis-driven, but rather exploratory.  Yet we hope that 
the reported statistics will help in formulating a theory that explains how 
function arises from structure. 

\section*{Materials and Methods}

\subsection*{Data Acquisition}

We began assembling the wiring diagram by consolidating existing data from 
both published and unpublished sources.  Using J.~G.\ White \emph{et al}.'s The Mind of a Worm (MOW) \cite{WhiteSTB1986}
as the starting point, we extracted wiring data from diagrams, figures, 
tables, and text (for example, see \cite[Appendix A, pp. 118--122]{WhiteSTB1986} 
on neuron AVAL/R).  Connectivity of each neuron, its synaptic partner, and synaptic 
type (chemical, gap junction, neuromuscular) was manually entered into an electronic 
database.  In the ventral cord of the worm, this level of synaptic specification 
was complicated by the fact that connections were recorded by neuron class.  For 
example, bilateral neurons PVCL and PVCR were simply listed as PVC.  We were 
able to assign proper connections to the appropriate left/right neuron by 
referring to White and coworker's original laboratory notebooks and original electron 
micrographs.  In some cases, the number of synapses for a given neuron class in 
MOW differed from the sum of connections for the bilateral pairs in the 
notebooks and/or electron micrographs.  The synaptic value of these neurons was 
determined by appropriating the value in MOW according to proportionality from 
the notebooks and/or electron micrographs.  

From here, we incorporated R.~M.~Durbin's data, which was applicable to the anterior 
portion of the worm, reconstructed from the animal \emph{N2U} \cite{Durbin1987}.  
For neurons that projected beyond the nerve ring, only the anterior connections 
needed to be updated. Since data from MOW did not specify the location of synapses, 
integration proved difficult.  For these neurons, we obtained positional information 
by cross-referencing Durbin's data against original electron micrographs and 
his handwritten annotations in White's laboratory notebooks.  Only synapses located 
in regions addressed by Durbin were included.  Connections in the middle and tail 
regions of the worm were mostly unaffected by these updates.  

With the advent of green fluorescent protein (GFP) reporters, researchers are able to visualize the 
neuroanatomy of individual neurons.  Studies based on this technology mostly confirmed the electron micrograph 
reconstructions described in MOW.   A few differences between GFP-stained neurons 
and White's work were observed \cite{HobertHu}.  Notably, the anterior processes 
of DVB and PVT could have been mistakenly switched in MOW \cite{WhiteSTB1986}.  
Based on these findings, we reversed the connections for neurons DVB and PVT anterior 
to the vulva.

Most published works have focused in the neck and tail regions of \emph{C.\ elegans}, 
where the majority of neuron cell bodies reside. Reconstructions of neurons in the 
mid-body of the worm, on the other hand, are scant and incomplete.  From a combination 
of published works \cite{WhiteSTB1976,WhiteSTB1986, Durbin1987, HallR1991}, we found 
that wiring data for $64$ neurons had large gaps or were missing entirely.  Sixty-one 
of these are motor neurons in the ventral cord.  Two are excretory neurons (CANL/R) 
that do not appear to make any synapses.  The remaining neuron, RID, is the only 
process in the dorsal cord that extends over the length of the animal.

At the \emph{C.\ elegans} archive (Albert Einstein College of Medicine), we uncovered 
a large number of reconstruction records in White \emph{et al.}'s laboratory notebooks.  
These notebooks identified neurons by different color code labels depending on the animal, 
the location of the neurite (ventral or dorsal), and magnification of the electron micrograph.  
To ascertain the identity of the neurons, we relied on a combination of color code tables 
and comparisons of common anatomical structures between electron micrograph prints.  
In the end, we identified notes for full reconstructions of $24$ of the aforementioned 
neurons.  Partial connectivity data for the remaining $38$ were also available where $22$ 
neurons have partial/missing dorsal side connections and $6$ neurons have partial ventral 
side connections.  We checked the connections of all (both published and unpublished) 
neurons in the ventral cord against electron micrographs used by White and coworkers.  
Over $600$ updates were made to the original notes and published reconstructions.  Many 
of these updates were additions of previously missed neuromuscular junctions between 
ventral cord motor neurons and body wall muscles.  

While conducting this work, we found that a large section of the worm on the 
dorsal side, from just anterior to the vulva to the pre-anal ganglion, was never 
imaged at high power magnification with an electron microscope.  This lack of 
electron micrographs was the reason why so many neurons were missing dorsal side 
reconstructions.  Using original thin sections for the \emph{N2U} worm prepared by 
White \emph{et al.}, we produced new high power electron micrographs of this dorsal region.  
Due to the condition of the sections, only one of every $2$--$3$ sections was imaged.  
These new electron micrographs extended nearly $9$$\mu$m on the dorsal side.  New dorsal 
side data for 3 neurons (DA5, DB4, DD3) were obtained from these electron micrographs.  
Resource constraints prevented us from covering the entire dorsal gap. 

From our compilation of wiring data, including new reconstructions of ventral cord 
motor neurons, we applied self-consistency criteria to isolate neurons with 
mismatched reciprocal records.  The discrepancies were reconciled by checking against 
electron micrographs and the laboratory notebooks of White \emph{et al}.  Connections in 
the posterior region of the animal were also cross-referenced with reconstructions 
published in \cite{HallR1991}.  Reconciliation involved $561$ synapses for $108$ 
neurons ($49$\% chemical ``sends,'' $31$\% chemical ``receives,'' and $20$\% electrical 
junctions).

\subsection*{Giant Component for Random Networks}

For a random network with $N$ neurons and probability $p$ of a connection being 
present, if the constant $c  = Np > 1$, then the size of the giant component 
is asymptotically normal with mean $N\alpha (c)$ and variance $N\beta (c)$ 
\cite[p. 120]{Kolchin1998}.  These quantities are given by 
\begin{equation}
\alpha (c) = 1 - \frac{\gamma}{c} \mbox{ and } \beta (c) = \frac{\gamma (1-\tfrac{\gamma}{c})}{c(1-\gamma)^2}\mbox{,}
\end{equation}
where
\begin{equation}
\gamma = -W\left(-\frac{c}{e^c}\right)\mbox{,}
\end{equation}
and $W(\cdot)$ is the Lambert $W$-function.  If we take $N$ to be $279$ and $p$
to be $514/\binom{279}{2}$, then $c = 3.698$.  Using the asymptotic approximation, 
the size of the giant component is distributed approximately normally  
with mean $271$ and variance $9.22$.  Thus the probability of having a giant 
component of size $248$, which is over $7$ standard deviations from the mean, is 
about $10^{-14}$.  If a precise evaluation of this infinitesimal value is desired, 
large deviations techniques, rather than the asymptotic approximation may be more 
valid \cite{EngelHM2004}.

To apply this method to the weakly connected component of a directed network, we 
are interested in the undirected network formed by adding a connection between two 
neurons if there is a connection in either direction.  For a random directed network 
with probability $q$ of presence of a directed connection, the probability of a connection 
existing in either direction is $p = q^2 + 2q(1-q)$.  Taking $q$ to be $2194/279/278 = 0.0283$, 
$p$ is $0.0558$.  Then for an undirected random network with $N = 279$ and the specified $p$, 
$c$ is $15.56$.  Then the size of the giant component is distributed approximately  
normally with mean $279$ and variance $0.0000487$.  Thus the probability of 
the giant weakly connected component containing all the neurons in such a random network 
is overwhelming.  Again, large deviations techniques should be used to get a precise 
evaluation of the probability of being on the order of $10000$ standard deviations
away from the mean.

\subsection*{Giant Component for Random Networks with Given Degree Distribution}
Consider the ensemble of random networks with a given degree 
distribution \cite{NewmanSW2001}.  For the gap junction network, the generating 
function corresponding to the measured degree distribution is
\begin{align*}
G_0(x) &= \left(\tfrac{1}{279}\right)\left[26 + 39x + 59x^2 + 43x^3 + 46x^4 + 23x^5 + 15x^6 + 5x^7 + 8x^8 + 4x^9 + 3x^{11} + 2x^{14} \right. \\ \notag
&\left. + 2x^{15} + x^{24} + x^{29} + x^{34} + x^{40}\right] \mbox{,}
\end{align*}
with derivative
\begin{align*}
G_0^{\prime}(x)  &= \left(\tfrac{1}{279}\right)\left[39 + 118x + 129x^2 + 184x^3 + 115x^4 + 90x^5 + 35x^6 + 64x^7 + 36x^8 + 33x^{10}  \right. \\ \notag
&\left. + 28x^{13} + 30x^{14} + 24x^{23} + 29x^{28} + 34x^{33} + 40x^{39}\right] \mbox{.}
\end{align*}
Therefore $G_0^{\prime}(1) = \tfrac{1028}{279}$.  The generating function $G_1$ is then
\begin{align*}
G_1(x) &= \left(\tfrac{1}{1028}\right)\left[39 + 118x + 129x^2 + 184x^3 + 115x^4 + 90x^5 + 35x^6 + 64x^7 + 36x^8 + 33x^{10} + 28x^{13} \right. \\ \notag 
&\left. + 30x^{14} + 24x^{23} + 29x^{28} + 34x^{33} + 40x^{39}\right] \mbox{.}
\end{align*}
As shown in \cite{NewmanSW2001}, the expected fraction of the network 
taken up by the giant component, $S$, is $S = 1 - G_0(u)$, where $u$ is
the smallest non-negative solution to $u = G_1(u)$.  In our case, we find
$u = 0.043$, and so $S = 0.90$.  That is to say, one would
expect the giant component to consist of $251$ neurons.

Using the computed $S$ and $G_0^{\prime}(1)$, we can find
the average component size excluding the giant component, which turns out to
be $1.58$.

For the symmetrized chemical network, the generating function corresponding
to the measured degree distribution is
\begin{align*}
H_0(x) &= \left(\tfrac{1}{279}\right)\left[2x + 6x^2 + 8x^3 + 6x^4 + 14x^5 + 14x^6 + 19x^7 + 20x^8 + 19x^9 + 20x^{10} + 17x^{11} + 18x^{12}\right.\\ \notag
&\left. + 14x^{13} + 9x^{14} + 10x^{15} + 9x^{16} + 4x^{17} + 9x^{18} + 7x^{19} + 3x^{20} + 9x^{21} + 8x^{22} + 3x^{23} + 4x^{24} \right. \\ \notag
&\left. + 3x^{25} + 2x^{26} + 3x^{27} + 2x^{29} + x^{31} + x^{32} + 2x^{33} + x^{34} + x^{36} + x^{42} + x^{48} + x^{49} + 2x^{50} \right. \\ \notag
&\left. + x^{51} + x^{52} + x^{53} + x^{56} + x^{83} + x^{85}\right] \mbox{,}
\end{align*}
with derivative
\begin{align*}
H_0^{\prime}(x) &= \left(\tfrac{1}{279}\right)\left[2 + 12x + 24x^2 + 24x^3 + 70x^4 + 84x^5 + 133x^6 + 160x^7 + 171x^8 + 200x^9 + 187x^{10}  \right. \\ \notag
&\left. + 216x^{11} + 182x^{12} + 126x^{13} + 150x^{14} + 151x^{15} + 68x^{16} + 162x^{17} + 133x^{18} + 60x^{19} + 189x^{20}  \right. \\ \notag
&\left. + 176x^{21} + 69x^{22} + 96x^{23} + 75x^{24} + 52x^{25} + 81x^{26} + 58x^{28} + 31x^{30} + 32x^{31} + 66x^{32} \right. \\ \notag
&\left. + 34x^{33} + 36x^{35} + 42x^{41} + 48x^{47} + 49x^{48} + 100x^{49} + 51x^{50} + 52x^{51} + 53x^{52} + 56x^{55} + 83x^{82} \right. \\ \notag
&\left. + 85x^{84} \right] \mbox{.}
\end{align*}
Therefore $H_0^{\prime}(1) = \tfrac{3929}{279}$.  The generating function $H_1$ is then
\begin{align*}
H_1(x) &= \left(\tfrac{1}{3929}\right)\left[2 + 12x + 24x^2 + 24x^3 + 70x^4 + 84x^5 + 133x^6 + 160x^7 + 171x^8 + 200x^9 + 187x^{10} \right. \\ \notag
&\left. + 216x^{11} + 182x^{12} + 126x^{13} + 150x^{14} + 151x^{15} + 68x^{16} + 162x^{17} + 133x^{18} + 60x^{19} + 189x^{20} \right. \\ \notag
&\left. + 176x^{21} + 69x^{22} + 96x^{23} + 75x^{24} + 52x^{25} + 81x^{26} + 58x^{28} + 31x^{30} + 32x^{31} + 66x^{32}  \right. \\ \notag
&\left. + 34x^{33} + 36x^{35} + 42x^{41} + 48x^{47} + 49x^{48} + 100x^{49} + 51x^{50} + 52x^{51} + 53x^{52} + 56x^{55} + 83x^{82} \right. \\ \notag
&\left. + 85x^{84} \right] \mbox{.}
\end{align*}

The expected fraction of the network taken up by the giant component, $S$, is $S = 1 - H_0(u)$, where $u$ is
the smallest non-negative solution to $u = H_1(u)$.  Here $u$ is found to be $0.00051$, and so $S = 0.999996$.  
That is to say, one would expect the giant component to consist of $278.9990$ neurons.

\subsection*{Path Length for Random Networks with Given Degree Distribution}
Continuing from the previous subsection, we find the derivative of the
generating function $G_1$ for the gap junction network to be
\begin{align*}
G_1^{\prime}(x) &= \left(\tfrac{1}{1028}\right)\left[118 + 258x + 552x^2 + 460x^3 + 450x^4 + 210x^5 + 448x^6 + 288x^7 + 330x^{9} + 364x^{12} \right. \\ \notag
&\left. + 420x^{13} + 552x^{22} + 812x^{27} + 1122x^{32} + 1560x^{38}\right]\mbox{.}
\end{align*}
Thus $G_1^{\prime}(1) = \tfrac{1986}{257}$.  Letting $z_1 = G_0^{\prime}(1) = \tfrac{1028}{279}$
and $z_2 = G_0^{\prime}(1)G_1^{\prime}(1) = \tfrac{2648}{93}$, it is shown 
in \cite[(53)]{NewmanSW2001}, that the expected path length is
\begin{equation}
L = \frac{\ln\left[(N-1)(z_2 - z_1) + z_1^2\right] - \ln z_1^2}{\ln\left[z_2/z_1\right]} = 3.05 \mbox{.}
\end{equation}

\subsection*{Fitting Tails of Distributions}
To find functional forms of the tails of various distributions, we follow the procedure outlined in 
\cite{ClausetSN2009}.  For the candidate functional forms---say, the power law 
$p(d) \sim d^{-\gamma}$ and the exponential decay $p(d) \sim \exp(-\lambda d)$---we perform 
the following steps. First, we find the optimal parameter of the fit by maximizing 
the log-likelihood and the optimal starting point of the fit by minimizing the
Kolmogorov-Smirnov statistic. Second, we evaluate the goodness
of fit by calculating the $p$-value that the observed data was generated by the
optimized distribution using $p>0.1$ as a criterion for plausibility. Finally, if 
several distributions pass the $p$-value test we compare their log-likelihoods 
to find the most probable one. 

\subsection*{Circuits in Eigenmodes}
Let us bound the probability of finding an eigenmode that comprises a random set of neurons.  
Let $N$ be the number of neurons in the network being analyzed.  Let $K_i$ be the number of neurons
that appear strongly in the $i$th eigenmode and let $K = \max_i K_i$.  Furthermore let $M$ be the 
number of neurons in the random set, which one might endeavor to investigate as a putative functional 
circuit derived from an eigenmode.

Now go through each eigenmode and add to a list all possible unordered $M$-tuples of strong neurons. Even if all
of these are unique, the size of the list is upper-bounded by $\sum_{i=1}^N \binom{K_i}{M}$ which itself is upper-bounded 
by $N \binom{K}{M}$.  

Additionally, we can compute the number of all unordered $M$-tuples of neurons.  This number is $\binom{N}{M}$.

Thus, if a random set of neurons was selected from all possible sets of neurons, the probability $p$ that there would be
an eigenmode containing all of them is upper-bounded as
\[
p \le \frac{\sum_{i=1}^N \binom{K_i}{M}}{\binom{N}{M}} \le \frac{N \binom{K}{M}}{\binom{N}{M}} = \frac{NK!}{(K-M)!M!}\frac{M!(N-M)!}{N!} = \frac{K(K-1)\cdots(K-M+1)}{(N-1)(N-2)\cdots(N-M+1)} \le \frac{K^M}{N^{M-1}} \mbox{.}
\]

Suppose we are interested in putative functional circuits of size $M=6$ in the giant component of the gap 
junction network, which has $N = 248$ and from Figure~\ref{fig:eigs_gap} take $K = 20$.  Then even the loosest upper-bound
yields
\[
p \le \frac{K^M}{N^{M-1}} = \frac{20^6}{248^5} = 6.8 \times 10^{-5} \mbox{,}
\]
and so finding a random set of neurons in an eigenmode is unlikely.

Suppose we know $L$ functional circuits of size $M$ through molecular biology 
and want to know the probability of at least one of them appearing in the eigenmodes by chance.
By the union bound (Boole's inequality), this probability is less than $pL$.  If we take $L = 20$
and $M = 6$, the probability of a known functional circuit appearing in the eigenmodes by chance 
is less than $1.4 \times 10^{-3}$ for the giant component of the gap junction network.

\subsection*{Gap Junction--Chemical Synapse Likelihoods}
The likelihood ratios shown in Figure~\ref{fig:likeli2} are the following quantities,
empirically estimated from either all neuron pairs or pairs with a GABAergic presynaptic neuron.  
The first is 
\[
\frac{\Pr[\mbox{chem.\ absent}| \mbox{gap absent}] }{\Pr[\mbox{chem.\ absent}| \mbox{gap present}]} \mbox{,}
\]
The second is
\[
\frac{\Pr[\mbox{chem.\ unidirectional}| \mbox{gap absent}] }{\Pr[\mbox{chem.\ unidirectional}| \mbox{gap present}]} \mbox{,}
\]
and the third is
\[
\frac{\Pr[\mbox{chem.\ bidirectional}| \mbox{gap absent}] }{\Pr[\mbox{chem.\ bidirectional}| \mbox{gap present}]} \mbox{.}
\]

\subsection*{Degree Correlation Coefficients}
Table~\ref{tab:corrcoef} shows the correlation coefficients between neuron degree sequences.
The average correlation coefficients of randomly permuted degree sequences 
from $10000$ trials are also shown for comparison.  The standard deviation is also shown since
the distributions of the three randomized correlation coefficients were all nearly symmetric about zero.

\subsection*{$\epsilon$-Pseudospectrum Computation}
We used the MATLAB package EigTool \cite{Wright2002} to compute pseudospectra.

\subsection*{MATLAB Code and Data}
Note that MATLAB code for computing several network properties is available at \\
{\tt http://mit.edu/lrv/www/elegans/}.  
This collection of software may be used not only to reproduce most of the figures 
in this paper, but also for future connectomics analyses. 

The collected data is available from the WormAtlas \cite{Wormatlas} as well as from
the same website as the MATLAB code.

\section*{Acknowledgment}
We thank John White and Jonathan Hodgkin for the generous donation of the 
MRC/LMB archival documents and experimental materials to Hall's laboratory at 
AECOM, without which this study would not have been possible.
We also thank Markus Reigl for providing some of the software used in this study.  

We thank Sanjoy K.\ Mitter, Scott Emmons, Leon Avery, Mark Goldman, Cori Bargmann, 
Alexander Teplyaev, Shawn Lockery and Gonzalo de Polavieja for helpful discussions 
and for commenting on the manuscript.

\bibliographystyle{plos} 
\bibliography{elegans}

\appendices
\section{Algorithm for Directed Network Drawing}
\label{app:drawing}

To visualize a directed neuronal network we modify an approach 
suggested in \cite{CarmelHK2004,Koren2005}. In this approach, the vertical and 
the horizontal coordinates are chosen independently. The arrangement 
of neurons along the vertical axis conveys information about the directionality 
of the signal flow in the network and the arrangement of neurons along
the horizontal axis or axes conveys information about the strength of 
connectivity regardless of directionality. 

To find the vertical coordinate, $z$, we try to arrange the neurons so that
for every synaptically connected pair of neurons, the difference in $z$
between a presynaptic neuron $i$ and a postsynaptic neuron $j$
is as close to one as possible. Specifically, we minimize the following 
energy function:
\begin{equation}
E = \frac{1}{2} \sum_{i,j=1}^n W_{ij} \left(z_i-z_j-\text{sgn}(A_{ij}-A_{ji})\right)^2
\end{equation}
of the connectivity matrix $A_{ij}$, which is the sum of the gap junction 
and chemical connectivity matrices, and the symmetrized connectivity matrix 
$W_{ij}$, which satisfies $W_{ij}=(A_{ij}+A_{ji})/2$. By setting the derivative of this expression to zero, we find:
\begin{equation}
Lz=b\mbox{,}
\end{equation}
where $b_i= \sum_{j=1}^n W_{ij}  \text{sgn}(A_{ij}-A_{ji})$ and the Laplacian $L=D-W$ is 
defined in terms of a diagonal matrix $D$ that contains the number of synaptic terminals on corresponding neurons,
\begin{equation}
D_{ij}=\delta_{ij} \sum_{k=1}^n W_{ik}\mbox{.}
\end{equation}
A unique solution to this equation can be found by using the pseudoinverse.

To find the horizontal coordinates, we use the Laplacian, $L$, normalized by the number-of-terminals matrix $D$,
\begin{equation}
Q=D^{-1/2}LD^{-1/2} \mbox{.}
\end{equation}
The eigenmodes corresponding to the second and third lowest eigenvalues of $Q$ are denoted
$v_2$ and $v_3$.  Then, the horizontal coordinates are
\begin{equation}
x=D^{-1/2}v_2 \quad \mbox{and} \quad y=D^{-1/2}v_3\mbox{.}
\end{equation}
This method produces an aesthetically appealing drawing because each neuron is placed
in the weighted centroid of its neighbors. Thus strongly coupled neurons tend to be colocated.  

\section{Algebraic Form of Survival Functions}
\label{app:survival}

Here we consider several commonly encountered distributions and their
survival functions. If a 
distribution were to follow a power law, $p(d) \sim d^{-\alpha}$, 
then the survival function (under a continuous approximation) also follows a power law:
\begin{equation}
P(d) = \sum_{k=d}^{\infty} k^{-\alpha} \sim d^{-(\alpha-1)} \mbox{.}
\end{equation}
Similarly, if a distribution follows an exponential decay, 
$p(d) \sim e^{-d/\kappa}$, then the survival function also has an 
exponential decay, with the same exponent:
\begin{equation}
P(d) = \sum_{k=d}^{\infty} e^{-k/\kappa} \sim e^{-d/\kappa} \mbox{.}
\end{equation}
If a distribution were to follow the (continuous) stretched exponential 
distribution, $p(d) \sim (d/\beta)^{\gamma-1} e^{-(d/\beta)^\gamma}$, then the survival
function would have a decay given by a stretched exponential function
with the same stretch factor $\gamma$:
\begin{equation}
P(d) = \sum_{k=d}^{\infty} (k/\beta)^{\gamma-1} e^{-(k/\beta)^\gamma} \sim e^{-(d/\beta)^\gamma} \mbox{.}
\end{equation}

\section{Further Spectral Properties of the Gap Junction Network}
\label{app:eigenratio}

The spectral radius of the Laplacian plays a significant role in the performance 
of linear systems with dynamics that are slightly different from charge equilibration,
but which have been used to describe the synchronization of networks of oscillators
and the operation of distributed control systems in engineering 
\cite{BarahonaP2002, KarAM2008}.  The spectral radius is the largest eigenvalue
and is denoted by $\lambda_N$ for networks of size $N$.  For these
dynamics, the ratio of the spectral radius and the algebraic connectivity,  $\lambda_N/\lambda_2$
determines the rate of convergence of synchronization.
From Figure~\ref{fig:LSdist}, it may be computed that the eigenratio for the giant
component of the actual gap junction network is $1026$.

There is a general lower bound for the eigenratio \cite{Goldberg2006}:
\begin{equation}
\frac{\lambda_N}{\lambda_2} \ge \frac{d_{\max}+1}{d_{\min}} \mbox{,}
\end{equation}
where $d_{\max}$ is the maximum degree of all the neurons in the network.  There 
are networks which achieve this bound.  For a network with maximum degree $40$ and 
minimum degree $1$, as in the giant component of the gap junction network, 
this bound is $41$.  We see that the eigenratio for the actual gap junction 
network is $1026$, off from the optimal. 

Another quantity that often arises in discussions of signal propagation 
in networks is the magnification coefficient, $c$ \cite{Bien1989, HooryLW2006,Estrada2006, Estrada2006b}.
Networks that have large magnification coefficients transmit signals quickly.  
The magnification coefficient is difficult to compute, but can be approximated 
by the algebraic connectivity through an unexpected connection between local connectivity
properties and spectral properties.  A large algebraic connectivity implies a 
large magnification coefficient.  In particular, there are upper and lower bounds that relate 
the two \cite{Bien1989}.  
\begin{equation}
c \ge \frac{2\lambda_2}{d_{\max}} + 2\lambda_2 \mbox{,}
\end{equation}
and
\begin{equation}
\lambda_2 \ge \frac{c^2}{4} + 2c^2 \mbox{.}
\end{equation}

The algebraic connectivity may be compared to a general upper bound \cite{Fiedler1973}:
\begin{equation}
\lambda_2 \le \frac{Nd_{\min}}{N-1} \mbox{,}
\end{equation}
where $d_{\min}$ is the minimum degree of all the neurons in the network.  
There are classes of networks, called Ramanujan graphs that have constant 
degree and that can get close to the bound.  For a network with $248$ neurons, 
$511$ connections, and minimum degree $1$, as in the giant component of the gap 
junction network, the bound is $1.00$.  For 
the gap junction giant component, the algebraic connectivity is $0.12$, so we 
see that it is not very close to the upper bound, but is not too far either; 
the algebraic connectivity is reasonably large.  The main cause for deviating 
from the bound is non-constant degree distribution.  

Since the algebraic connectivity is fairly large, 
the gap junction network also has a fairly large magnification coefficient.

\section{Eigendecomposition}
\label{app:eigendecomposition}

Physical systems are often represented by
linear, constant-coefficient differential equations.
Differential equations provide an implicit specification of the system,
giving the relationship between input and output, rather than an explicit
expression for the system output as a function of the input.  After specifying
initial conditions, differential equations can be solved to find explicit
expressions for the output.

Dynamical systems that can store energy in only one form and location are called
\emph{first-order}, since the equation describing time evolution can be written 
only in terms of a single variable and its first derivative.
Storing energy is a form of short-term memory.  For a single state variable $V_i$,
a canonical first-order, linear, constant-coefficient differential 
equation is
\[
\tau \frac{d V_i(t)}{dt} + L_{ii} V_i(t) = M(t) \mbox{,}
\]
where $\tau$ and $L_{ii}$ are fixed constants and $M(t)$ is some
signal.  

The natural (unforced) response of a system corresponds
to $M(t) = 0$ and is completely determined by the system's
\emph{eigenvalue}.  In particular, solving
\[
\tau \frac{d V_i(t)}{dt} + L_{ii} V_i(t) = 0
\]
with initial condition $V_i(t=0) = V_0$, yields
\[
V_i(t) = V_0 e^{(-L_{ii}/\tau)t} \mbox{,}
\]
where $-L_{ii}/\tau$ is the eigenvalue.

The forced response occurs when some exogenous perturbation is applied to the system.
For example if a scaled step function $M_0 u(t)$ is applied, then the differential equation
\[
\tau \frac{d V_i(t)}{dt} + L_{ii} V_i(t) = M_0 u(t)
\]
with initial condition $V_i(t=0) = V_0$ has solution
\[
V_i(t) = \left\{ \frac{M_0}{L_{ii}} + \left[ V_0 - \frac{M_0}{L_{ii}}\right] e^{(-L_{ii}/\tau)t}\right\} \mbox{, } t > 0 \mbox{.}
\]

The response of a first-order system to a unit impulse is identical to
its natural response; the impulse generates the initial condition in such a short time 
that it has no other effect on the system.  That is, the system is jarred to the initial
position by the impulse.

Generally when a forcing function is applied to a linear constant-coefficient dynamic 
system, the response will consist of the superposition of
the forced response (a modification of the input signal) and 
the natural response governed by the system's eigenproperties.

Thus far, we have considered a single state variable $V_i(t)$, but in neuronal
networks we actually have a vector of states, $V(t) = \begin{bmatrix}V_1(t) & V_2(t) & \cdots & V_N(t)\end{bmatrix}^T$,
governed by a system of linear constant-coefficient differential equations.
A canonical form is
\begin{align*}
\tau \frac{d V_1(t)}{dt} + L_{11} V_1(t) + L_{12} V_2(t) + \cdots + L_{1N} V_N(t) &= M_1(t) \\
\tau \frac{d V_2(t)}{dt} + L_{21} V_1(t) + L_{22} V_2(t) + \cdots + L_{2N} V_N(t) &= M_2(t) \\
\vdots \qquad \qquad \qquad \qquad \qquad &= \quad \vdots \\
\tau \frac{d V_N(t)}{dt} + L_{N1} V_1(t) + L_{N2} V_2(t) + \cdots + L_{NN} V_N(t) &= M_N(t)
\end{align*}
which can be written in matrix-vector form as
\[
\tau \frac{d V(t)}{dt}  + L V(t) = M(t) \mbox{.}
\]

The natural response of such a system with initial condition $V(t=0) = V_0$
is the vector 
\[
V(t) = V_0 e^{(-L/\tau)t} \mbox{.}
\]
Although this is in principle the solution to the system of differential 
equations, it is difficult to examine.  Study of system behavior is complicated 
by the fact that each of the equations is coupled to the others through the off-diagonal
elements of $L$.  It would be desirable to find a new coordinate
system in which all equations are decoupled (such that the coefficient matrix is diagonal).

A vector $v$ is called an eigenmode of a matrix $L$ if it
satisfies 
\[
Lv = \lambda v
\]
for some number $\lambda$, which is called the eigenvalue.  
Decomposing the coefficient matrix into its eigendecomposition,\footnote{Note
that not all matrices have an eigendecomposition.  Instead, the Jordan decomposition
should be used for these non-diagonalizable matrices \cite{NishikawaM2006}.  The three
matrices we consider, $L$, $A^T$, and $\Phi$ are diagonalizable and so the eigendecomposition
is identical to the Jordan decomposition.  

Another decomposition that has been proposed
for use in systems neuroscience is the Schur decomposition \cite{Goldman2009}.  Since
the gap junction network is undirected, the Schur decomposition is also identical 
to the eigendecomposition.  For the chemical and combined networks, the Schur modes
may provide additional insights, but we do not consider them in this work.}
\[
L = \begin{bmatrix}v_1 &v_2 &\cdots &v_N \end{bmatrix}
\begin{bmatrix}\lambda_1 & 0 & \cdots & 0 \\ 0 & \lambda_2 & \cdots & 0 \\ \vdots & \vdots & \ddots & \vdots \\ 0 & 0 & \cdots & \lambda_N \end{bmatrix}
\begin{bmatrix}v_1 & v_2 & \cdots & v_N \end{bmatrix}^{-1}
\]
allows us to write the natural response as
\[
V(t) = \sum_{i=1}^N v_i \alpha_i e^{(-\lambda_i/\tau) t} \mbox{,}
\]
where $\alpha_i$ is the projection of the initial condition vector $V_0$ onto $v_i$.

The essential idea of the eigenmode decomposition is that the natural response of the system
can be viewed as the superposition of a number of distinct types of dynamics---the eigenmodes---each
one associated with a particular natural frequency of the system.  The natural
frequencies, $-\lambda_i/\tau$, of the system are determined by the eigenvalues $\lambda_i$ of $L$.
Each mode involves excitation of one and only one natural frequency of the system.  

If an eigenmode is real, then the dynamics associated with the solution can be described 
by a straight line in state space.  The system moves in the direction of the eigenmode.  For
example, moving in the direction of the eigenmode $\begin{bmatrix}+1 & -1 & 0 & 0 & 0 & \cdots & 0\end{bmatrix}^T$ would equalize 
the values of $V_1$ and $V_2$ but not affect $V_3,\ldots,V_N$.  A more complicated eigenmode
would involve all state variables that are non-zero.

Beyond their simple geometric interpretation in state space,
the eigenmodes also have a simple representation as time functions, since
each one involves a single exponential rather than a mixture of several
exponentials with different exponents.  The exponent $-\lambda_i/\tau$
determines how quickly the system response in the direction of eigenmode
$v_i$ decays.  For fixed $\tau$, the larger the eigenvalue $\lambda_i$,
the more quickly the eigenmode decays. 

The forced response of a network proceeds in the same way as the forced
response of a scalar system.  Further details on linear system analysis 
with eigenmodes can be found, e.g.~in the textbooks \cite{EdwardsP2000,Stern1965}.

\newpage

\section*{Supplemental Material}
\label{app:sup}
\renewcommand{\thefigure}{S\arabic{figure}}
\setcounter{figure}{0}
\renewcommand{\thetable}{S\arabic{table}}
\setcounter{table}{0}

\begin{table}[h]
  \caption{Connected components of the gap junction network.  Note the single giant 
	component and the large number of disconnected/isolated neurons.}
  \label{tab:3}
  \begin{tabular}{l}
  \bf{Giant Component (248 neurons)}
  \end{tabular} \\
  \begin{tabular}{lllllllll}
  \hline
  ADAL/R & ALNL & AVG & DD01-05 & PDA & PVR & RIVL/R & SABVL/R & URYVL/R \\ 
  ADEL/R & AQR & AVHL/R & DVA & PDB & PVT & RMDDL/R & SDQL/R & VA01-12 \\
  ADFL/R & AS01-11 & AVJL/R & DVB & PDEL/R & PVWL/R & RMDL/R & SIADL/R & VB01-11 \\
  ADLL/R & ASGL/R & AVKL/R & DVC & PHAL/R & RIBL/R & RMDVL/R & SIAVL/R & VC01-05 \\
  AFDL/R & ASHL/R & AVL & FLPL/R & PHBL/R & RICL/R & RMED & SIBDL/R & VD01-10,13 \\
  AIAL/R & ASIL/R & AVM & IL1DL/R & PHVL/R & RID & RMEL/R & SIBVL/R \\	
  AIBL/R & ASKL/R & AWAL/R & IL1L/R & PLML/R & RIFL/R & RMEV & SMBDL/R \\	
  AIML & AUAL/R & AWBL/R & IL1VL/R & PQR & RIGL/R & RMFL & SMBVL/R \\	
  AINL/R & AVAL/R & BAGL/R & IL2L/R & PVCL/R & RIH & RMGL/R & SMDDL/R \\	
  AIYL/R & AVBL/R & CEPDL/R & LUAL/R & PVM & RIML/R & RMHL/R & SMDVL/R \\	
  AIZL/R & AVDL/R	& CEPVL/R & OLLL/R & PVNL & RIPL/R & SAADL/R & URBL/R \\	
  ALA & AVEL/R & DA01-09 & OLQDL/R & PVPL/R & RIR & SAAVL/R & URXL/R \\	
  ALML/R & AVFL/R & DB01-07 & OLQVL/R & PVQL/R & RIS & SABD & URYDL/R \\	
  \end{tabular}
  \begin{tabular}{l}
  \\
  \bf{First Small Component (2 neurons)}
  \end{tabular} \\
  \begin{tabular}{lllllllll}
  \hline
  {ASJL/R} & \hspace{7mm} &&&&&&& \\
  \end{tabular} \\
  \begin{tabular}{l}
  \\
  \bf{Second Small Component (3 neurons)} 
  \end{tabular} \\
  \begin{tabular}{lllllllll}
  \hline
  HSNL/R & PVNR &&&&&&& \\
  \end{tabular}

  \begin{tabular}{l}
  \\
  \bf{Neurons with no gap junctions (26 neurons)} 
  \end{tabular} \\
  \begin{tabular}{lllllllll}
  \hline
  AIMR & ASEL/R & BDUL/R & IL2DL/R & PLNL/R & RIAL/R & URADL/R & VD11-12 & \\
  ALNR & AWCL/R & DD06   & IL2VL/R & PVDL/R & RMFR   & URAVL/R &         & \\   \end{tabular}
\end{table} 

\begin{table}[h]
  \centering
  \caption{(A) Number of gap junction contacts between different neuron categories.  (B) Percent of 
	gap junctions on neurons of the row category that connect to neurons of the column 
	category.
  }
  \begin{tabular}{|l|l|l|l|}
  \hline
  \bf{A} & Sensory & Inter- & Motor \\ \hline
  Sensory & $108$ & $119$ & $26$ \\ \hline
  Inter- & $119$ & $368$ & $342$ \\ \hline
  Motor & $26$ & $342$ & $324$ \\ \hline
  \end{tabular}
  \hspace{8mm}
  \begin{tabular}{|l|l|l|l|}
  \hline
  \bf{B} & Sensory & Inter- & Motor \\ \hline
  Sensory & $42.7$\% & $47.0$\% & $10.3$\% \\ \hline
  Inter- & $14.4$\% & $44.4$\% & $41.3$\% \\ \hline
  Motor & $3.8$\% & $49.4$\% & $46.8$\% \\ \hline
  \end{tabular}
  \label{tab:1}
\end{table}

\begin{figure}[h]
  \centering
\subfigure[]{
	\includegraphics[width=2in]{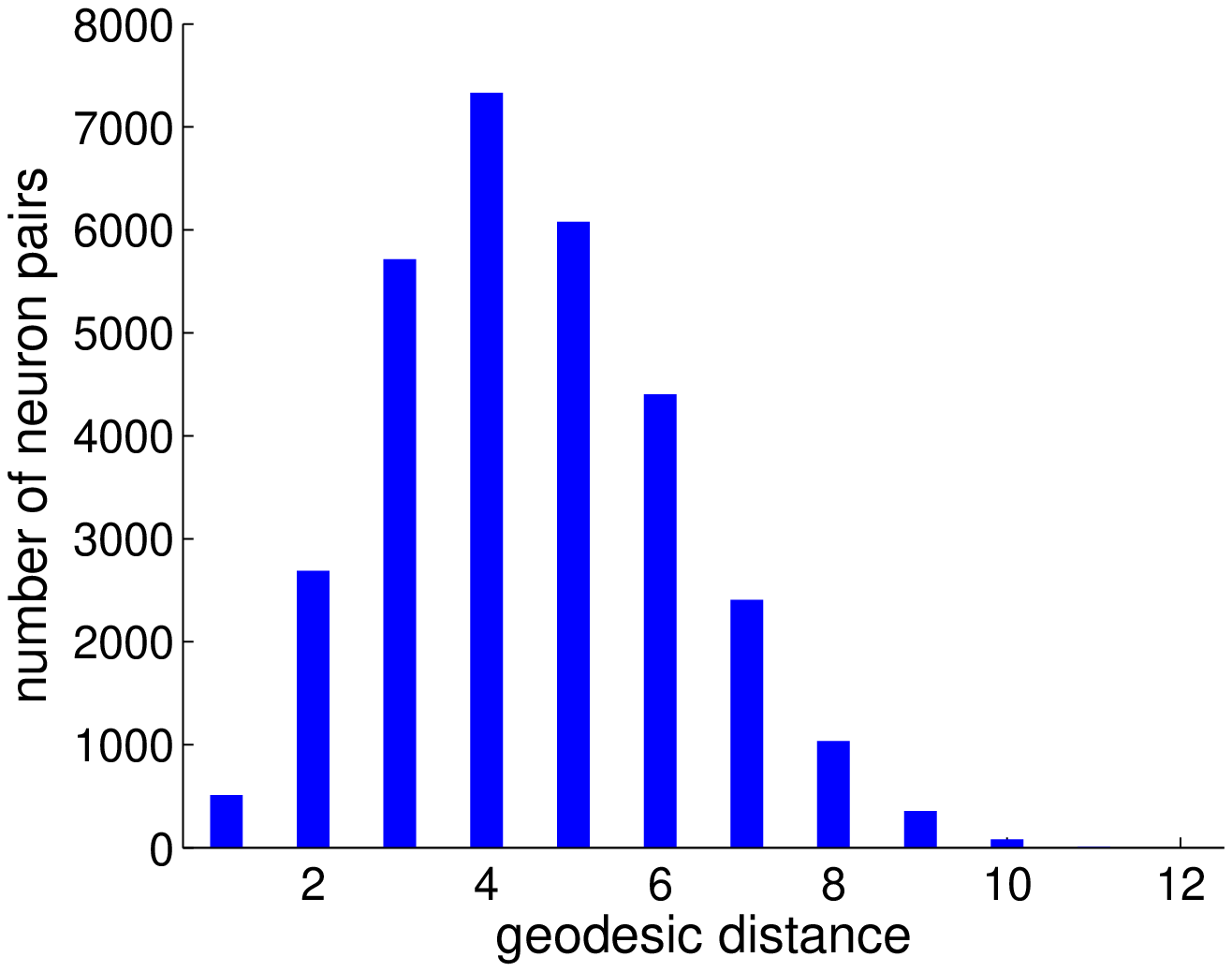} 
	\label{fig:6}
	}
\subfigure[]{
	\includegraphics[width=2in]{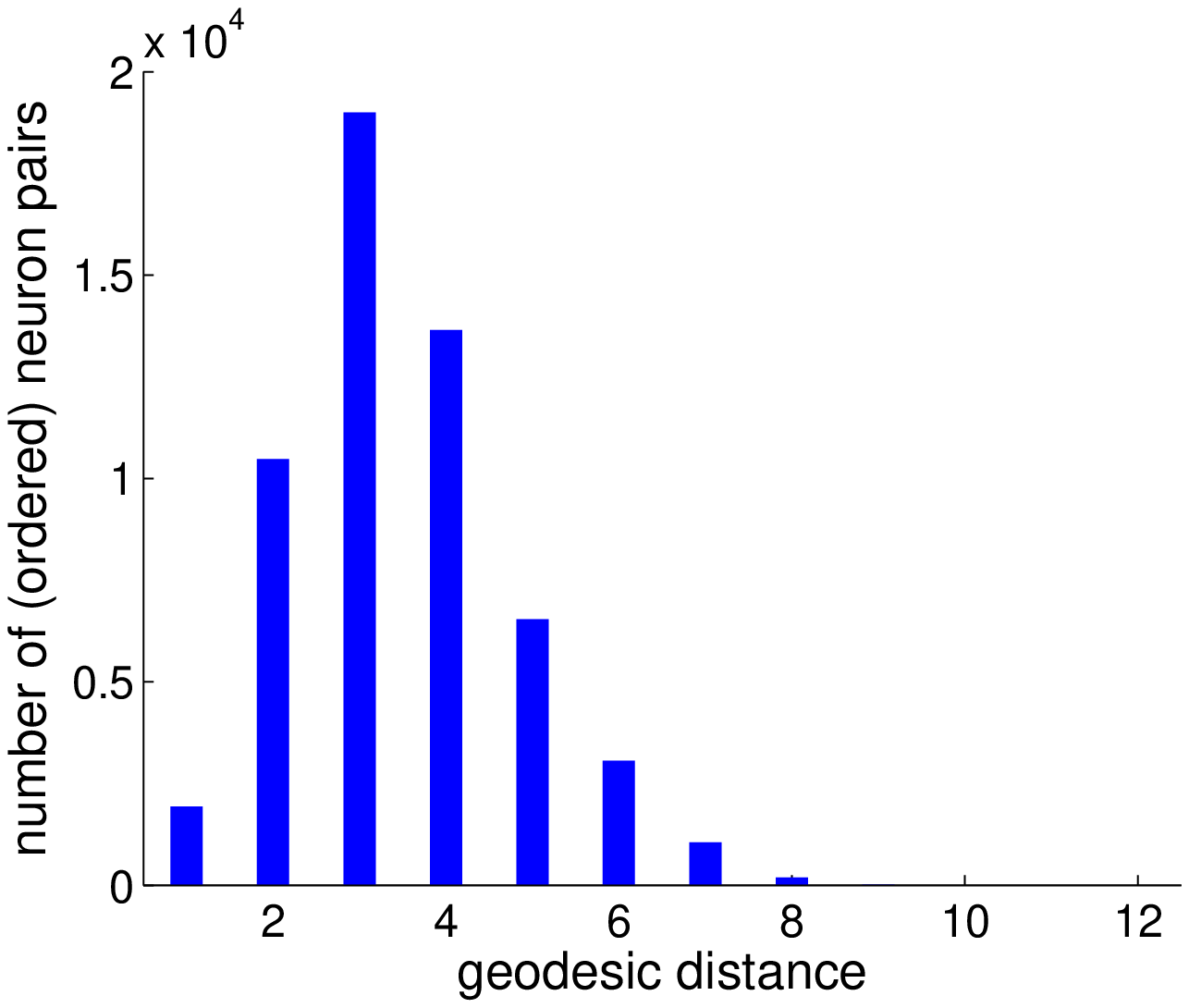} 
	\label{fig:gdDc}
	}
\subfigure[]{
	\includegraphics[width=2in]{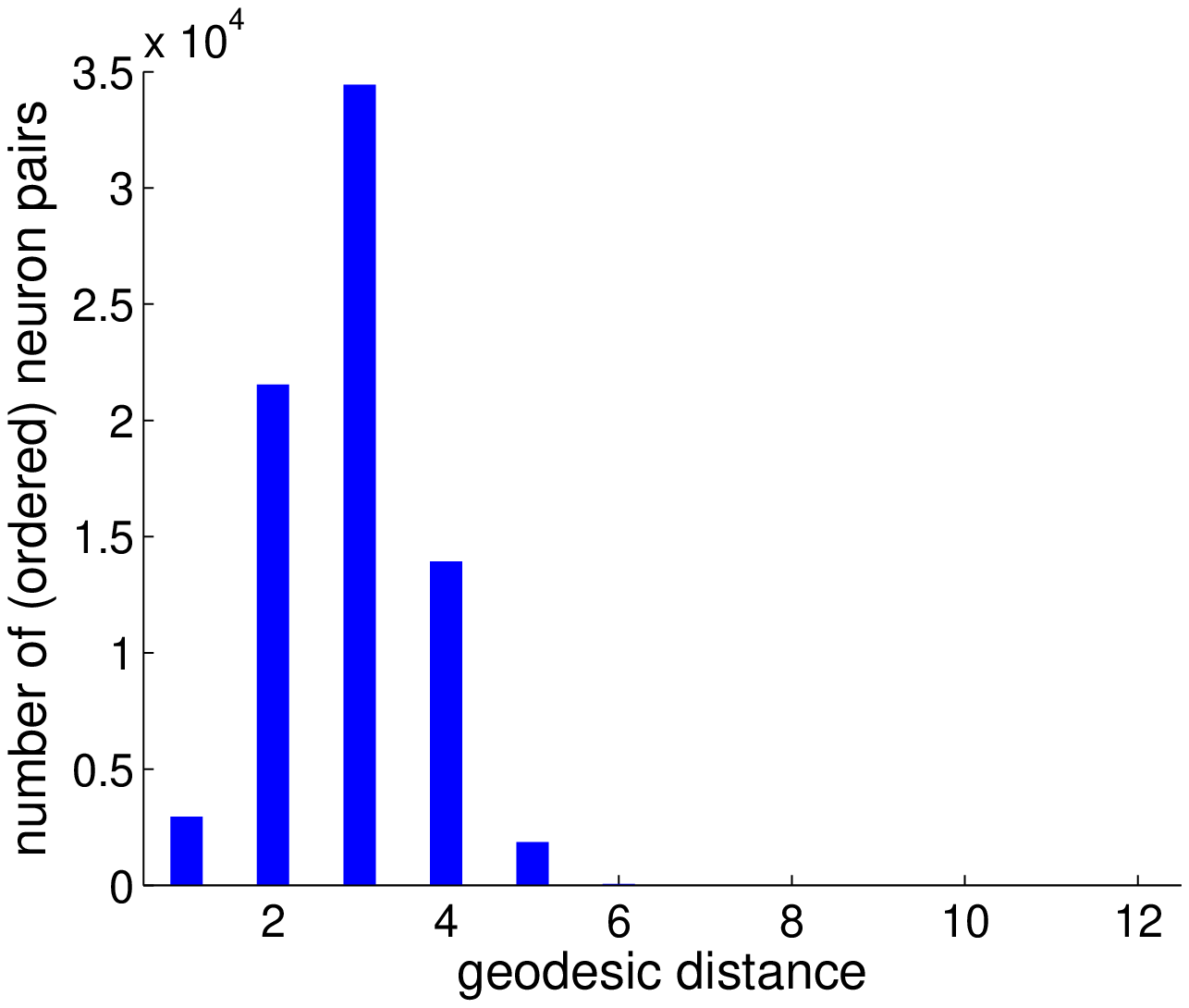}
	\label{fig:geoDb}
	}
  \caption{ Geodesic distance distributions. \subref{fig:6}. Giant component of gap junction network.
	\subref{fig:gdDc}.  Giant component of chemical network.
	\subref{fig:geoDb}.  Giant component of combined network.
}
\end{figure}

\begin{table}
  \centering
  \caption{Comparison of clustering coefficient and characteristic path length 
	of the giant component of the \emph{C.\ elegans} gap junction network and 
	several other networks that have been classified as small world networks.  
	The clustering coefficient of an equivalent Erd\"{o}s-R\'{e}nyi random 
	network is indicated in parentheses.  This is calculated using the Watts 
	and Strogatz approximations to $L$ and $C$ by finding 
	$C_r \approx \tfrac{1}{N}\exp(\tfrac{\ln(N)}{L})$.    
  }
  \begin{tabular}{|l|l|l|l|}
  \hline
  Network & $N$ & $C$ ($C_r$) & $L$ \\ \hline
  Giant component of gap junction network & $248$ & $0.21$ ($0.014$) & $4.52$ \\ \hline
  Analog electronic circuit \cite{CanchoJS2001} & $329$ & $0.34$ ($0.019$) & $3.17$ \\ \hline
  Class dependency graph of Java computer language \cite{ValverdeCS2002} & $1376$ & $0.06$ ($0.002$) & $6.39$ \\ \hline
  Film Actors \cite{WattsS1998} & $225226$ & $0.79$ ($0.00013$) & $3.65$ \\ \hline
  Power Grid \cite{WattsS1998} & $4941$ & $0.080$ ($0.00032$) & $18.7$ \\ \hline
  \end{tabular}

  \label{tab:SMN}
\end{table} 

\begin{sidewaysfigure}
  \centering
  \includegraphics[width=9in]{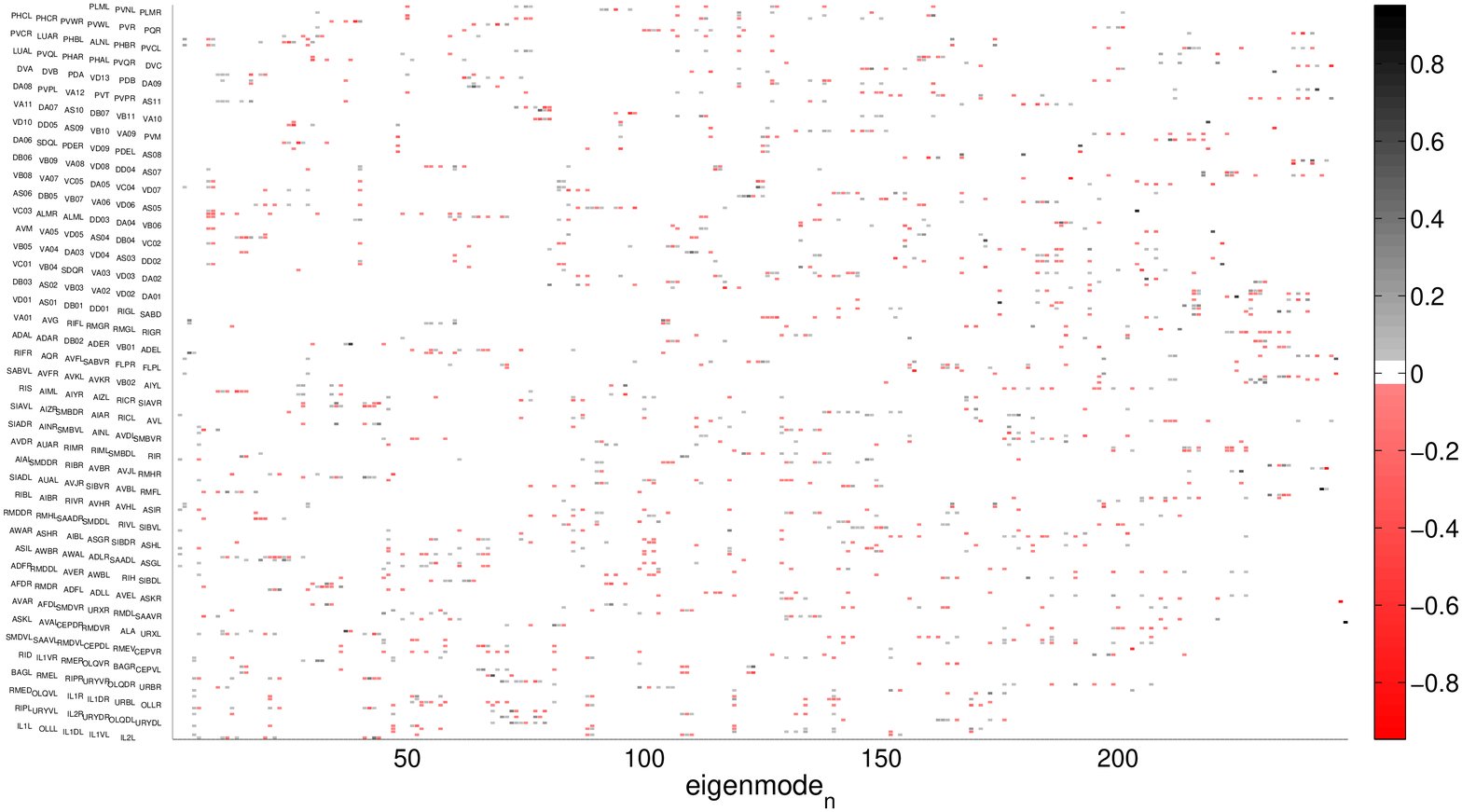}
  \caption{Eigenmodes of Laplacian for giant component of gap junction network.
	}
  \label{fig:eigs_gap}
\end{sidewaysfigure}

\begin{figure}
  \centering
  \includegraphics[width=7in]{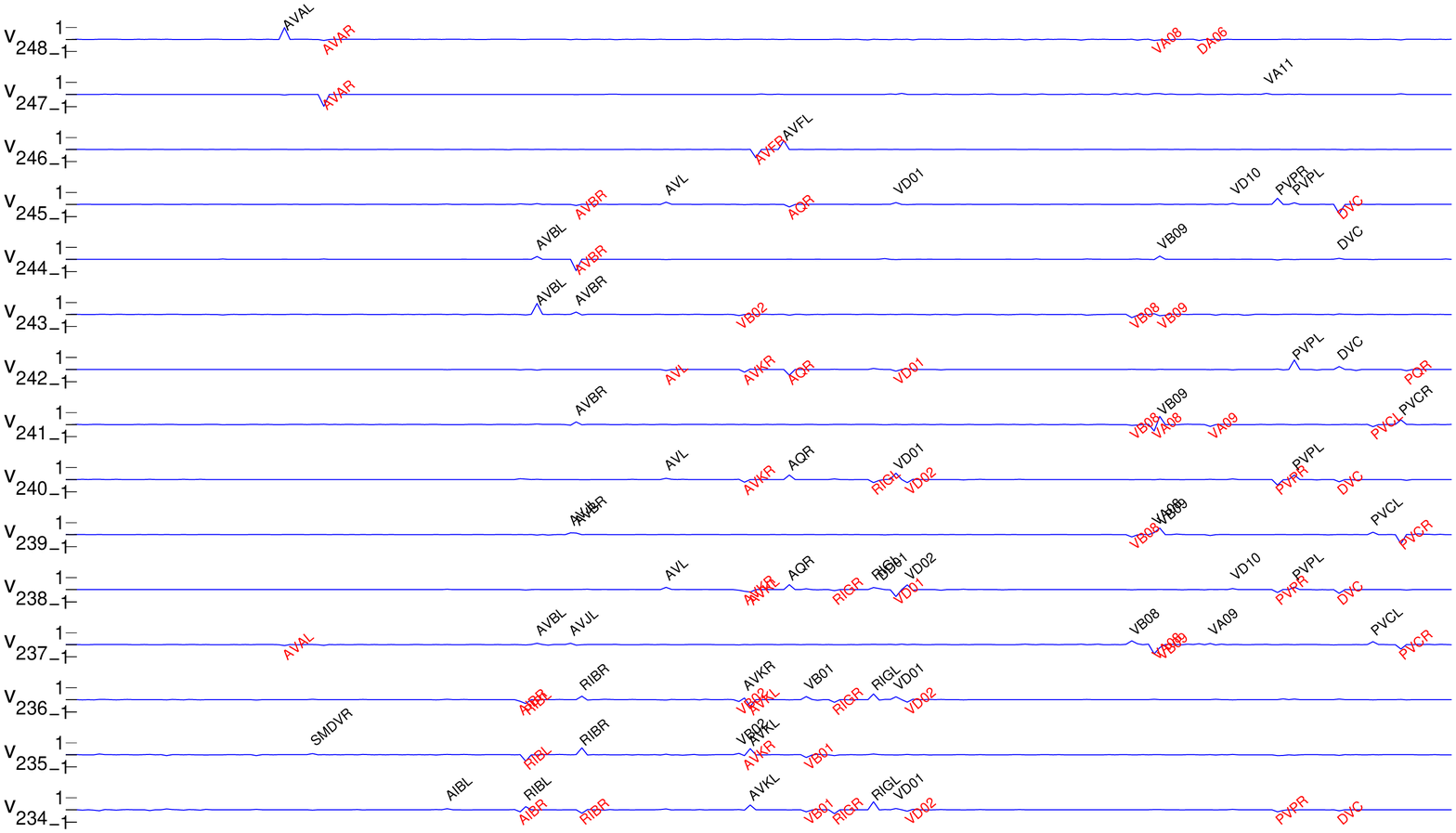}
  \caption{Fastest eigenmodes of Laplacian for giant component of gap junction network.  Eigenmodes
		corresponding to $\lambda_{248}, \lambda_{247},\ldots,\lambda_{234}$ are shown.  The eigenmodes
		are labeled with neurons that take value above a fixed absolute value threshold.  Neurons
		with negative values are in red, whereas neurons with positive values are in black.
	}
  \label{fig:fasteigs_gap}
\end{figure}

\begin{table}
  \centering
  \caption{(A) Number of chemical synapse contacts from row category to column category.  (B) 
	Percent of synapses in row category that synapse to column category.  
  }
  \begin{tabular}{|l|l|l|l|}
  \hline
  \bf{A} & Sensory & Inter- & Motor \\ \hline
  Sensory & $474$ & $1434$ & $353$ \\ \hline
  Inter- & $208$ & $1359$ & $929$ \\ \hline
  Motor & $30$ & $275$ & $1332$ \\ \hline
  \end{tabular}
  \hspace{8mm}
  \begin{tabular}{|l|l|l|l|}
  \hline
  \bf{B} & Sensory & Inter- & Motor \\ \hline
  Sensory & $21.0$\% & $63.4$\% & $15.6$\% \\ \hline
  Inter- & $8.3$\% & $54.5$\% & $37.2$\% \\ \hline
  Motor & $1.8$\% & $16.8$\% & $81.4$\% \\ \hline
  \end{tabular}
  \label{tab:2}
\end{table}

\begin{table}
  \caption{Strongly connected components of the chemical network.  Note the 
	single giant component and the large number of isolated neurons.
  }
  \begin{tabular}{l}
  \bf{Giant Component (237 neurons)}
  \end{tabular} \\
  \begin{tabular}{lllllllll}
  \hline
  ADAL/R & ALNL/R & AVFL/R & CEPVL/R & LUAL/R & PVM & RIH & RMHL/R & URYDL/R \\
  ADEL/R & AQR & AVG & DA01-06,09 & OLLL/R & PVNL/R & RIML/R & SAADL/R & URYVL/R \\
  ADFL/R & AS01-06,09,11 & AVHL/R & DB01-04,07 & OLQDL/R & PVPL/R & RIPL/R & SAAVL/R & VA01-06,08-09,11-12 \\
  ADLL/R & ASEL/R & AVJL/R & DD01-02,05 & OLQVL/R & PVQL/R & RIR & SABD & VB01-06,08-11 \\
  AFDL/R & ASGL/R & AVKL/R & DVA & PDA/B & PVR & RIS & SDQL & VC01-05 \\
  AIAL/R & ASHL/R & AVL & DVC & PDEL/R & PVT & RIVL/R & SMBDL/R & VD01-03,05-06,08,10-13 \\
  AIBL/R & ASJL/R & AVM & FLPL/R & PHAL/R & PVWL/R & RMDDR & SMBVL/R \\	
  AIML/R & ASKL/R & AWAL/R & HSNL/R & PHBL/R & RIAL/R & RMDL/R & SMDDL/R \\	
  AINR & AUAL/R & AWBL/R & IL1DL/R & PLMR & RIBL/R & RMDVL & SMDVL/R \\	
  AIYL/R & AVAL/R & AWCL/R & IL1L/R & PLNL & RICL/R & RMED & URADL/R \\	
  AIZL/R & AVBL/R & BAGL/R & IL1VL/R & PQR & RID & RMEV & URAVL/R \\	
  ALA & AVDL/R & BDUL/R & IL2L/R & PVCL/R & RIFL/R & RMFL/R & URBL/R \\	
  ALML/R & AVEL/R & CEPDL/R & IL2VL/R & PVDL & RIGL/R & RMGL/R & URXL/R \\
  \end{tabular}
  \begin{tabular}{l}
  \\
\bf{Small Component (2 neurons)}
  \end{tabular} \\
  \begin{tabular}{lllllllll}
  \hline
  RMDVR & RMDDL &&&&&&& \\
  \end{tabular} \\
\begin{tabular}{l}
\\
  \bf{Isolated neurons in chemical network (40 neurons)} 
  \end{tabular} \\
  \begin{tabular}{lllllllll}
  \hline
  AINL & DA07-08 & DVB & PLML & RMEL/R & SDQR & SIAVL/R & SIBVL/R & VB07 \\
  ASIL/R & DB05-06 & IL2DL/R & PLNR & SABVL/R & SIADL/R & SIBDL/R & VA07,10 & VD04,07,09 \\
  AS07,08,10 & DD03-04,06 & PHCL/R & PVDR
  \end{tabular}
  \label{tab:NSCCc}
\end{table}

\begin{figure}[h]
  \centering
\subfigure[]{
  \includegraphics[width=3in]{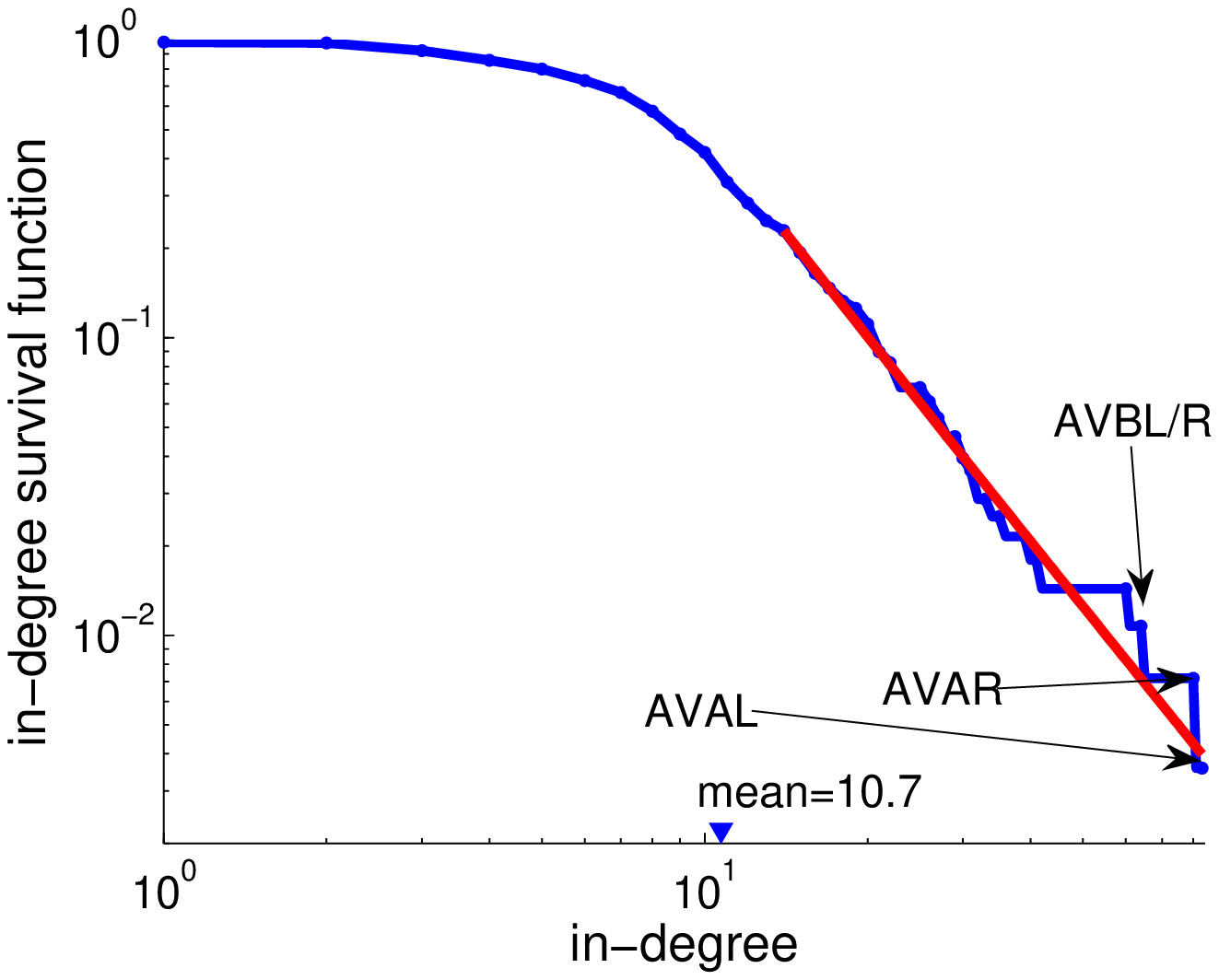}
  \label{fig:inDb}
	}
\subfigure[]{
  \includegraphics[width=3in]{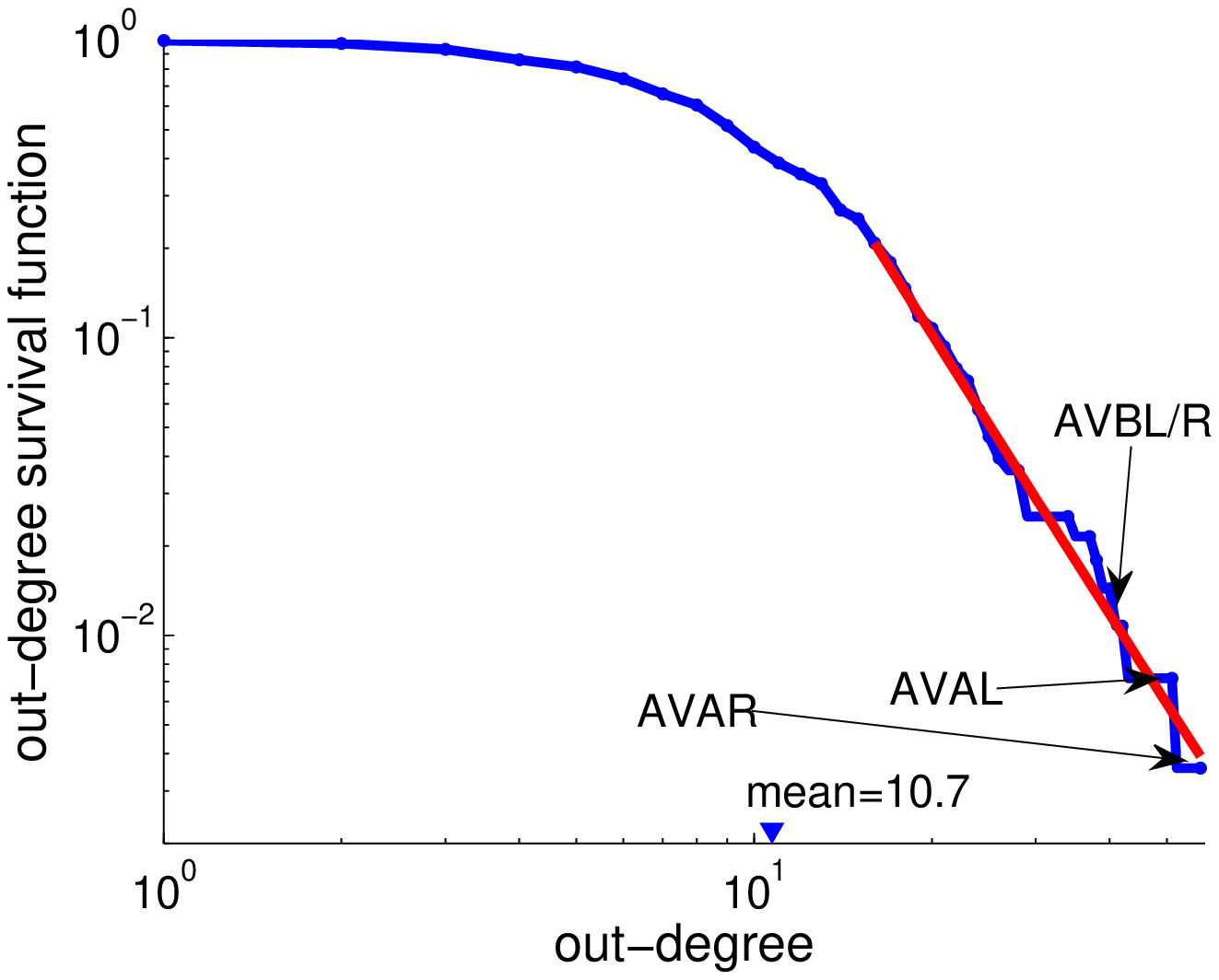}
  \label{fig:outDb}
	}
  \caption{ Survival functions of the in-degree \subref{fig:inDb} and 
    out-degree \subref{fig:outDb} distributions in the combined network.
    The tails of the distributions can be fit with power laws. 
}
\end{figure}

\begin{table}[h]
  \centering
  \caption{Some structural properties of the \emph{C.\ elegans} gap junction network, 
  randomly edited networks (E$_{\rm{gap}}$), and the AY network \cite{AchacosoY1992}. 
  }
  \begin{tabular}{|l|l|l|l|}
  \hline
   & \emph{C.\ elegans} & AY's \emph{C.\ elegans} \cite{AchacosoY1992} & E$_{\rm{gap}}$\\ \hline
  $d_{\text{edit}}$ & --- & $454$ & $177 \pm 18.5$  \\ \hline
  giant component~size & $248$ & $253$ & $261 \pm 3.41$ \\ \hline
  giant component~pathlength & $4.52$ & $4.71$ & $4.09 \pm 0.078$ \\ \hline
  giant component~clust.~coef. & $0.21$ & $0.23$ & $0.14 \pm 0.011$ \\ \hline
  \end{tabular}
  \label{tab:robustness_gap}
\end{table}

\begin{table}
  \centering
  \caption{Some structural properties of the \emph{C.\ elegans} chemical network, 
  randomly edited networks (E$_{\rm{chem}}$), and the AY network \cite{AchacosoY1992}. 
  }
  \begin{tabular}{|l|l|l|l|}
  \hline
   & \emph{C.\ elegans} & AY's \emph{C.\ elegans} \cite{AchacosoY1992} & E$_{\rm{chem}}$ \\ \hline
  $d_{\text{edit}}$ & --- & $3546$ & $638 \pm 33.2$  \\ \hline
  weak giant component~size & $279$ & $279$ & $279 \pm 0.07$  \\ \hline
  strong giant component~size & $237$ & $239$ & $267 \pm 3.19$  \\ \hline
  strong giant component~pathlength & $3.48$ & $3.99$ & $3.12 \pm 0.028$  \\ \hline
  strong giant component~clust.~coef. & $0.22$ & $0.20$ & $0.16 \pm 0.006$ \\ \hline
  \end{tabular}
  \label{tab:robustness_chem}
\end{table}

\end{document}